\documentclass[3p,10pt]{elsarticle}
\usepackage{amssymb}
\usepackage{textcomp,gensymb} 
\usepackage[labelfont=bf]{caption} 
\usepackage{float} 
\usepackage{easyReview}
\usepackage{subcaption}
\usepackage{amsthm}
\usepackage{amsmath}
\usepackage{natbib}
\biboptions{sort&compress}
\usepackage{hyperref}
\usepackage{cleveref}
\usepackage[T1]{fontenc}

\graphicspath{{figures/}}

\makeatletter
\def\ps@pprintTitle{%
    \let\@oddhead\@empty
    \let\@evenhead\@empty
    \def\@oddfoot{\footnotesize\itshape
         {} \hfill {}}%
    \let\@evenfoot\@oddfoot
    }
\makeatother

\begin{document}
\newcommand{\reviewerO}[1]{{{#1}}}
\newcommand{\reviewerT}[1]{{{#1}}}
\begin{frontmatter}

\title{Correlations for aerodynamic force coefficients of non-spherical particles in compressible flows}

\author[affil1]{Christian Gorges}
\author[affil1]{Victor Ch\'eron}
\author[affil1]{Anjali Chopra}
\author[affil2]{Fabian Denner}
\author[affil1]{Berend van Wachem\corref{cor1}}
\ead{berend.vanwachem@multiflow.org}
\address[affil1]{Chair of Mechanical Process Engineering,
Otto-von-Guericke-Universit\"{a}t Magdeburg,\\ Universit\"atsplatz 2, 39106 Magdeburg, Germany}
\address[affil2]{Department of Mechanical Engineering, Polytechnique Montr\'eal,\\ Montr\'eal, H3T 1J4, QC, Canada}

\cortext[cor1]{Corresponding author}

\begin{abstract}
This study presents particle-resolved direct numerical simulations using three-dimensional body-fitted hexahedral meshes to investigate the aerodynamic force and torque coefficients of non-spherical particles in compressible flows. The simulations focus on three particle shapes: a prolate spheroid, an oblate spheroid, and a rod-like particle, across a range of Mach numbers (0.3 to 2.0), angles of attack (0\degree to 90\degree), and particle Reynolds numbers (100 to 300). Results show that the particle shape significantly impacts the aerodynamic forces on a particle in a compressible flow, with oblate spheroids exhibiting the highest drag, lift, and torque values. Correlations for these aerodynamic coefficients of the particles in a compressible flow are developed and validated. These correlations advance multiphase flow modeling by improving the accuracy of point-particle simulations for non-spherical particles in compressible flows.
\end{abstract}

\end{frontmatter}

\section{Introduction}
\label{intro}

Particle-laden flows are of significant scientific interest due to their ubiquity and application in many industrial processes. In nature, examples of particle-laden flows are dust storms, volcanic eruptions, sediments in water bodies, eroding soils, the dispersion of aerosols, and blood flow. They are also applied in many modern technologies in industries such as aerospace, medicine and engineering. Some examples are solid-propellant rocket combustion \cite{Carlson1964}, transdermal injections \cite{Hogan2015}, inhalers \cite{Capecelatro2022}, and spray coating \cite{Villafuerte2015}. Such flows involve complex particle-fluid and particle-particle interactions at varying time and length scales, which makes it essential to understand the flow physics in order to design systems that are efficient and cost-effective.

\reviewerO{Particle}-resolved direct numerical simulations (PR-DNS) of particle-laden flows require tremendous computational resources. However, a point-particle approach can significantly reduce the computational power required to model particle-laden flows. This approach requires force models to determine the forces exerted on the particles by the surrounding fluid, as well as by contact with other particles \cite{Wachem2022}. In this context, several force models have been developed which are applicable for various flow regimes and volume fractions \cite{Tenneti2011,Zastawny2012,Loth2008,Loth2021,Sanjeevi2018,Sanjeevi2022,Clift1971,Henderson1976,Parmar2010,Cheron2024}. In order to derive these correlations, it is essential to first understand the behaviour of an isolated particle and the forces experienced by the particle in various flow regimes. These force correlations can then be extended to multi-particle systems.

Forces acting on an isolated particle depend on several factors, such as the particle Reynolds number, the Mach number, the fluid properties, the particle shape, and angle of attack between the mean fluid flow and the axis of the particle. Predicting the forces that act on a particle in a fluid flow and predicting its behaviour have been the subject of studies for decades \cite{Stokes1851,Wadell1934,Henderson1976,Rosendahl2000,Holzer2008,Mando2010,Tenneti2011,Nagata2016,Nagata2018,Nagata2020,vanWachem2015}. One of the first analytical models to predict the drag of a spherical particle for particle Reynolds numbers in the Stokes regime ($\mathrm{Re}_\mathrm{p} \ll 1$) was given by \citet{Stokes1851} as $C_\mathrm{D} = 24/\mathrm{Re_p}$ where $C_\mathrm{D}$ is the drag coefficient. This model is the basis for later proposed drag models \cite{Kaskas1964,Carlson1964,Happel1981,Salman1988}, applicable to various flow regimes and practical applications.

Since particles are not limited to spherical shapes, the proposed models also included non-spherical particle shapes. Some of the most notable correlations for non-spherical particle shapes were those presented by \citet{Zastawny2012}, \citet{Ouchene2015, Ouchene2016}, \citet{Sanjeevi2018,Sanjeevi2022}, and \citet{Cheron2024,Cheron2024b}. \citet{Zastawny2012} proposed shape-specific correlations for prolate and oblate spheroids, as well as rod-like particles using PR-DNS. They are the most widely used correlations for non-spherical particles in incompressible flows. A detailed analysis of various correlations for prolate spheroids has been performed by \citet{Ouchene2015}. They conclude that, while the correlation of \citet{Zastawny2012} predict the aerodynamic coefficients for the ellipsoidal particles accurately, the correlations of \citet{Zastawny2012} only consider ellipsoidal particles with two different aspect ratios. To address this, \citet{Ouchene2016} propose new correlations for drag, lift and  torque, which are valid for aspect ratios ranging from $1-32$ of prolate particles. Another recent study on non-spherical particles has been conducted by \citet{Sanjeevi2018} using the Lattice-Boltzmann method. The shape-specific correlations proposed by them are applicable for particle Reynolds numbers $\mathrm{Re_p}$ ranging from the Stokes limit to 2000, and depend on the particle Reynolds number and the angle of attack of the particle. They extended the correlation to include prolate particles with different aspect ratios, the upper limit of the aspect ratio being 32 \cite{Sanjeevi2022}. More recently, shape-dependent correlations have been extended to accurate correlations to predict the interaction forces between non-spherical particles and locally uniform and non-uniform fluid flow, including the aspect ratio \cite{Frohlich2020,Ouchene2020,Sanjeevi2022,Cheron2024}.

Drag predictions for non-spherical particles in the incompressible regime have received significant attention in recent years, and the proposed correlations are becoming more and more accurate by considering locally non-uniform fluid flow due to shear effects \cite{Dabade2016,Cheron2024} and wall effects \cite{Fillingham2021,Bhagat2022,Cheron2024b}. However, correlations for compressible particle-laden flows have not seen much development, as the prediction of aerodynamic coefficients is much more complex in this case~\cite{Capecelatro2024}. In compressible flows, the particle behaviour depends not only on the particle Reynolds number, the particle shape, and the angle of attack, but also on the Mach number of the flow relative to the particle. The forces experienced by the particles are affected by high-speed flow phenomena, such as the formation of shock waves, and their interaction with the particle~\cite{Capecelatro2024}. Moreover, non-continuum effects due to rarefaction of the gas must also be considered, which occur when the mean free path of the gas molecules is comparable to the characteristic length of the particle. This requires a deeper understanding of the flow physics around the particle in a compressible regime.

\citet{Bashforth1870} describes experiments with cannonballs, and shows the effect of compressibility on the drag force in the subsonic, transonic and supersonic regimes ($0.3 \leq \mathrm{Ma} \leq 2.0$). In this study, they found the drag force to be proportional to the square of the velocity in the subsonic and supersonic regimes and, proportional to the cube of velocity in the transonic regime. This is an important observation in regard to the effect of compressibility on the drag force. Later, \citet{Carlson1964} propose a drag correlation for spherical particles in a compressible flow, also including rarefaction effects. This model is applicable for $\mathrm{Ma} \leq 2$ and $10^{-1} \leq \mathrm{Re_p} \leq 10^{3}$. Subsequently, \citet{Henderson1976} also propose a Reynolds number and Mach number dependent correlation for spherical particles, valid for a larger range of $\mathrm{Re_p} \leq 2 \times 10^4$ and $\mathrm{Ma} \leq 6$ based on existing experimental data. \citet{Loth2008} show that $\mathrm{Re_p}$$\approx45$ separates the compression-dominated ($\mathrm{Re_p} \gtrapprox 45$) and rarefaction-dominated ($\mathrm{Re_p} \lessapprox 45$) regimes. 

Another noteworthy study on the effects of compressibility for an isolated spherical particle is presented in \citet{Nagata2016}.
They perform an extensive PR-DNS study to quantify drag, lift and torque for Reynolds numbers in the range of $50-300$ and Mach numbers in the range of $0.3-2.0$. They show that increasing the Mach number leads to an increase of the drag coefficient due to the formation of a detached shock wave in the supersonic regime. A comparison of the drag coefficients obtained using PR-DNS with the existing drag correlations~\cite{Carlson1964,Clift1971,Henderson1976} reveal that there exists a significant variation among the correlations in transonic and supersonic regimes. In order to investigate the effect of the Mach number and the particle Reynolds number on the flow field, \citet{Nagata2020} present an experimental study using Schlieren images for $10^3 \leq \mathrm{Re_p}\leq10^5$ and $0.9\leq \mathrm{Ma} \leq1.6$. They found that the oscillations of the wake and its size decrease with decreasing $\mathrm{Re_p}$ for the considered range of Reynolds numbers. This behaviour is opposite to the one observed for $\mathrm{Re_p} \leq10^3$. Moreover, the wake structure amplitude decreases with increasing Mach numbers. Using PR-DNS, \citet{Nagata2020a} examines the change in drag coefficient due to changing $\mathrm{Re_p}$ and $\mathrm{Ma}$ based on the location of the separation point on the sphere.

Subsequently, \citet{Loth2021} propose a drag model for spherical particles derived by combining existing experimental and numerical data from various sources \cite{Loth2008,Parmar2010,Jacobs1929,Roos1971,Short1967,Spearman1993,Theofanous2018,Nagata2016,Nagata2020a}. This model encompasses a wide range of Reynolds and Mach numbers to take into account the continuum- and rarefaction-dominated regimes, with a focus on the quasi-nexus at $\mathrm{Re_p}\approx45$. The model seems to be the most accurate drag model developed so far that predicts drag of a sphere in a compressible flow in various regimes, including the quasi-nexus that occurs when the flow transitions from the rarefaction-dominated to the compression-dominated regime. All available drag models have recently been summarised and reviewed in detail by \citet{Capecelatro2021}, including theoretical and experimental work in the field of gas-solid flows with spherical particles. They discuss the challenges that lie in the realm of modelling compressible flows, such as additional time and length scales due to the presence of shock waves. The comprehensive review sheds light on the unavailability of drag correlations for multi-particle systems in compressible flows. They suggest using a modified Loth model for isolated particles, which incorporates the volume fraction as presented in \citet{Tenneti2011}.
A more recent development with respect to the drag model for compressible flows is presented in \citet{Singh2022}, which aims to make existing drag correlations for spherical particles more accurate. The proposed drag model takes into account the surface temperature, the specific heat ratio, and other thermodynamic parameters to incorporate the effects of the gas surrounding the particle. Moreover, their model is applicable for all flow regimes classified according to the Knudsen number, that is, from the continuum regime with $\mathrm{Kn} < 0.01$ to the rarefied regime with $\mathrm{Kn} > 10$, including intermediate slip and transitional flow regimes ($0.01 < \mathrm{Kn} < 10 $).

Research to date on the forces on particles in compressible flows has only considered perfectly spherical particles, and the effect of the particle shape on the aerodynamic coefficients has not yet been investigated. The available correlations, therefore, miss important aspects of the flow physics for non-spherical particles. For example, no transverse force acts on spherical particles, which is not the case for non-spherical particles. Such correlations for lift, drag and torque prediction using such an assumption cannot be used to represent an array of various particle shapes that exist in practical applications. 

In this study, we address this gap by conducting PR-DNS of three shapes of non-spherical particles in compressible flows, in the range $100 \leq \mathrm{Re}_\mathrm{p} \leq 300$ and $0.3 \leq \mathrm{Ma} \leq 2$. 
The particle shapes and their aspect ratios ($\beta=a/b$) considered in the current study are: a prolate spheroid ($\beta=5/2$), an oblate spheroid ($\beta=5$), and a rod-like particle ($\beta=5$). We have also carried out simulations with spherical particles for validation purposes.
Based on our PR-DNS results, we derive correlations to predict the lift, drag and torque for non-spherical particles in compressible flows, considering the Mach number, the particle Reynolds number, the particle shape, and the angle of attack. The current study focuses on the compression-dominated flow regime, which corresponds to $\mathrm{Re_p}>50$ and $\mathrm{Kn}<0.01$.

\section{Methodology}
\label{method}
PR-DNS are performed with a 3-dimensional body-fitted mesh for the various particle shapes using a finite-volume based in-house compressible flow solver. The fully-coupled pressure-based algorithm described in \citet{Xiao2017} and \citet{Denner2020} is used to solve the equations governing a compressible flow as follows:
\begin{equation}
\frac{\partial \rho}{\partial t} + \frac{\partial (\rho u_i)}{\partial x_i} = 0,
\label{eq:continuity_full}
\end{equation}
\begin{equation}
\frac{\partial (\rho u_i)}{\partial t} + \frac{\partial (\rho u_i u_j)}{\partial x_j} = -\frac{\partial p}{\partial x_i} + \frac{\partial \tau_{ij}}{\partial x_j},
\label{eq:momentum_full}
\end{equation}
\begin{equation}
\frac{\partial (\rho h)}{\partial t} + \frac{\partial (\rho h u_i)}{\partial x_i} = \frac{\partial p}{\partial t} - \frac{\partial q_i}{\partial x_i} + \frac{\partial (\tau_{ij} u_j)}{\partial x_i},
\label{eq:energy_h}
\end{equation}
where $\rho$ is the density, $\mathbf{u}$ is the fluid velocity, $p$ is the pressure, $\boldsymbol{\tau}$ is the shear stress tensor, $h$ is the specific total enthalpy, and $\mathbf{q}$ is the conductive heat flux. This set of equations is closed by the ideal gas equation of state, and the constitutive equation for a Newtonian fluid,
\begin{equation}
\tau_{ij} = -p \delta_{ij} + \mu\left[\frac{\partial u_{i}}{\partial x_j} + \frac{\partial u_{j}}{\partial x_i} \right] - \frac{2}{3}\mu \frac{\partial u_k}{\partial x_k} \delta_{ij} 
\end{equation}
where $\mu$ is the fluid viscosity. 
The reference temperature is set to $\theta_0=300$ K and the reference pressure to $p_0=1.0\times10^5$ Pa. Based on the properties of air, the ratio of specific heats is $\gamma = 1.4$ and the specific heat capacity at constant pressure is $c_p = 1008$ J/(kg K). The thermal conductivity and viscosity of the fluid are calculated for various $\mathrm{Ma}$ and $\mathrm{Re_p}$ such that the Prandtl number, $\mathrm{Pr}$, of the gas is maintained at $0.71$. Furthermore, the viscosity is assumed to be constant in the domain. The free stream density is set to $\rho_\infty = 1.1574$ $\mathrm{kg}/\mathrm{m}^3$, whereas the free stream fluid velocity is determined based on the prescribed $\mathrm{Ma}$ and $\mathrm{Re_p}$ for each simulation.

Both the transient and the spatial terms are discretised using second-order accurate schemes. The second-order accurate backward Euler scheme is used for the temporal discretisation and the TVD Minmod scheme is used for spatial discretisation within the finite-volume framework. The fully-coupled pressure-based algorithm ensures a strong pressure-velocity and pressure-density coupling, and allows simulating a wide range of Mach numbers and does not require an under-relaxation parameter to ensure convergence \cite{Xiao2017, Denner2020}.

\begin{figure}
 \centering
 \includegraphics[width=0.6\linewidth]{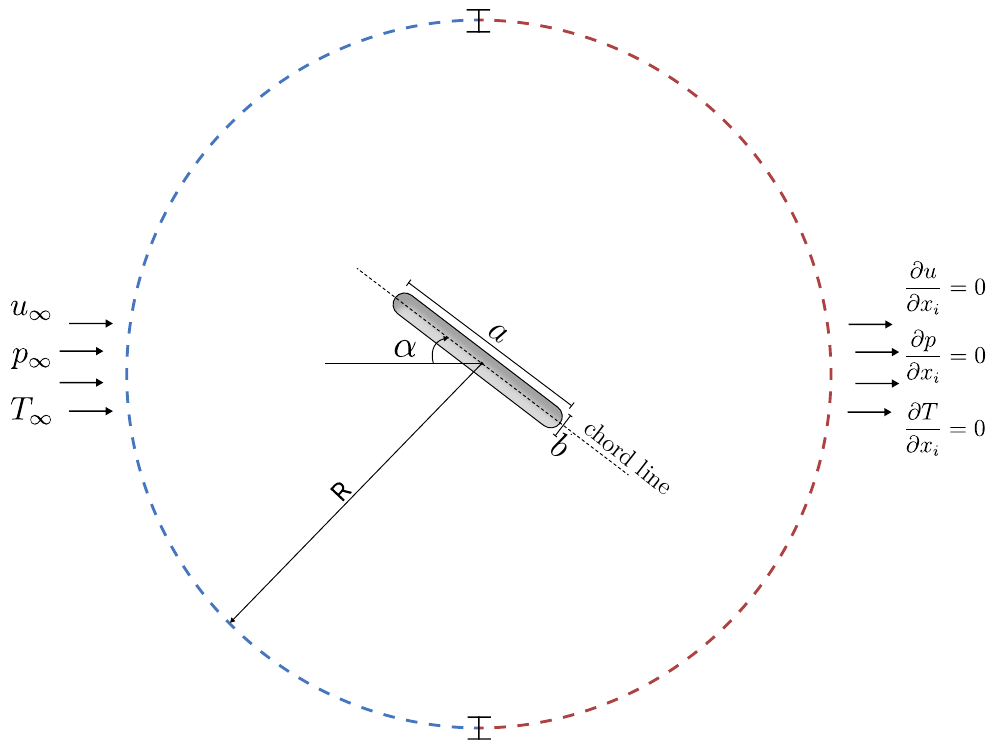}
  \caption[domain]{An illustration of the domain and the boundary conditions for the simulations in the current study. The particle is fixed at the centre of the spherical domain with hemispherical caps as the inlet (blue) and outlet (red) boundaries. The angle of attack of the particle is changed by rotating the hemispherical caps. It should be noted that the illustration is not to scale, as $R\gg \{a,b\}$.}
\label{fig:domain}
\end{figure}

\begin{figure}
  \centering
  \includegraphics[width=0.9\linewidth]{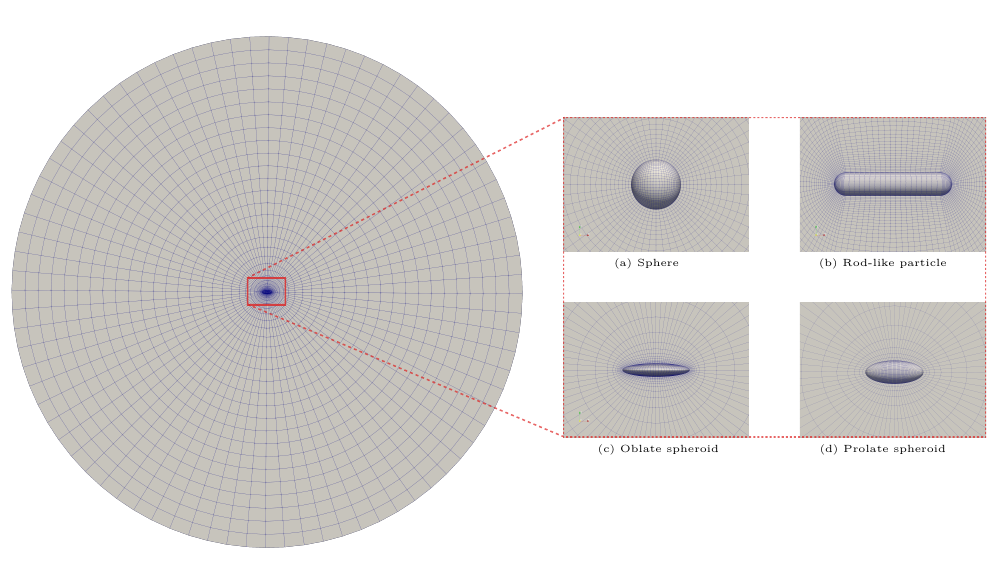}
  \caption[mesh]{A slice of the body-fitted hexahedral base mesh for the particle shapes used in the current study.}
\label{fig:mesh-whole}
\end{figure}

The fluid flow over a fixed, isolated particle is simulated in a spherical domain with radius $R$, where the particle lies at the centre of the domain. A spherical domain is chosen because the orientation of the particle relative to the flow can be changed without having to reconstruct the mesh. The desired angle of attack, $\alpha$, which is the angle between the inflow velocity vector and the chord line of the particle, is obtained by rotating the domain boundaries.
The hemispherical caps of the domain are set to be the inlet and outlet boundaries, as illustrated in Figure~\ref{fig:domain}. For compressible flows, the inlet hemisphere is prescribed with Dirichlet boundary conditions for velocity, pressure, and temperature, whereas the outlet hemisphere is preescribed with Neumann boundary conditions. The particle boundary is simulated as a no-slip wall \reviewerO{with an adiabatic boundary condition for temperature}.
The computational domain is discretised using a hexahedral non-uniform body-conforming mesh, with smaller cells near the particle and increasingly larger cells towards the domain boundary. The radius of the domain is $R=50c$, where $c$ is the semi-major axis of the particle. The particle shapes and their aspect ratios ($\beta=a/b$) considered in the current study are: a prolate spheroid ($\beta=5/2$), an oblate spheroid ($\beta=5$), and a rod-like particle ($\beta=5$). The discretised domains for the considered particle shapes are shown in Figure~\ref{fig:mesh-whole}. 

\reviewerT{In \ref{Appendix A}, Figures \ref{fig:Convergence_Prolate} to \ref{fig:Convergence_Rodlike} show convergence studies for the three considered non-spherical particle shapes for a Reynolds number of 300, Mach numbers of 0.3 and 2, and an angle of attack of 0\degree. Based on the results of the convergence studies in \ref{Appendix A}, the medium-sized mesh is chosen for all simulations of all three non-spherical shapes as it gives the best trade-off between accuracy and computational cost. Compared to the fine mesh, the relative error of the drag coefficient for the medium is below 5\% for all convergence studies, except for the prolate spheroid at $\mathrm{Ma}=2$ where the error is 6\%. The ratio of the equivalent diameter of a sphere of the same volume to the smallest mesh spacing at the particle surface is 75 ($\Delta x = 0.072 \, \mathrm{m}$) for the prolate spheroid, 115 ($\Delta x = 0.051 \, \mathrm{m}$) for the oblate spheroid and 50 ($\Delta x = 0.0019 \, \mathrm{m}$) for the rod-like particle, which leads to a resolved boundary layer for all 3 non-spherical particle shapes.}

\reviewerT{The force $\mathbf{F}$ and the torque $\mathbf{T}$ acting on the surface of the particles are computed by
\begin{equation}
	  {F_i =\int_S \tau_{ij} n_{j} \mathrm{d}S},
   \label{eq:force_particles}
 \end{equation}
 \begin{equation}
  T_i = \int_S \epsilon_{ijk} r_j \tau_{km} n_m \, \mathrm{d}S,
  \end{equation}
where $S$ is the surface of the particle, $\mathbf{n}$ is the normal vector to the particle surface, $\mathbf{r}$ is the vector from the particle center to the point where the force is applied on the particle surface, and $\epsilon$ is Levi-Civita symbol for the cross product.
The corresponding coefficients are defined as 
\begin{equation}
  C_\mathrm{D} = \frac{F_\mathrm{D}}{\frac{1}{2} \rho_{\infty} u_{\infty}^2 A_\mathrm{p}} , \,\, C_\mathrm{L} = \frac{F_\mathrm{L}}{\frac{1}{2} \rho_{\infty} u_{\infty}^2 A_\mathrm{p}} , \,\, C_\mathrm{M} = \frac{T_\mathrm{p}}{\frac{1}{2} \rho_{\infty} u_{\infty}^2 \pi d_\mathrm{p}^3/8},
 \label{eq:force_coeffs}
\end{equation} 
where $F_\mathrm{D}$ and $F_\mathrm{L}$ are the forces acting parallel and normal to the flow direction, respectively, $T_\mathrm{p}$ is the pitching torque acting on the particle, $\rho_\infty$ is the free stream density, $u_{\infty}$ is the free stream velocity, and $A_\mathrm{p} = \pi d_\mathrm{p}^2/4$ is the reference area of the particle based on the equivalent diameter of a sphere of the same volume, $d_\mathrm{p}$.
}

The time step is constrained by setting the maximum Courant-Friedrichs-Lewy (CFL) number to 0.3. The simulations are run until a steady state is observed in the temporal evolution of the aerodynamic force coefficients. The residuals have a convergence criterion of $O(10^{-8})$ and a total conservation error of $O(10^{-3})$ is achieved. All aerodynamic force coefficients shown in the results section are mean values of \reviewerT{at least} the last 1000 time steps.

\section{Results and discussion}
\label{results}

In this section, the PR-DNS results are first validated with results for particles in incompressible and compressible flows reported in the literature. Subsequently, the flow regimes and the aerodynamic force coefficients obtained for the considered non-spherical particles are presented and discussed.

\subsection{Validation}
\label{Validation}
The methodology in the current study is validated with the available data in the literature for a spherical particle in various flow regimes (from the incompressible limit at $\mathrm{Ma}=0.3$ to supersonic flow at $\mathrm{Ma}=2$), and non-spherical particles in the incompressible limit.

\begin{figure}
\centering
\includegraphics[width=1.0\linewidth]{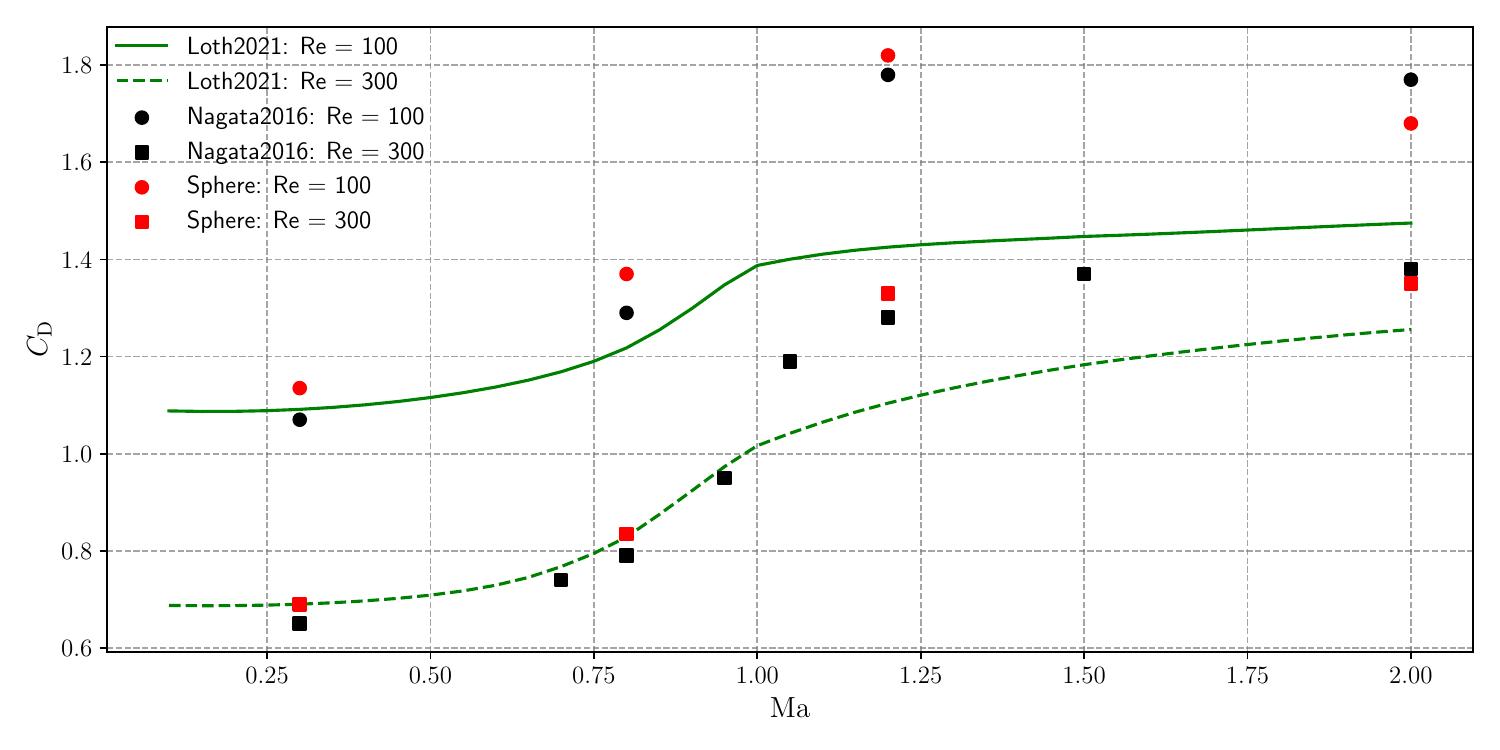}
 \caption{The drag coefficient as a function of the Mach number for an isolated sphere in a compressible flow for particle Reynolds numbers of 100 and 300, compared against the PR-DNS data from \citet{Nagata2016} and the correlation of \citet{Loth2021}.}
\label{fig:sphere-validation-plot}
\end{figure}

\begin{figure}
\includegraphics[width=1.0\linewidth]{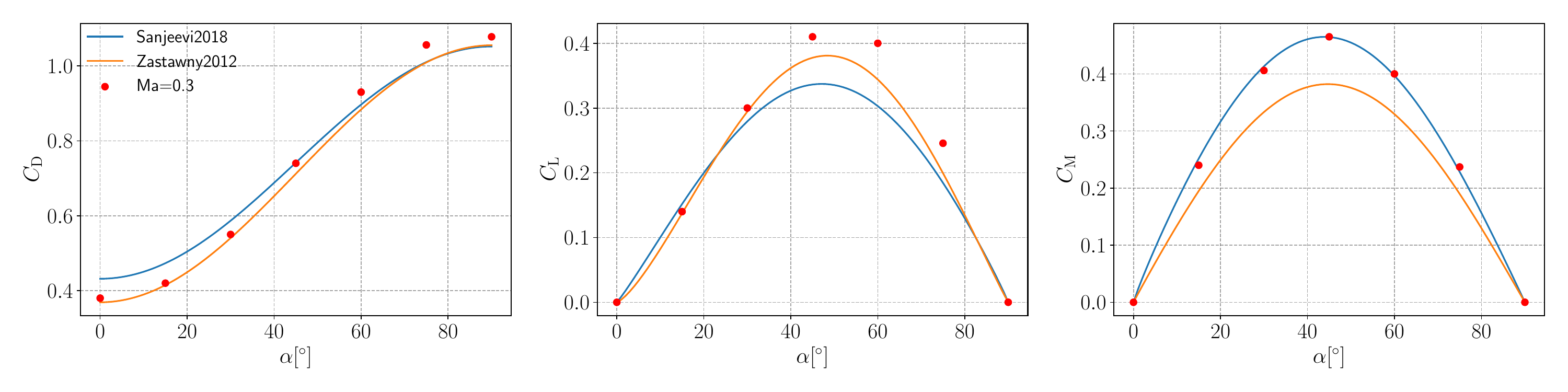}
 \caption{Comparison of our data with the correlations of \citet{Sanjeevi2018} and \citet{Zastawny2012} for the prolate spheroid in an incompressible flow at the incompressible limit of $\mathrm{Ma}=0.3$ at a particle Reynolds number of 300 for all three aerodynamic force coefficients.}
 \label{fig:Validation_Zastawny_Sanjeevi_Ellipsoid}
\end{figure}

\begin{figure}
\includegraphics[width=1.0\linewidth]{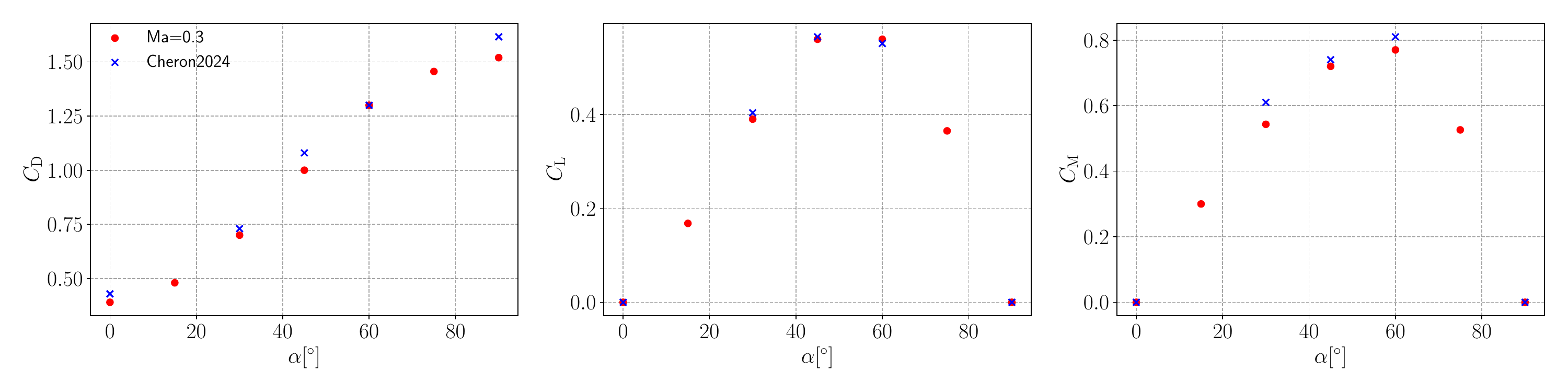}
 \caption{Comparison of our data with the correlations of \citet{Cheron2024} for the rod-like particle in incompressible flow to the incompressible limit simulations of $\mathrm{Ma}=0.3$ at a particle Reynolds number of 300 for all three aerodynamic force coefficients.}
 \label{fig:Validation_Cheron_Fibroid}
\end{figure}

The drag experienced by an isolated spherical particle for particle Reynolds numbers of 100 and 300 is compared with the PR-DNS data of \citet{Nagata2016} and the compressible drag model of \citet{Loth2021} as shown in \Cref{fig:sphere-validation-plot}. \reviewerT{For all Mach numbers, results from the current simulations show good agreement with the PR-DNS data of \citet{Nagata2016} and reasonable agreement with the model of \citet{Loth2021}, while the drag coefficient for supersonic Mach numbers predicted by our simulations agree better with the data of \citet{Nagata2016}.} It is, however, important to note that the transonic regime is not very well documented in the literature and that the existing data shows a lot of variation. Overall, the results for the sphere in compressible flow obtained in this work agree well with the existing literature.

\Cref{fig:Validation_Zastawny_Sanjeevi_Ellipsoid} shows a comparison of the simulation results for the prolate spheroid in the incompressible limit ($\mathrm{Ma}=0.3$) for a particle Reynolds number of 300 with the correlations of \citet{Sanjeevi2018} and \citet{Zastawny2012} for a prolate spheroid in incompressible flow. Furthermore, \Cref{fig:Validation_Cheron_Fibroid} shows a comparison of the simulation results for the rod-like particle in the incompressible limit ($\mathrm{Ma}=0.3$) for a Reynolds number of 300 with the \reviewerT{data} of \citet{Cheron2024} for the same rod-like particle in incompressible flow. The results obtained for the rod-like particle match very well with the \reviewerT{data} of \citet{Cheron2024}. The results obtained for the prolate spheroid match the correlations of \citet{Zastawny2012} and \citet{Sanjeevi2018} \reviewerT{very well}.

\begin{figure}
 \includegraphics[width=0.5\linewidth, height=5cm]{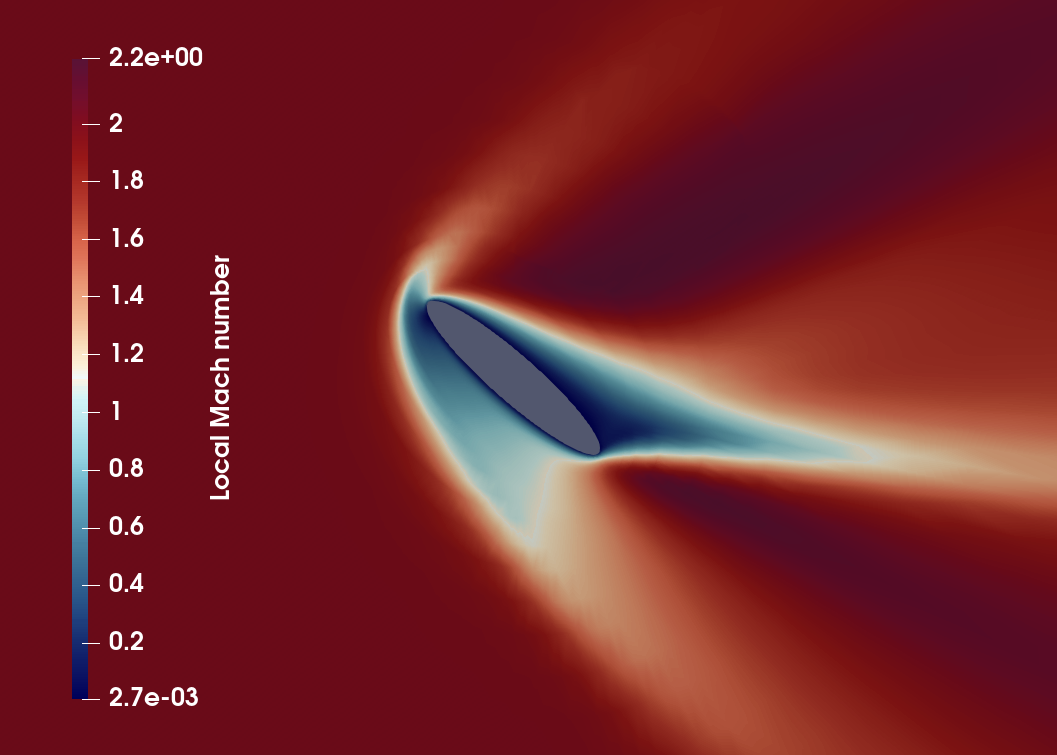}
 \includegraphics[width=0.5\linewidth, height=5cm]{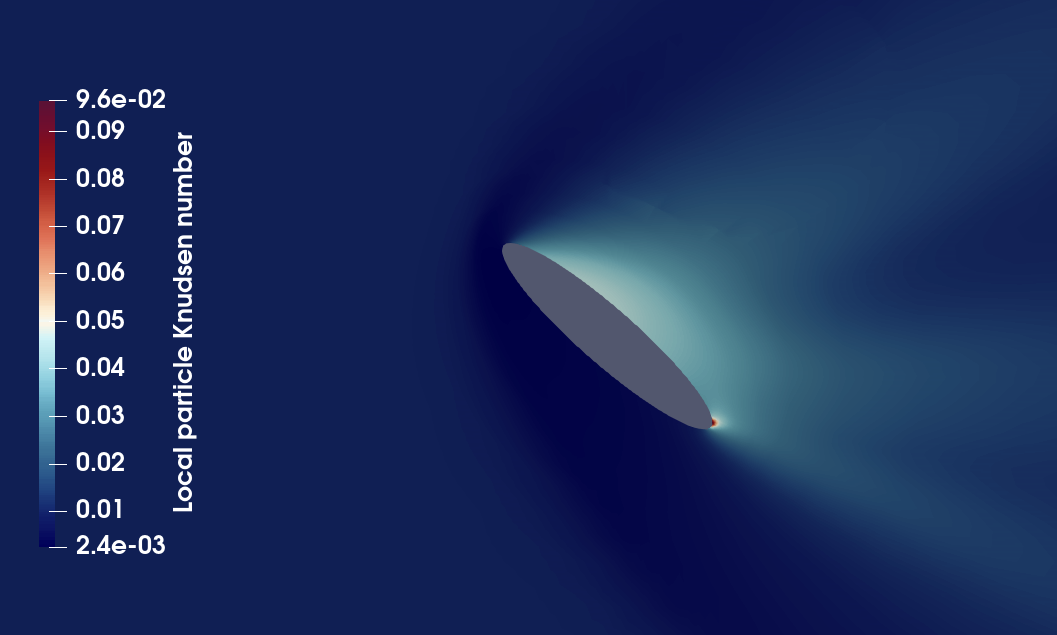}
 \caption{\reviewerO{Contour plots of the local Mach number (left) and of the local particle Knudsen number (right) for the oblate spheroid at $\mathrm{Re_p}=300$ and $\alpha = 45\degree$.}}
 \label{fig:Knudsen_Mach_Validation}
\end{figure}

\reviewerO{Figure \ref{fig:Knudsen_Mach_Validation} shows the local Mach number and the local Knudsen number for the oblate spheroid at $\mathrm{Re_p}=300$ and $\alpha = 45\degree$. The oblate spheroid in this flow configuration is one of the most extreme flow conditions in this work. As visible from Figure \ref{fig:Knudsen_Mach_Validation}, the local Mach number reaches a maximum of 2.2 and the local particle Knudsen number, computed as \cite{Loth2021}
\begin{equation}
    \mathrm{Kn_{local}} = \sqrt{\frac{\pi \gamma}{2}} \left( \frac{\mathrm{Ma_{local}}}{\mathrm{Re_{local}}}\right)
\end{equation}
reaches a maximum of 0.096 at the lower tip of the particle surface. Although a part of the flow in the wake of the particle is in the slip regime ($0.01 < \mathrm{Kn} < 0.1$), the flow is still dominated by the continuum regime ($\mathrm{Kn} \ll 0.1$). The continuum assupmtion, thus, forms an adequate basis for the conducted simulations.}

\subsection{Non-spherical particles in compressible flow}
\label{non-sphe-compr}

In this section, the results obtained from the PR-DNS with non-spherical particles in a compressible flow are shown. First, some representative contour plots of the velocity field are shown, which are analysed for the different particle shapes and configurations, and compared with each other. Subsequently, the results for the aerodynamic force coefficients are presented and discussed.

\begin{figure}
	\centering
    \subcaptionbox{$\mathrm{Re} = 300$, $\mathrm{Ma} = 0.8$, $\alpha = 0\degree$}{\includegraphics[width=5cm, height=4cm]{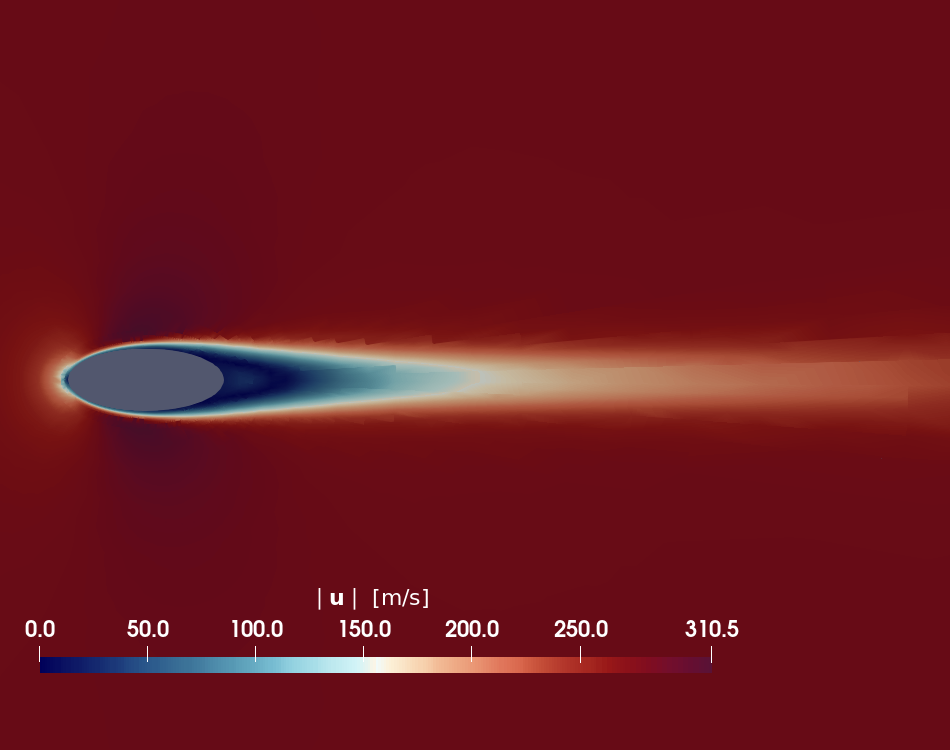}}
    \subcaptionbox{$\mathrm{Re} = 300$, $\mathrm{Ma} = 0.8$, $\alpha = 45\degree$}{\includegraphics[width=5cm, height=4cm]{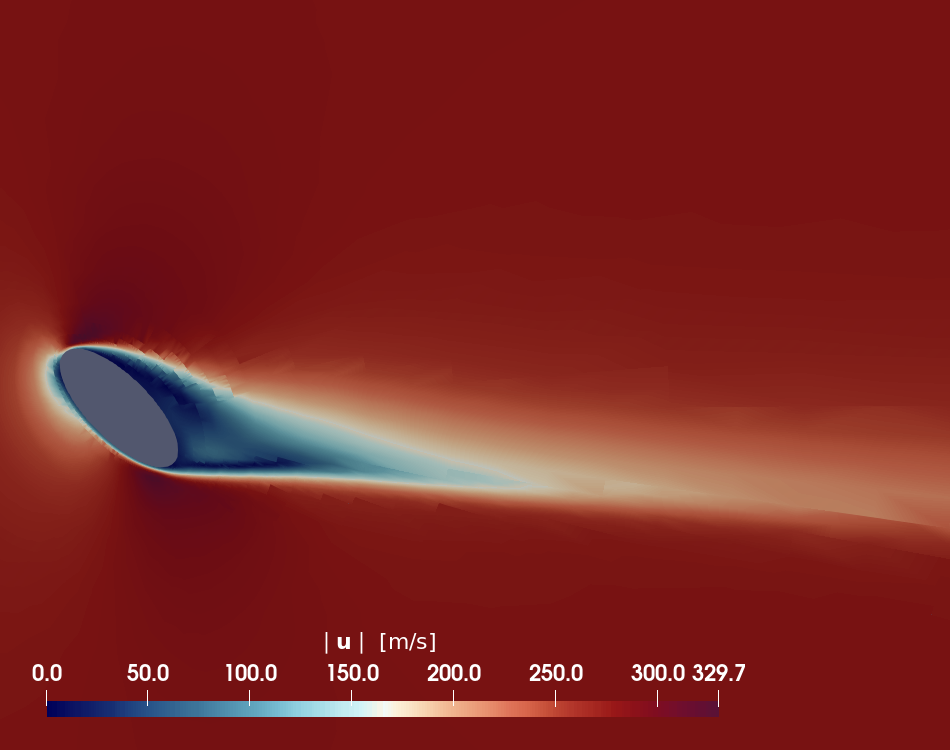}}
    \subcaptionbox{$\mathrm{Re} = 300$, $\mathrm{Ma} = 0.8$, $\alpha = 90\degree$}{\includegraphics[width=5cm, height=4cm]{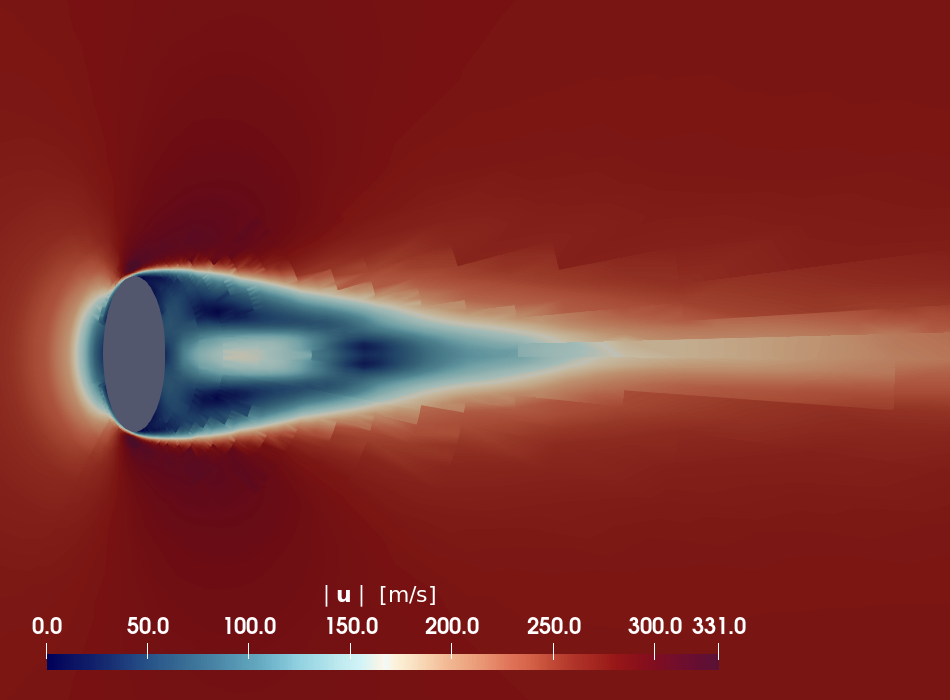}}\\
	\subcaptionbox{$\mathrm{Re} = 300$, $\mathrm{Ma} = 2$, $\alpha = 0\degree$}{\includegraphics[width=5cm, height=4cm]{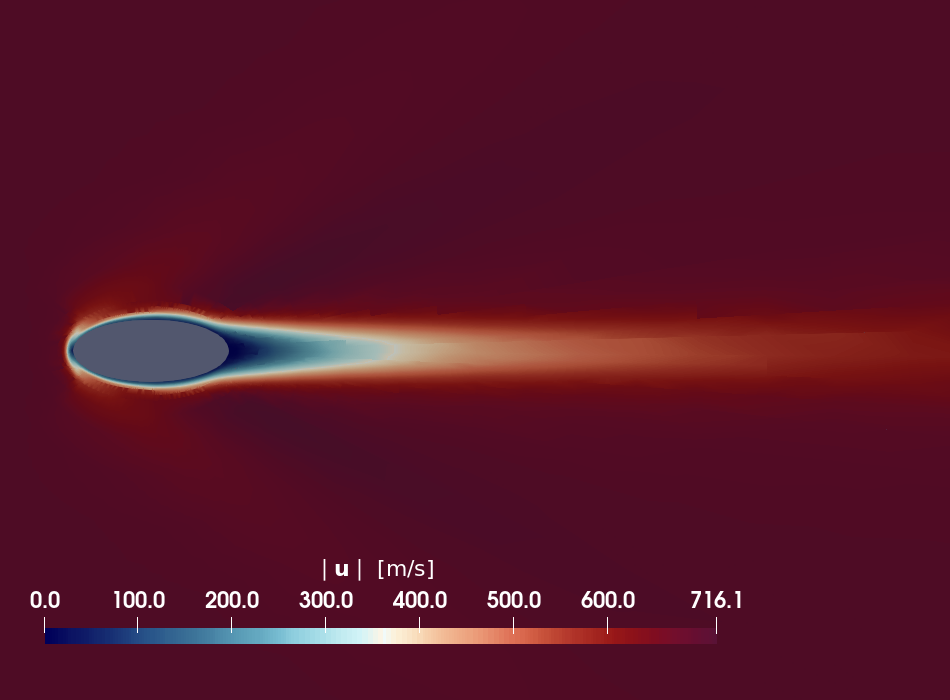}}
    \subcaptionbox{$\mathrm{Re} = 300$, $\mathrm{Ma} = 2$, $\alpha = 45\degree$}{\includegraphics[width=5cm, height=4cm]{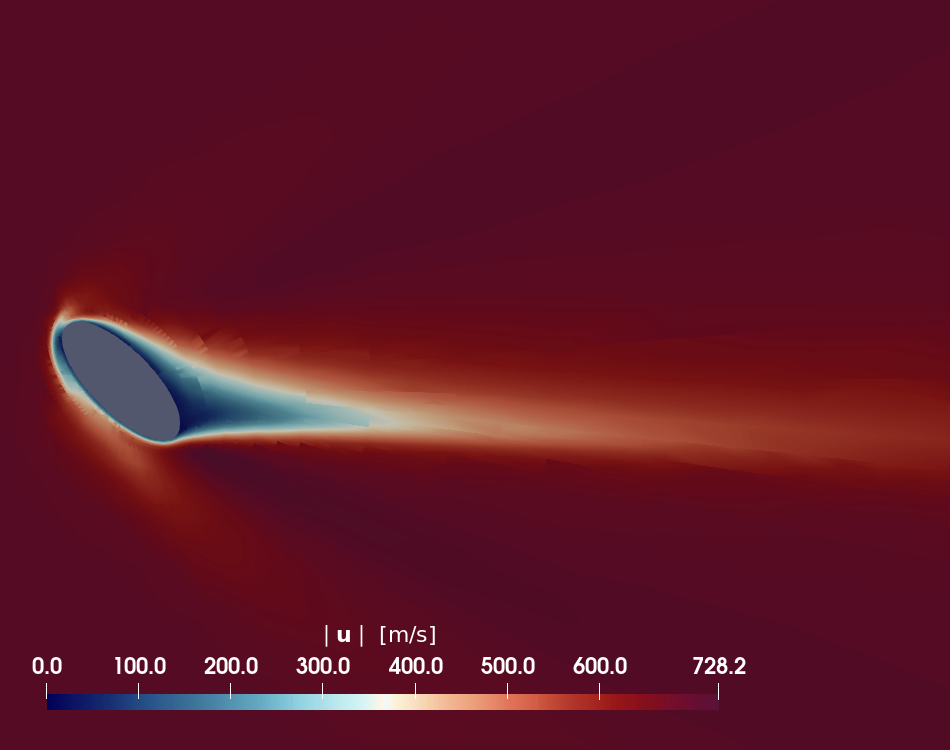}}
    \subcaptionbox{$\mathrm{Re} = 300$, $\mathrm{Ma} = 2$, $\alpha = 90\degree$}{\includegraphics[width=5cm, height=4cm]{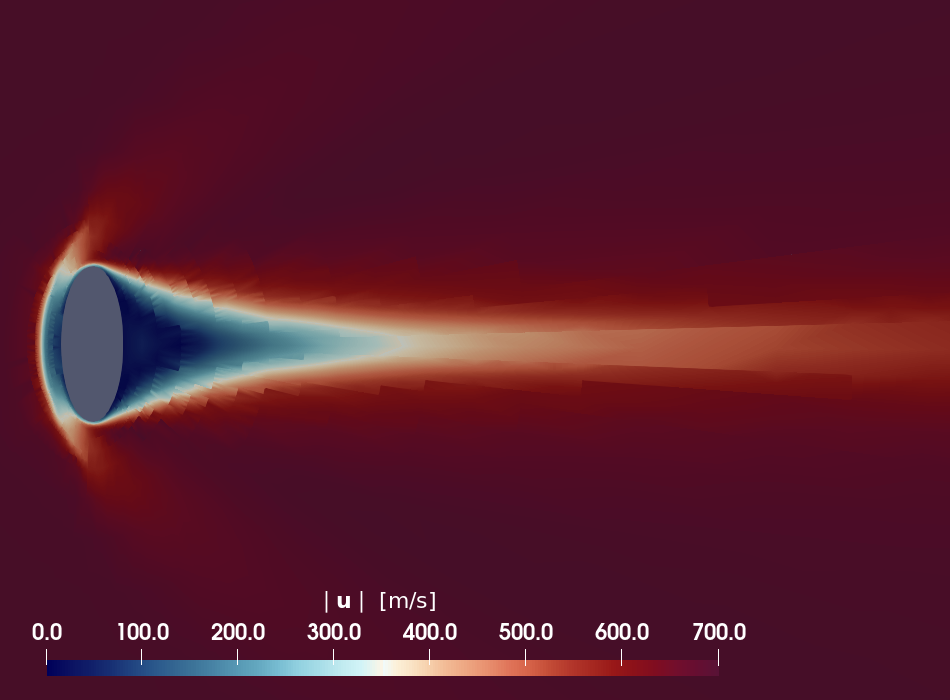}}
\caption{Velocity fields of the prolate spheroid for Reynolds number 300, Mach numbers of 0.8 and 2 and angles of 0\degree, 45\degree, 90\degree.}
\label{fig:Vel_cont_ellipsoid}
\end{figure}

\begin{figure}
	\centering
    \subcaptionbox{$\mathrm{Re} = 300$, $\mathrm{Ma} = 0.8$, $\alpha = 0\degree$}{\includegraphics[width=5cm, height=4cm]{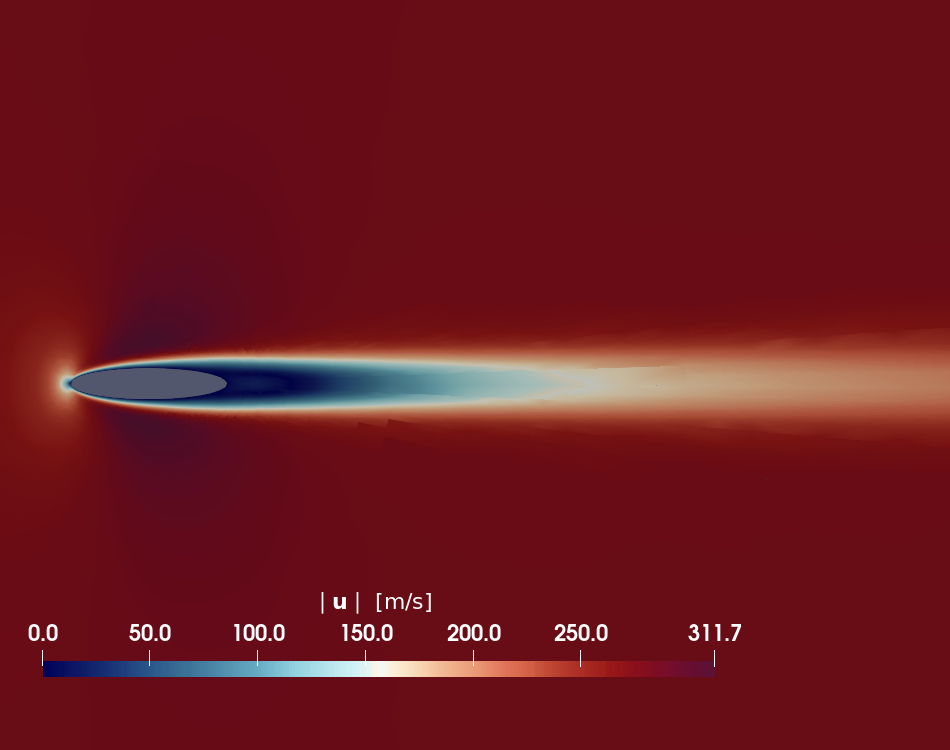}}
    \subcaptionbox{$\mathrm{Re} = 300$, $\mathrm{Ma} = 0.8$, $\alpha = 45\degree$}{\includegraphics[width=5cm, height=4cm]{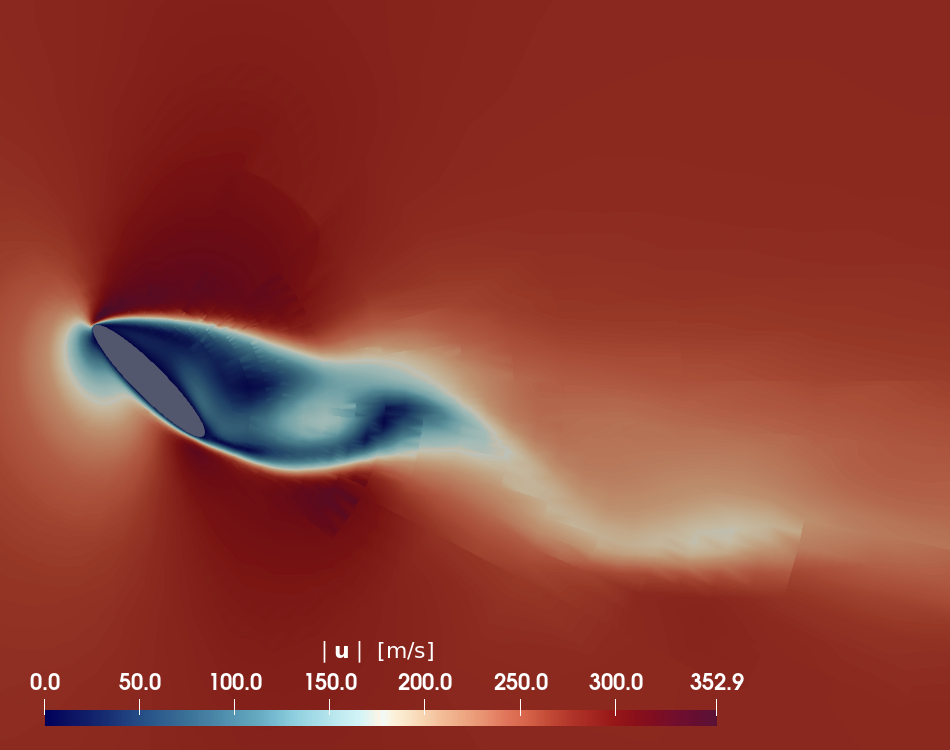}}
    \subcaptionbox{$\mathrm{Re} = 300$, $\mathrm{Ma} = 0.8$, $\alpha = 90\degree$}{\includegraphics[width=5cm, height=4cm]{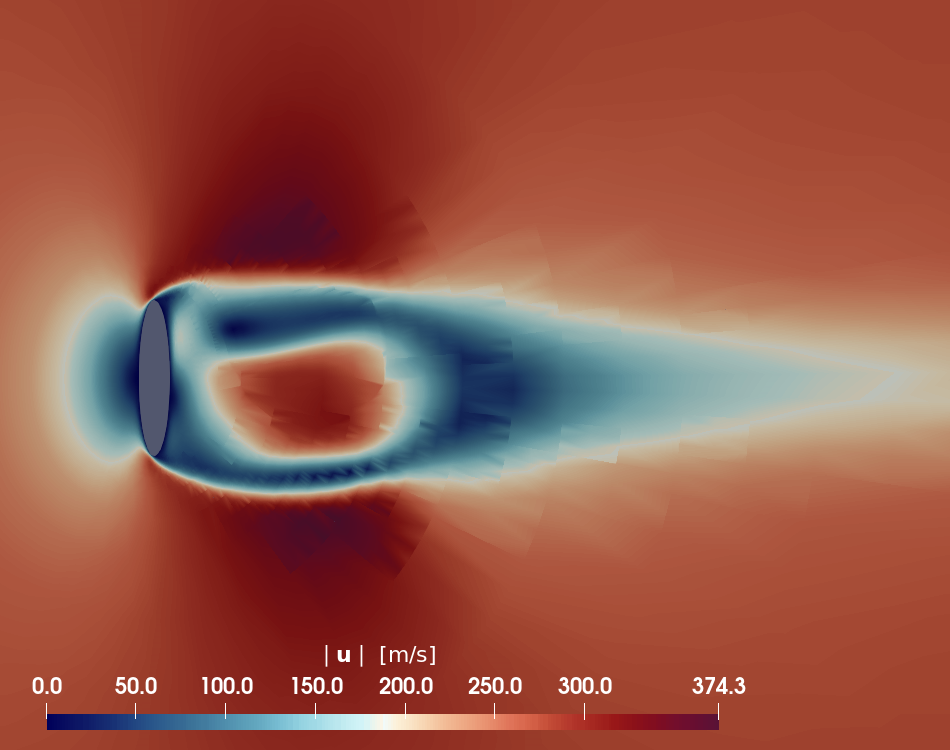}}\\
	\subcaptionbox{$\mathrm{Re} = 300$, $\mathrm{Ma} = 2$, $\alpha = 0\degree$}{\includegraphics[width=5cm, height=4cm]{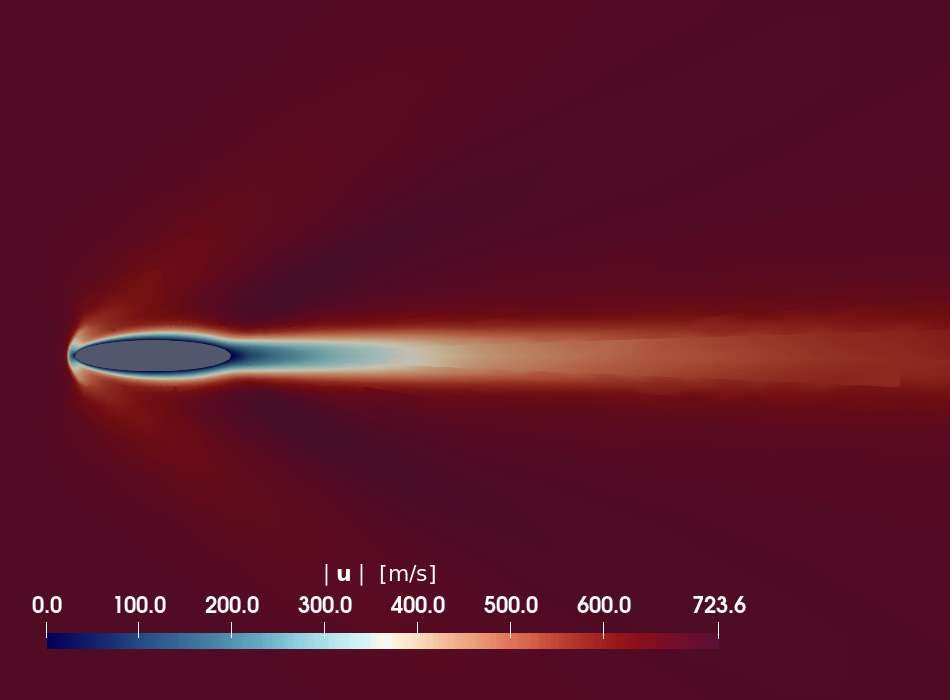}}
    \subcaptionbox{$\mathrm{Re} = 300$, $\mathrm{Ma} = 2$, $\alpha = 45\degree$}{\includegraphics[width=5cm, height=4cm]{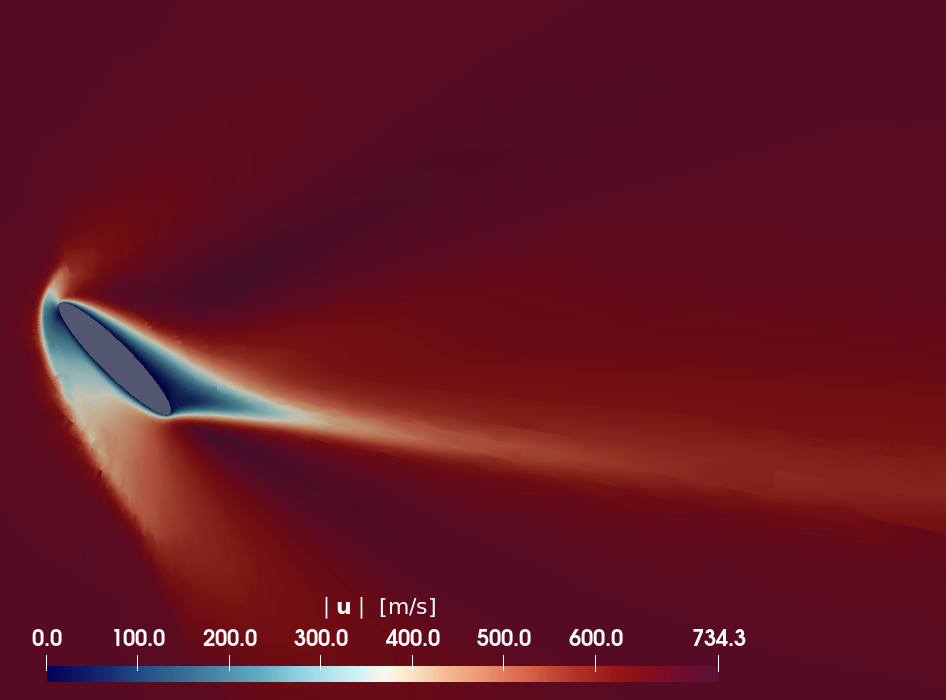}}
    \subcaptionbox{$\mathrm{Re} = 300$, $\mathrm{Ma} = 2$, $\alpha = 90\degree$}{\includegraphics[width=5cm, height=4cm]{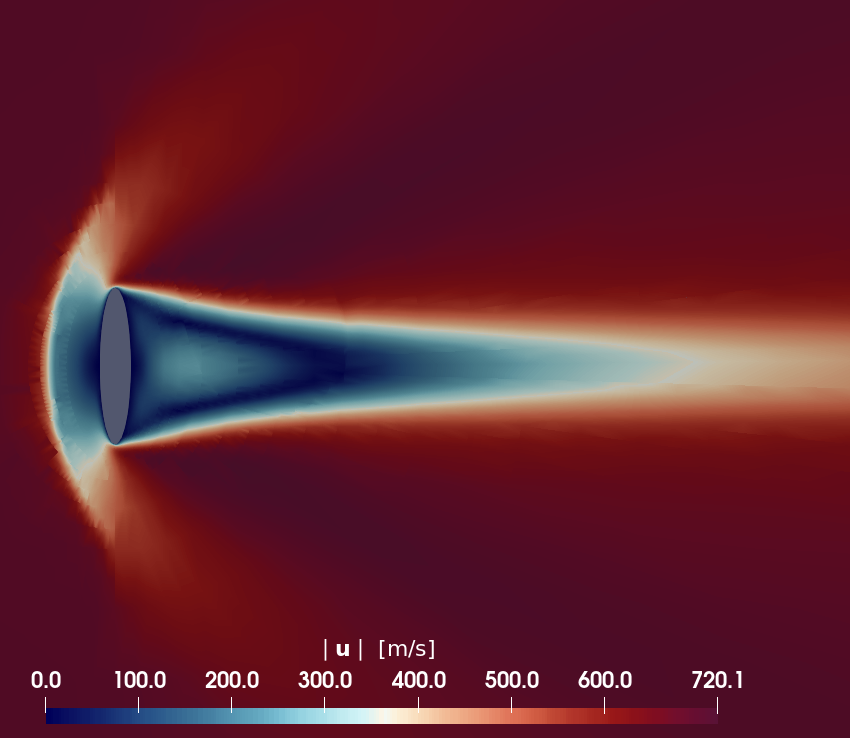}}
\caption{Velocity fields of the oblate spheroid for Reynolds number 300, Mach numbers of 0.8 and 2 and angles of 0\degree, 45\degree, 90\degree.}
\label{fig:Vel_cont_disc}
\end{figure}

\begin{figure}
	\centering
    \subcaptionbox{$\mathrm{Re} = 300$, $\mathrm{Ma} = 0.8$, $\alpha = 0\degree$}{\includegraphics[width=5cm, height=4cm]{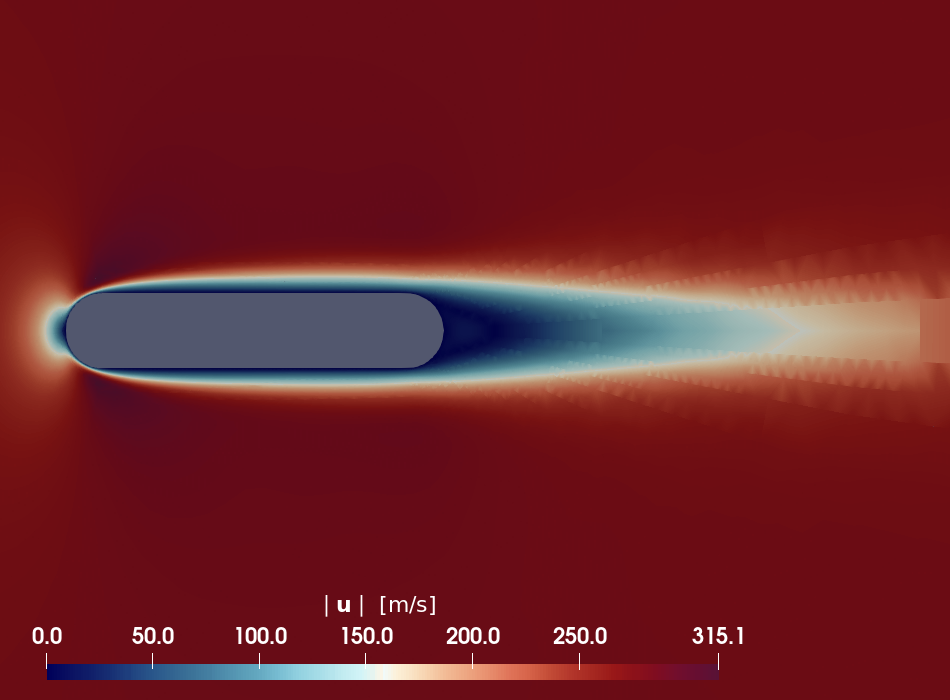}}
    \subcaptionbox{$\mathrm{Re} = 300$, $\mathrm{Ma} = 0.8$, $\alpha = 45\degree$}{\includegraphics[width=5cm, height=4cm]{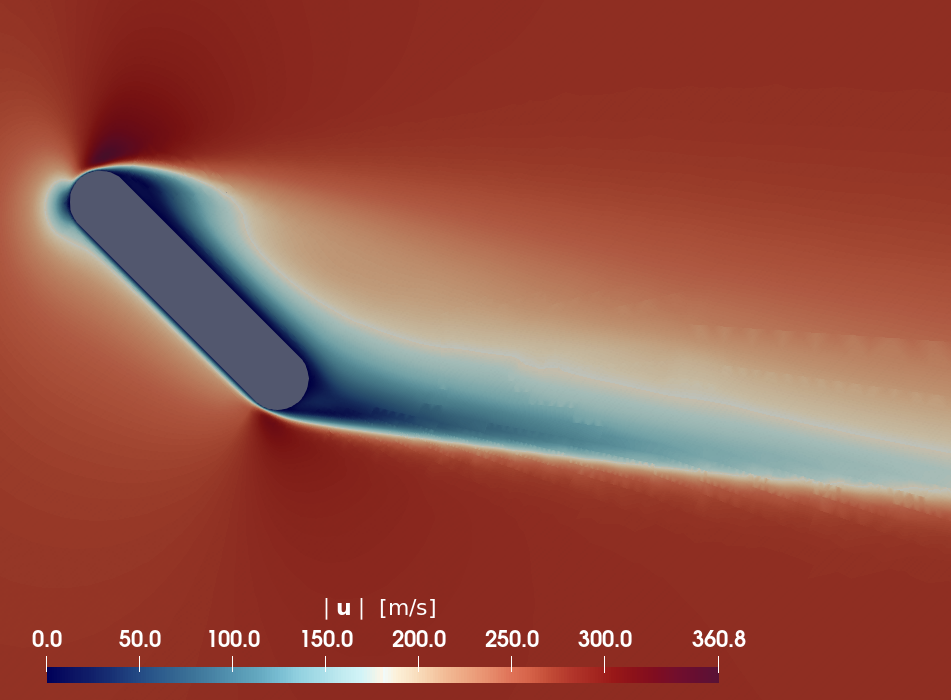}}
    \subcaptionbox{$\mathrm{Re} = 300$, $\mathrm{Ma} = 0.8$, $\alpha = 90\degree$}{\includegraphics[width=5cm, height=4cm]{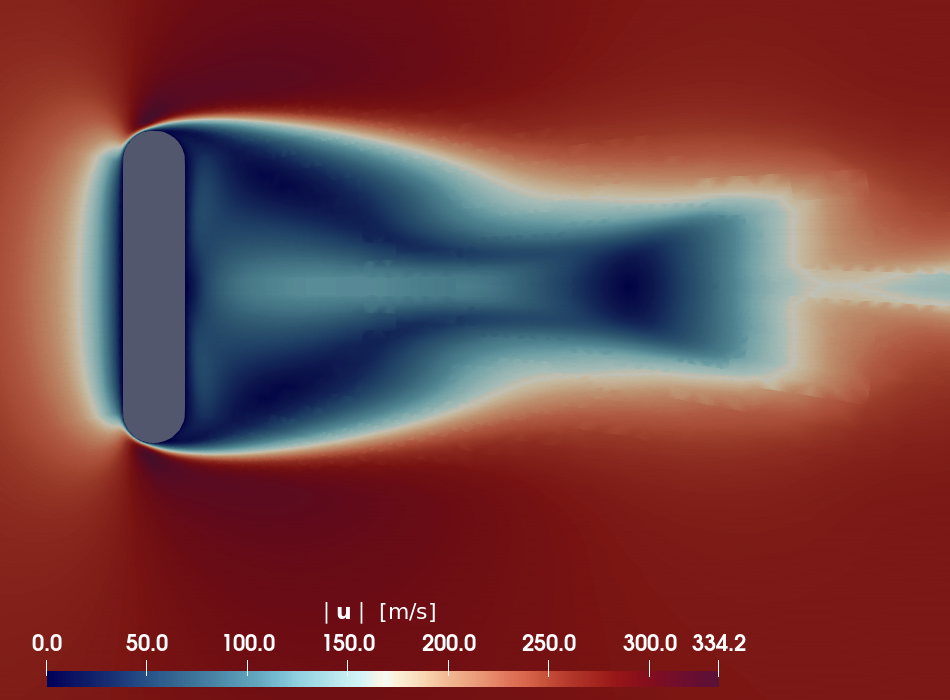}}\\
	\subcaptionbox{$\mathrm{Re} = 300$, $\mathrm{Ma} = 2$, $\alpha = 0\degree$}{\includegraphics[width=5cm, height=4cm]{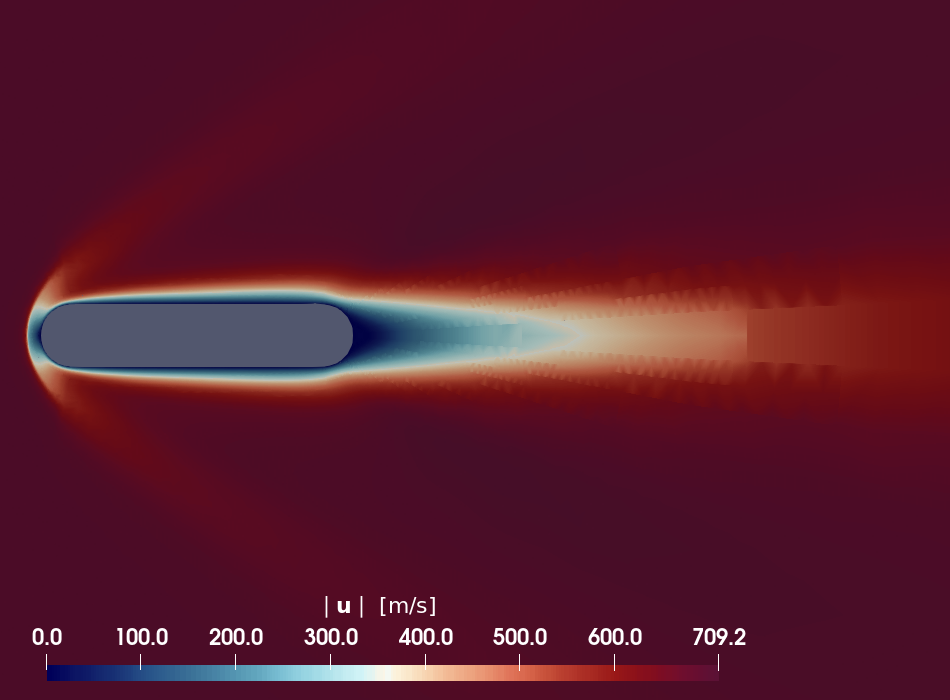}}
    \subcaptionbox{$\mathrm{Re} = 300$, $\mathrm{Ma} = 2$, $\alpha = 45\degree$}{\includegraphics[width=5cm, height=4cm]{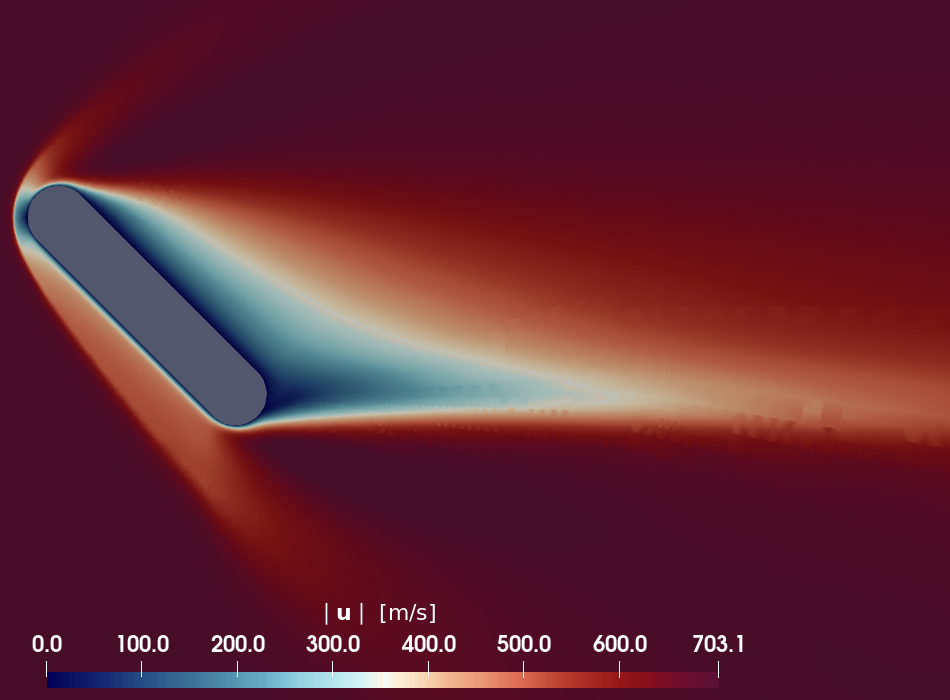}}
    \subcaptionbox{$\mathrm{Re} = 300$, $\mathrm{Ma} = 2$, $\alpha = 90\degree$}{\includegraphics[width=5cm, height=4cm]{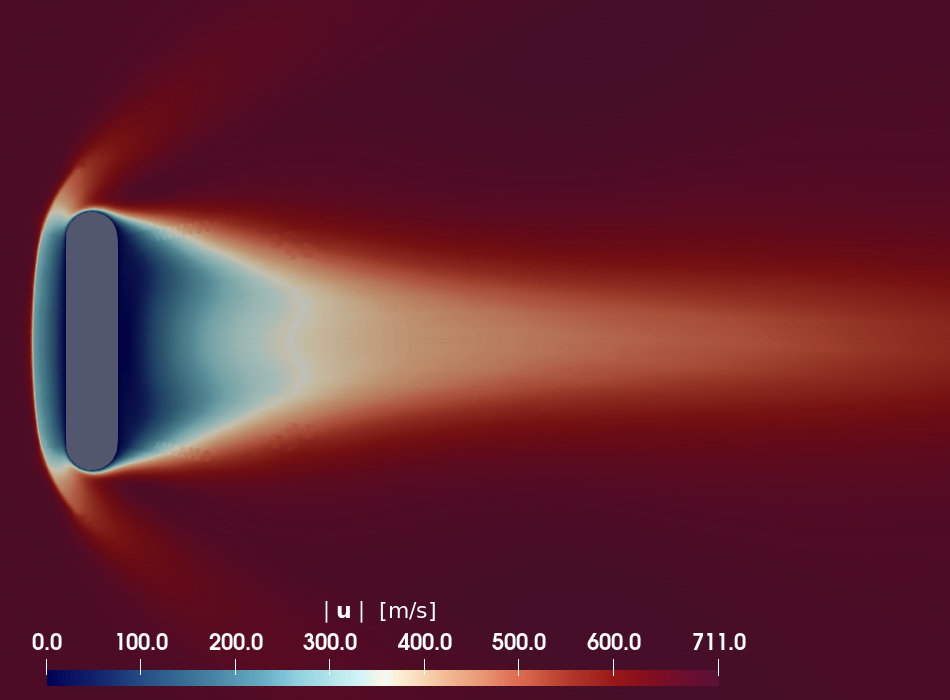}}
\caption{Velocity fields of the rod-like particle for Reynolds number 300, Mach numbers of 0.8 and 2 and angles of 0\degree, 45\degree, 90\degree.}
\label{fig:Vel_cont_fibroids}
\end{figure}

Figures \ref{fig:Vel_cont_ellipsoid}, \ref{fig:Vel_cont_disc} and \ref{fig:Vel_cont_fibroids} show the steady-state velocity flow fields for all three non-spherical particle shapes for a Reynolds number of 300 and Mach numbers of 0.8 and 2, for angles of attack of 0\degree, 45\degree, and 90\degree. These two Mach numbers are representative examples of the flow fields in the transonic ($\mathrm{Ma}=0.8$) and supersonic ($\mathrm{Ma}=2$) regimes. The flow fields of the other Mach numbers do not differ significantly from the two shown. The same applies to the Reynolds number. The flow fields at a Reynolds number of 100 differ for the most part only in magnitude compared to a Reynolds number of 300.

When comparing the velocity fields for $\mathrm{Ma}=2$, as shown in Figures \ref{fig:Vel_cont_ellipsoid}, \ref{fig:Vel_cont_disc} and \ref{fig:Vel_cont_fibroids} it becomes clear that the oblate spheroid shape forms the largest bow shock in front of the particle, followed by the rod-like particle, and the prolate spheroid with the smallest bow shock for all three angles. This observation goes hand in hand with the aspect ratios or sphericity of the respective particles. The prolate spheroid shape has the lowest aspect ratio and the fewest ``sharp'' edges or regions of high curvature, whereas the oblate spheroid shape has the most obvious ``sharp'' edges.

Considering the transonic regime ($\mathrm{Ma}=0.8$) for all three particle shapes, one can observe that the oblate spheroid is the only shape where the velocity field shows an unsteady and oscillating behaviour in the wake behind the particle for angles of attack of 45\degree and 90\degree. This behaviour is not observed for the other two shapes and also not for the simulations of the oblate spheroid with Reynolds number 100. Furthermore, for all three shapes the bow shock in front of the particles starts to form but, is not fully formed in the transonic regime. Comparison of the transonic regime with the supersonic regime also shows that the wake behind the particles is longer in the transonic regime and shorter in the supersonic regime.

\begin{figure}
	\centering
    \subcaptionbox{Prolate spheroid\label{fig:CD_ProlateSpheroid}}{\includegraphics[width=16cm]{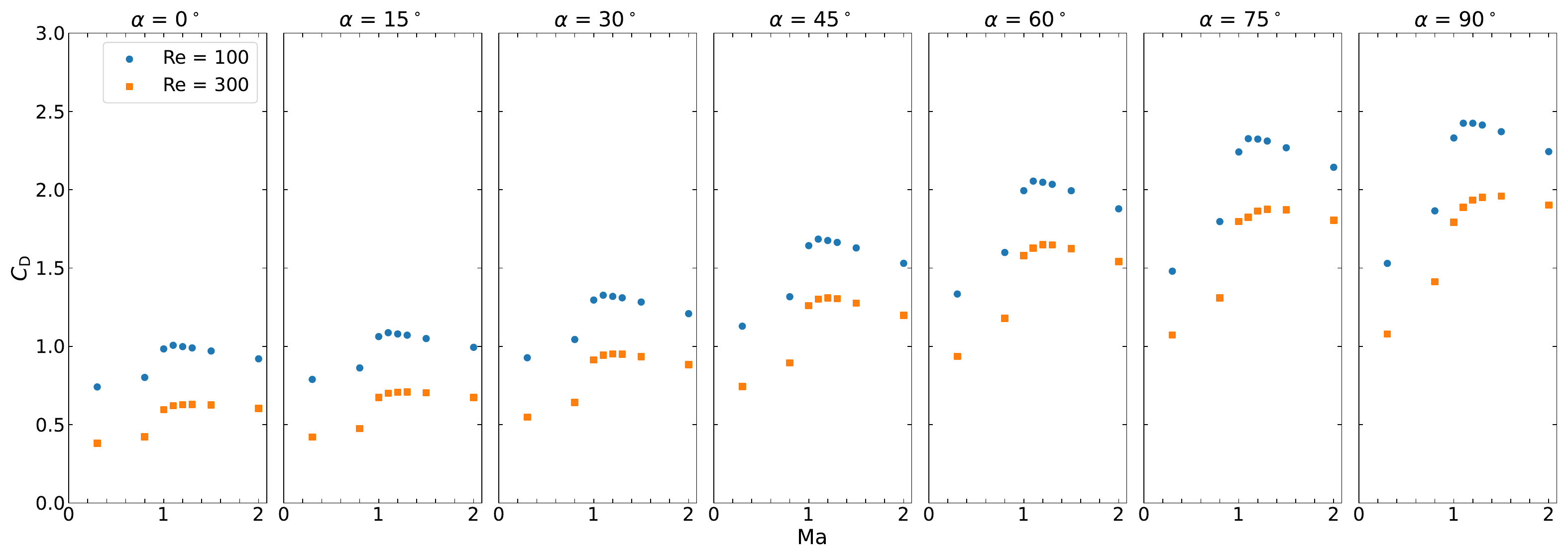}}\\
    \subcaptionbox{Oblate spheroid\label{fig:CD_OblateSpheroid}}{\includegraphics[width=16cm]{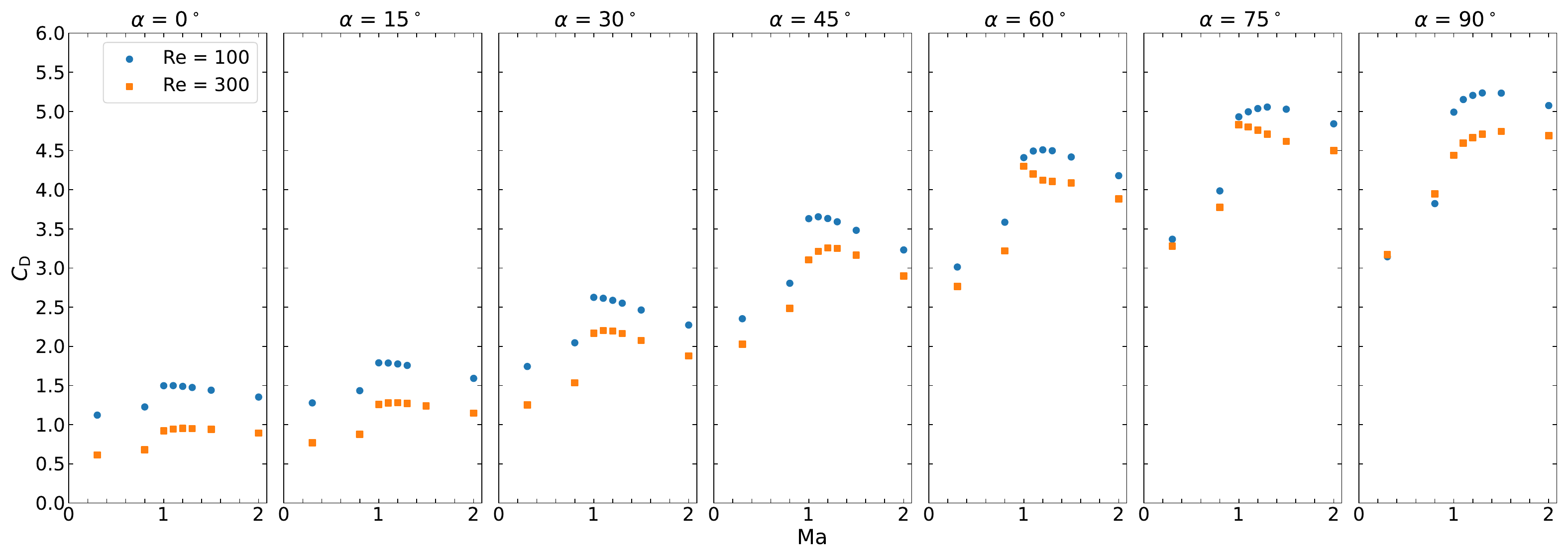}}\\
    \subcaptionbox{Rod-like particle\label{fig:CD_Fibroid}}{\includegraphics[width=16cm]{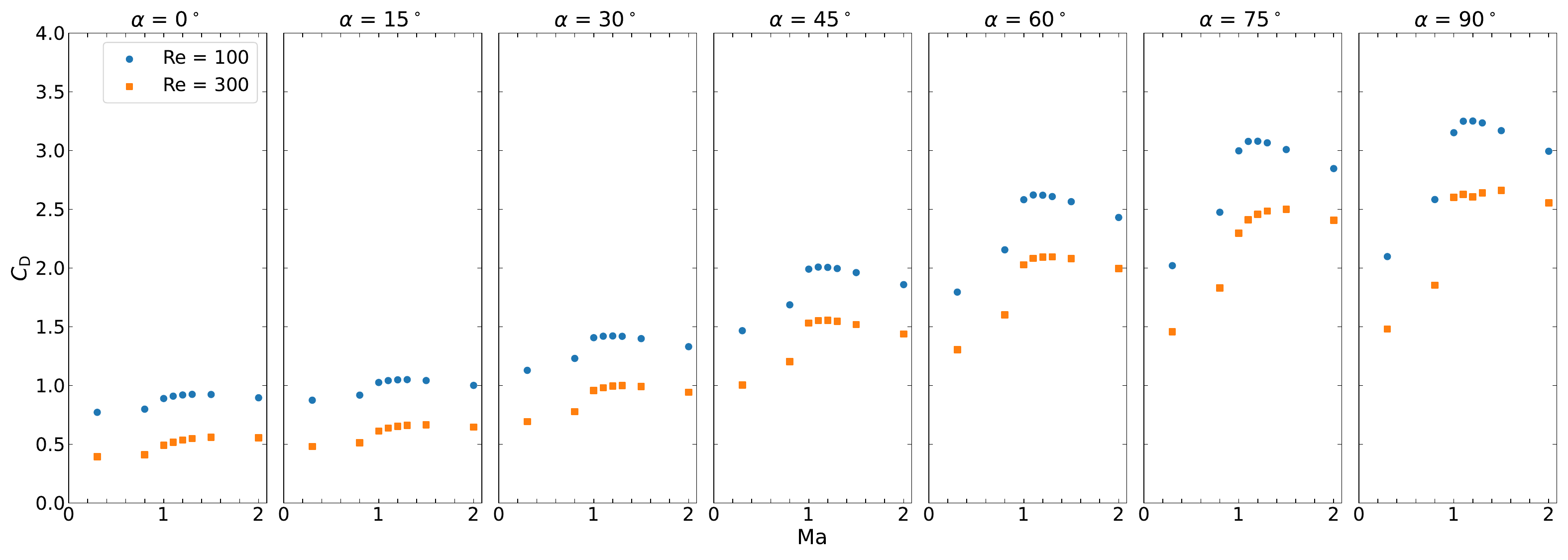}}   
\caption{Variation of the drag coefficient $C_{\mathrm{D}}$ with Mach number $\mathrm{Ma}$ for different angles of attack considered for the three considered non-spherical particle shapes.}
\label{fig:CD_over_M}
\end{figure}

\reviewerO{Figure \ref{fig:CD_over_M} presents the drag coefficients, $C_{\mathrm{D}}$, for all three non-spherical particle shapes as a function of the Mach number for all considered angles of attack. The lift and torque coefficients as a function of the Mach number for all three non-spherical particle shapes are shown in \ref{Appendix B}.}

For a given Mach number, the magnitudes of the drag coefficients for all three non-spherical particle shapes in Figure \ref{fig:CD_over_M} increase with increasing angle of attack. Furthermore, as already indicated by the velocity contour plots, the oblate spheroid with the strongest shock formation has the largest drag coefficients, followed by the rod-like particle, and the prolate spheroid with the weakest shock formation and the smallest drag coefficients. For all three shapes, a smaller particle Reynolds number yields a higher drag coefficient.

\reviewerT{For all three non-spherical particle shapes, the drag coefficient in the subsonic regime increases only slightly for angles of attack of 0\degree to 15\degree, but increases more significantly as the angle of attack increases, as a function of the Mach number. For all angles of attack, the drag coefficient increases even more rapidly in the transonic regime and, afterwards, decreases in the supersonic regime.} Intuition suggests that an increase in the Mach number leads to an increase in the drag coefficient, since the pressure contribution increases for high-speed flows, while the viscous contribution remains significantly smaller. However, in reality, this is only partially the case. The effect of compressibility depends on the Mach number regime of the flow. In the subsonic regime, the drag coefficient remains independent of the Mach number up to a critical Mach number, $\mathrm{Ma}_\mathrm{cr}$. This critical Mach number is defined as the lowest free-stream Mach number at which the flow around the particle is locally supersonic. If the free-stream Mach number is larger than $\mathrm{Ma}_\mathrm{cr}$, the drag coefficient starts to increase slowly. A sharp increase in the drag coefficient is observed in the transonic regime followed by a decrease in value for the supersonic regime. For a spherical particle, a similar sharp increase in the drag coefficient is observed for the transonic regime. However, there is little to no decline of the drag coefficient in the supersonic regimes. \reviewerT{The sharp increase of the drag coefficient occurs when the free-stream Mach number exceeds a critical value known as the drag divergence Mach number, which marks the onset of a rapid increase in aerodynamic drag due to the formation of shock waves and compressibility effects.} This termination of the supersonic flow region by a shock wave gives rise to an adverse pressure gradient. This causes a separation of the boundary layer and a dramatic increase in pressure drag \cite{Anderson2003}. The critical Mach number varies for different particle shapes, a blunt body such as a sphere having a relatively high critical Mach number compared to a slender body \cite{Anderson2003}. Moreover, the extent to which the drag coefficient decreases after the peak drag is also dependent on the Reynolds number. For a Reynolds number of 100, the drag coefficient has a steeper descent than for a Reynolds number of 300.

The lift and  torque coefficients shown in Figures \ref{fig:CL_over_M} and \ref{fig:CM_over_M} show a similar Mach number dependence as the previously analysed drag coefficient. The difference in the magnitudes of both force coefficients for different Reynolds numbers is not as large as is the case for the drag coefficient. Overall, the oblate spheroid shows the largest coefficients for lift and torque, followed by the rod-like particle and prolate spheroid. \reviewerT{The torque coefficient has, in comparison to the drag and lift coefficients, a much steeper descent in the supersonic regime and the torque coefficients in the supersonic flow regime are even smaller compared to the coefficients in the subsonic regime.} 

\section{Correlations for the aerodynamic force and torque coefficients}
\label{correlation}

The results from the PR-DNS are used to determine the aerodynamic forces and torque, which are used to develop correlations to predict the drag, lift, and torque coefficients for each particle as a function of the particle Reynolds number, $\mathrm{Re_p}$, the Mach number, $\mathrm{Ma}$, and the angle of attack between the flow and the major axis of the particle, $\alpha$. 
The proposed correlations are all valid for a particle Reynolds number that varies in the range $100 \leq \mathrm{Re_p} \leq 300$, a Mach number of $0 \leq \mathrm{Ma} \leq 2$, and an angle of attack of $0 \degree \leq \alpha \leq 90 \degree$. The correlations to account for compressible effects presented in this section are extensions of the correlations of the coefficients for a particle of the same shape in an incompressible flow, for the same particle Reynolds number and angle of attack \reviewerO{($C_\mathrm{D,L,M}\left(\mathrm{Re_p},\alpha,\mathrm{Ma}=0\right)$)}. 
\reviewerO{To this end, for instance, the correlations for the aerodynamic force coefficients for prolate and oblate spheroids presented by \citet{Zastawny2012} and \citet{Sanjeevi2018}, and the correlations for rod-like particles of \citet{Cheron2024} and \citet{Zastawny2012} could be used as a basis for $C_\mathrm{D,L,M}\left(\mathrm{Re_p},\alpha,\mathrm{Ma}=0\right)$. This assumes that all force coefficients are constant in the range from $\mathrm{Ma} = 0$ to $0.3$ for the flow conditions studied. In \ref{Appendix C} a summary of the full models for all three non-spherical particles shapes are given exemplary with the correlations of \citet{Zastawny2012} for the incompressible limit.}  
For each correlation, the fitting coefficients for a given function are estimated using curve-fitting algorithms using the Levenberg-Marquardt algorithm of the SciPy library \cite{Johansson2019}.
The accuracy of the new correlations are assessed by calculating the maximum and median errors of their prediction, $\epsilon_\mathrm{max}$ and $\epsilon_\mathrm{med}$. The error in the prediction is defined as
\begin{equation}
\epsilon_i = \frac{ \lVert f_{\mathrm{corr},i} - f_{\mathrm{PR-DNS},i} \rVert}{\mathrm{max}\left( \lVert f_{\mathrm{corr},i} \rVert , \lVert f_{\mathrm{PR-DNS},i} \rVert \right)},
\end{equation}
where $f_{\mathrm{PR-DNS},i}$ represents the value of a given force or torque coefficient from the $i$-th PR-DNS simulation, and $f_{\mathrm{corr},i}$ is the value predicted by the correlation for the given force or torque coefficient. The prediction error is scaled by the maximum value between the coefficient prediction by the curve fitting algorithms and the PR-DNS value because the force coefficient can be zero. This zero value occurs for the lift and torque coefficients of a particle with an angle of attack of $\alpha = 0 \degree$ or $90 \degree$. The fitting coefficients given in Tables \ref{table:dragcoefficients}-\ref{table:torquecoefficients} are rounded to three significant figures for readability, exact values are given in the attached data repository.

\subsection{Drag coefficient}
The proposed correlation for the drag coefficient of each studied particle shape is based on the drag coefficient in the incompressible flow regime and a compressible correction that is a function
of the particle Reynolds number, the Mach number, and the angle of attack. 
The general expression is defined as 
\begin{equation}\label{eq:General_Drag_Coefficient}
C_\mathrm{D}\left(\mathrm{Re_p},\alpha,\mathrm{Ma}\right) = C_\mathrm{D}\left(\mathrm{Re_p},\alpha,\mathrm{Ma}=0\right) \left[1  + \mathcal{F}_\mathrm{D}^{\mathrm{Ma}}\left(\mathrm{Re_p},\alpha,\mathrm{Ma}\right)\right]
\end{equation}
where $C_\mathrm{D}$ is the drag coefficient of a specific particle given as a function of the particle Reynolds number, angle of attack, and Mach number. This coefficient depends on the drag coefficient for equivalent particle Reynolds number and angle of attack in incompressible flow conditions, and an additional shape-dependent function, $\mathcal{F}_\mathrm{D}^{\mathrm{Ma}}$ to account for the change in the coefficient as a result of the compressibility of the flow. The expressions of these shape-dependent functions are given in the dedicated sections, along with the associated prediction errors. The fitting coefficients, $\vartheta_i$, for the drag coefficients of all three considered particle shapes are given in Table~\ref{table:dragcoefficients}.

\begin{table}
    \centering
     \caption{List of the fitting coefficients in Eqs.~\eqref{eq:prolate_drag_compressible}, ~\eqref{eq:oblate_drag_compressible}, and~\eqref{eq:fibroid_drag_compressible}, used in the correlation to predict the change in the drag coefficient in the case of compressible flow.}\label{table:dragcoefficients}
    \begin{tabular}{l || c c c c c c c c c c c}
     & $\vartheta_1$ & $\vartheta_2$ & $\vartheta_3$ & $\vartheta_4$ & $\vartheta_5$ & $\vartheta_6$ & $\vartheta_7$ & $\vartheta_8$ & $\vartheta_9$ & $\vartheta_{10}$ & $\vartheta_{11}$ \\
    \hline
    \hline
    \reviewerT{Eq.~\eqref{eq:prolate_drag_compressible}} & 1.55 $\times$ $10^{-3}$ & 0.201 & 0.492 & 15.6 & 0.878 & 0.183 & 1.21 & 0.142 & - & - & - \\
    \reviewerT{Eq.~\eqref{eq:oblate_drag_compressible}} & 6.07 & 0.269 & 4.40 $\times$ $10^{-5}$ & 2.10 & 3.90 & 21.3 & 0.853 & 0.130 & 1.17 & 0.252 & - \\
    \reviewerT{Eq.~\eqref{eq:fibroid_drag_compressible}} & 1.14 $\times$ $10^{-3}$ & 0.121 & 2.64 & 0.0983 & 1.40 & 14.6 & 0.861 & 3.94 & 0.631 & 1.20 & 0.330
    \end{tabular}
\end{table}

\begin{figure}
  \begin{center}
    \includegraphics[width=1.0\linewidth]{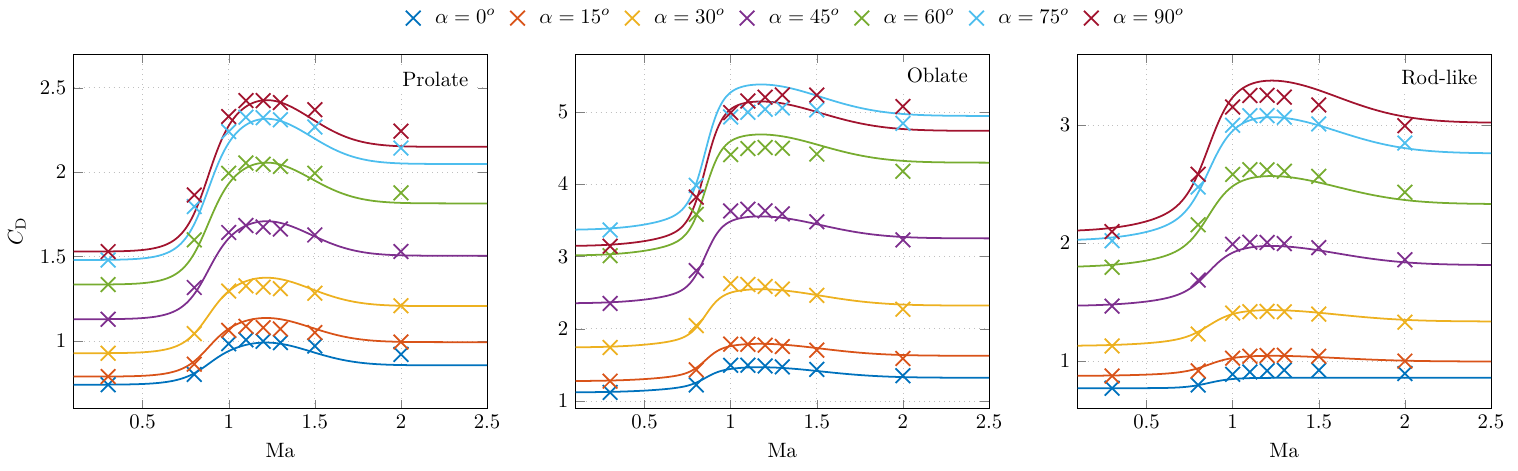}
    \includegraphics[width=1.0\linewidth]{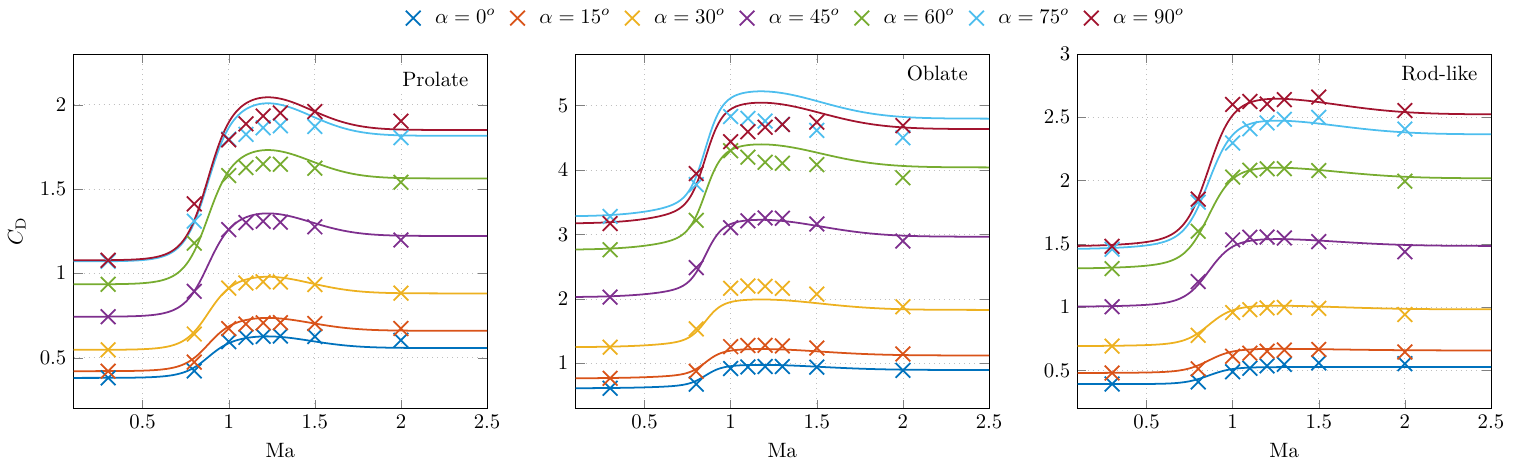}
  \end{center}
  \caption{The drag coefficient as a function of the Mach number for the three shapes studied, at different particle Reynolds numbers, for all considered angles of attack. From left column to right column: prolate spheroid, oblate spheroid, and rod-like particle. Top row: $\mathrm{Re_p} = 100$, bottom row: $\mathrm{Re_p} = 300$. The colour indicates the angle of attack, the marker indicates PR-DNS result, and the solid line shows the prediction of the drag coefficient given by the derived correlations, Eqs.~\eqref{eq:prolate_drag_compressible}, ~\eqref{eq:oblate_drag_compressible}, and~\eqref{eq:fibroid_drag_compressible}.}
\label{fig:drag_prediction}
\end{figure}

\subsubsection{Prolate spheroid}
To consider compressible effects in the prediction of the drag coefficient of an isolated prolate spheroid particle, the shape-dependent function, $\mathcal{F}_\mathrm{D}^{\mathrm{Ma}}$, in Eq.~\ref{eq:General_Drag_Coefficient} is given by
\begin{equation}\label{eq:prolate_drag_compressible}
\mathcal{F}_\mathrm{D}^{\mathrm{Ma}}\left(\mathrm{Re_p},\alpha,\mathrm{Ma}\right) =  \frac{\left(\vartheta_1 \mathrm{Re_p} + \vartheta_2 \alpha^{\vartheta_3}\right)}{\left[1 + \exp\left(-\vartheta_4 (\mathrm{Ma} - \vartheta_5)\right)\right]} + \vartheta_6 \exp\left(-\frac{(\mathrm{Ma} - \vartheta_7)^2}{\vartheta_8}\right)\, .
\end{equation}
The first term on the right hand-side of Eq.~\eqref{eq:prolate_drag_compressible} captures the sigmoid evolution of the drag coefficient as the Mach number transitions from the subsonic to the supersonic flow regime, where two plateaus are observed. In the transonic region, there is a noticeable increase in the drag coefficient, forming a bell-shaped peak. This peak, caused by the transition from subsonic to supersonic flow, is captured with the second term on the right-hand side of Eq.~\eqref{eq:prolate_drag_compressible} that is modelled based on a Gaussian function. Both terms vary as a function of the particle Reynolds number and the angle of attack; the closer the angle of attack is to $\alpha = 90 \degree$, the larger is the change in the drag coefficient.

The PR-DNS results along with the correlation for the prediction of the drag coefficient of the prolate spheroid are shown in the left of Figure~\ref{fig:drag_prediction}, as a function of the Mach number, for all angles of attack and particle Reynolds number values considered in this work. 
The correlation accurately recovers the change in the evolution of the drag coefficient in both transonic and supersonic regimes, for all particle Reynolds numbers and angles of attack. 
The maximum and median prediction errors between the correlation and the PR-DNS are \reviewerT{7.56\% and 1.77\%, respectively}.

\subsubsection{Oblate spheroid}
To consider compressible effects in the prediction of the drag coefficient of an isolated oblate spheroid particle, the shape-dependent function, $\mathcal{F}_\mathrm{D}^{\mathrm{Ma}}$, in Eq.~\ref{eq:General_Drag_Coefficient} is given by
\begin{equation}\label{eq:oblate_drag_compressible}
\mathcal{F}_\mathrm{D}^{\mathrm{Ma}}\left(\mathrm{Re_p},\alpha,\mathrm{Ma}\right) = \frac{\vartheta_1\big/{\log(\mathrm{Re_p})} \left(\vartheta_2{\alpha}\big/{\pi}\right)^{ \vartheta_3 \left({\mathrm{Re_p}}\right)^{\vartheta_4}} + {\log(\mathrm{Re_p})}\big/{\vartheta_5} - 1}{\left[1 + \exp\left(-\vartheta_6 (\mathrm{Ma} - \vartheta_7)\right)\right]} + \vartheta_8 \exp\left(-\frac{(\mathrm{Ma} - \vartheta_{9})^2}{\vartheta_{10}}\right)\, .
\end{equation}
The purpose of the first term on the right hand-side of Eq.~\eqref{eq:oblate_drag_compressible} is similar to the model for the prolate spheroid, it captures the sigmoid evolution of the drag coefficient as the Mach number increases from subsonic to supersonic regimes. However, in the case of the oblate spheroid, at high particle Reynolds number, the drag coefficient is no longer largest at $\alpha = 90 \degree$. This is modelled by a modification of the term multiplying the Reynolds number and the angle of attack in Eq.~\eqref{eq:oblate_drag_compressible}. The remainder of the expression is similar to the expression given for the prolate spheroid, where the second term on the right hand side also models the smooth increase in the evolution of the drag coefficient in the transonic regime. 

The correlation for the prediction of the drag coefficient for the oblate spheroid is shown in the centre of Figure~\ref{fig:drag_prediction}, along with the PR-DNS results.
For all angles of attack, except for $\alpha = 75\degree$, the correlation accurately recovers the change in the evolution of the drag coefficient in both the transonic and supersonic regimes, for all particle Reynolds numbers. 
For $\alpha = 75\degree$, the variation in the difference between the value of the drag coefficient in the subsonic and supersonic regimes deviates from other angles and remains challenging to capture. 
However, a very good agreement is observed as the maximum and median prediction errors between the correlation and the PR-DNS are \reviewerT{10.19\% and 2.1\%, respectively}.

\subsubsection{Rod-like particle}

For the rod-like particle, the shape-dependent function $\mathcal{F}_\mathrm{D}^{\mathrm{Ma}}$, in Eq.~\ref{eq:General_Drag_Coefficient}, used to model compressible effects, is given by 
\begin{equation}\label{eq:fibroid_drag_compressible}
  \reviewerT{\mathcal{F}_\mathrm{D}^{\mathrm{Ma}}\left(\mathrm{Re_p},\alpha,\mathrm{Ma}\right) = \frac{\vartheta_1 \cdot \mathrm{Re_p} + \vartheta_2 \cdot \left(\frac{\mathrm{Re_p}}{\vartheta_3}\right)^{\vartheta_4} \cdot \alpha^{\vartheta_5}}{\left[ 1.0 + \exp{\left(-\vartheta_6 \cdot (\mathrm{Ma} - \vartheta_7)\right)}\right]} 
  + \left(\frac{\vartheta_8 \cdot \alpha}{\mathrm{Re_p}}\right)^{\vartheta_9} \cdot \exp{\left(-\frac{(\mathrm{Ma} - \vartheta_{10})^2}{\vartheta_{11}}\right)}}
  \end{equation}
The first term on the right hand-side of Eq.~\eqref{eq:fibroid_drag_compressible} is similar to the expression given for the prolate spheroid, Eq.~\eqref{eq:prolate_drag_compressible}, as the observations are the same. The larger is the angle of attack, the larger is the difference between the drag coefficient in the subsonic and the supersonic flow regimes.
However, the second term on the right hand-side of Eq.~\eqref{eq:fibroid_drag_compressible}, which is also based on a Gaussian function, is scaled by the angle of attack. The rationale is that, for the rod-like particle, the bell shape that represents the change in the drag coefficient in the transonic regime is less pronounced for angles of attack close to $\alpha \rightarrow 0 \degree$, and is not present for a rod-like particle with an angle of attack $\alpha = 0 \degree$. This is shown in the right of Figure~\ref{fig:drag_prediction}, depicting the prediction of the drag coefficient of the rod-like particle with the correlation along with the PR-DNS results. For all flow cases, a good agreement is observed between the predictions and the PR-DNS results as a maximum and median prediction errors between the correlation and the PR-DNS are \reviewerT{7.06\% and 1.29\%, respectively}.
\subsection{Lift coefficient}

Following a similar approach as for the drag coefficients, the proposed correlation for the lift coefficient of each studied particle shape is based on the lift coefficient in the incompressible flow regime and a compressible correction that is a function
of the particle Reynolds number, the Mach number, and the angle of attack. 
The general expression is defined as 
\begin{equation}\label{eq:General_Lift_Coefficient}
C_\mathrm{L}\left(\mathrm{Re_p},\alpha,\mathrm{Ma}\right) = C_\mathrm{L}\left(\mathrm{Re_p},\alpha,\mathrm{Ma}=0\right)  + \mathcal{F}_\mathrm{L}^{\mathrm{Ma}}\left(\mathrm{Re_p},\alpha,\mathrm{Ma}\right)
\end{equation}
where $C_\mathrm{L}$ is the lift coefficient of a specific particle given as a function of the particle Reynolds number, angle of attack, and Mach number. This coefficient is obtained from the summation of the lift coefficient for equivalent particle Reynolds number and angle of attack in incompressible flow conditions, and an additional shape-dependent function, $\mathcal{F}_\mathrm{L}^{\mathrm{Ma}}$, to account for the change in the lift coefficient in case of compressible flow. The expressions of the shape-dependent function $\mathcal{F}_\mathrm{L}^{\mathrm{Ma}}$ are given in the dedicated sections, along with the associated prediction errors. The fitting coefficients, $\vartheta_i$, for the lift coefficients of all three considered particle shapes are given in Table~\ref{table:liftcoefficients}.

\begin{table}
    \centering
     \caption{List of the fitting coefficients in Eqs.~\eqref{eq:prolate_lift_compressible}, ~\eqref{eq:oblate_lift_compressible}, and~\eqref{eq:fibroid_lift_compressible}, used in the correlation to predict the change in the lift coefficient in case of compressible flow.}\label{table:liftcoefficients}
    \begin{tabular}{l || c c c c c c c c}
     & $\vartheta_1$ & $\vartheta_2$ & $\vartheta_3$ & $\vartheta_4$ & $\vartheta_5$ & $\vartheta_6$ & $\vartheta_7$ & $\vartheta_8$ \\
    \hline
    \hline
    \reviewerT{Eq.~\eqref{eq:prolate_lift_compressible}} & 1.17 & -0.735 & -21.8 & 0.870 & 0.277 & 1.12 & 0.217 & 2.00\\
    \reviewerT{Eq.~\eqref{eq:oblate_lift_compressible}} & 0.535 & 130 & 0.807 & 1.84 & 1.56 & -0.996 & 2.25 & 0.516 \\
    \reviewerT{Eq.~\eqref{eq:fibroid_lift_compressible}} & 2.43 & 1.20 & -24.8 & 0.807 & 2.43 & 1.24 & 0.237 & 4.22 \\
  \end{tabular}
\end{table}

\begin{figure}
  \begin{center}
    \includegraphics[width=1.0\linewidth]{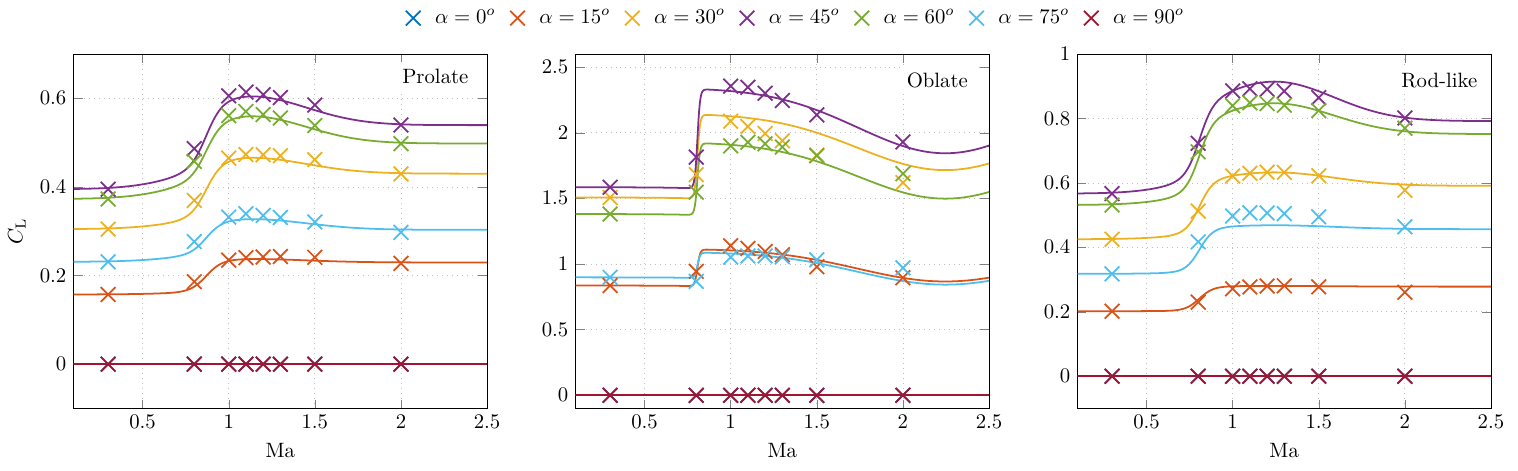}
    \includegraphics[width=1.0\linewidth]{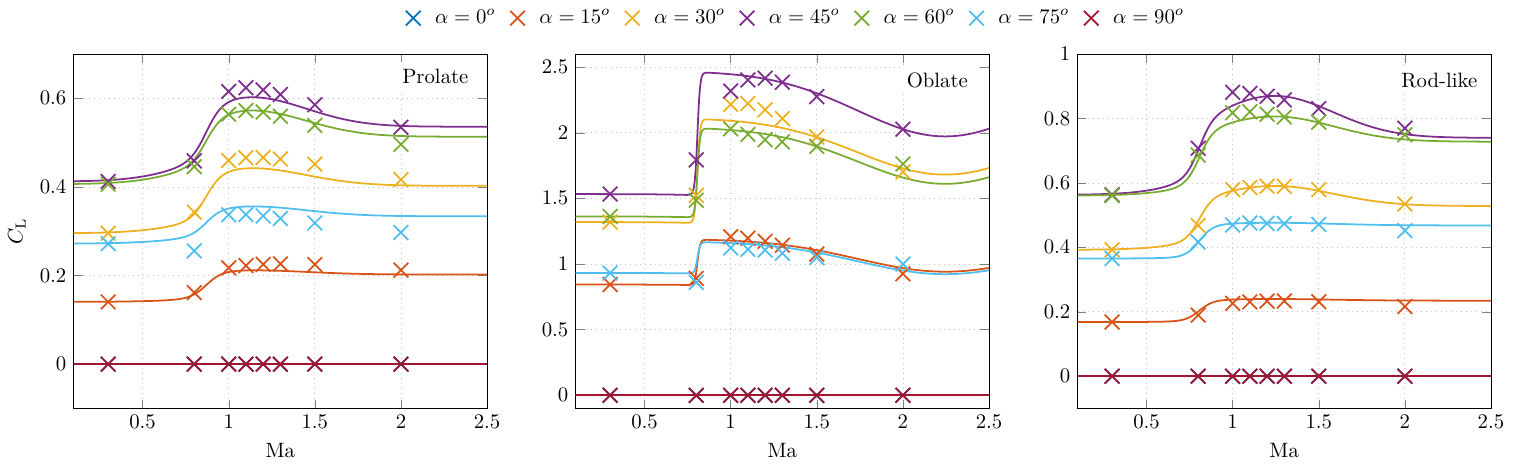}
  \end{center}
  \caption{The lift coefficient as a function of the Mach number for the three shapes studied, at different particle Reynolds numbers, for all considered angles of attack. From left column to right column: prolate spheroid, oblate spheroid, and rod-like particle. Top row: $\mathrm{Re_p} = 100$, bottom row: $\mathrm{Re_p} = 300$. The colour indicates the angle of attack, the marker indicates PR-DNS result, and the solid line shows the prediction of the lift coefficient given by the derived correlations, Eqs.~\eqref{eq:prolate_lift_compressible}, ~\eqref{eq:oblate_lift_compressible}, and~\eqref{eq:fibroid_lift_compressible}.}
\label{fig:lift_prediction}
\end{figure}

\subsubsection{Prolate spheroid}
To consider compressible effects in the prediction of the lift coefficient of an isolated prolate spheroid particle, the shape-dependent function, $\mathcal{F}_\mathrm{L}^{\mathrm{Ma}}$, in Eq.~\ref{eq:General_Lift_Coefficient}, is given by
\begin{align}\label{eq:prolate_lift_compressible}
\begin{split}
\mathcal{F}_\mathrm{L}^{\mathrm{Ma}}\left(\mathrm{Re_p},\alpha,\mathrm{Ma}\right) = & \frac{
\vartheta_1 \log\left(\mathrm{Re_p}\right)^{\vartheta_2}}{\left[1 + \exp\left(\vartheta_3 (\mathrm{Ma} - \vartheta_4)\right)\right]} \cos\left(\Psi_1\left(\alpha, \mathrm{Re_p}\right)\right) \sin\left(\Psi_1\left(\alpha, \mathrm{Re_p}\right)\right)\\ & + \vartheta_5 \exp\left(\frac{-(\mathrm{Ma} - \vartheta_6)^2}{ \vartheta_7}\right) \left[
\cos\left(\Psi_2\left(\alpha, \mathrm{Re_p}\right)\right)
\sin\left(\Psi_2\left(\alpha, \mathrm{Re_p}\right)\right)\right]^{\vartheta_8}\, .
\end{split}
\end{align}
As for the drag correlation derived for the prolate spheroid, Eq.~\eqref{eq:prolate_drag_compressible}, the correlation to predict the lift coefficient is split in two terms: a sigmoid term to model the change in the lift coefficient from subsonic to supersonic regimes, and a Gaussian function to model the smooth bell-shaped evolution of the coefficient from the subsonic to the supersonic flow regime. These two terms are multiplied by a cosine-sine factor so that the lift coefficients remain equal to ${C}_\mathrm{L} = 0$ for a prolate particle with angles of attack $\alpha = 0 \degree$ and $90 \degree$, and maximum for $\alpha = 45\degree$.
It is important to mention that for finite particle Reynolds numbers, such as those considered in the present study, the lift coefficient is no longer largest for a particle with an angle of attack of $\alpha = 45\degree$ \cite{Frohlich2020,Sanjeevi2022,Cheron2024}. This is also observed in this work and shifting functions are used to model the change in the maximum value of the lift coefficient as a function of the angle of attack. These functions are given by
\begin{equation}
    \Psi_1 = \cfrac{\pi}{2} \left(\alpha \cfrac{2}{\pi}\right)^{1 + 4.185 \times 10^{-5}\log{\left(\mathrm{Re_p}\right)}^{-10.5}}
\end{equation}
for the first cosine-sine term in Eq.~\eqref{eq:prolate_lift_compressible} and
\begin{equation}
    \Psi_2 = \cfrac{\pi}{2} \left(\alpha \cfrac{2}{\pi}\right)^{1 + 2.146 \log{\left(\mathrm{Re_p}\right)}^{-1.444}}
\end{equation}
for the second cosine-sine term in Eq.~\eqref{eq:prolate_lift_compressible}.

The PR-DNS results along with the correlation for the prediction of the lift coefficient of the prolate spheroid are shown in the left of Figure~\ref{fig:lift_prediction}, as a function of the Mach number, for all angles of attack and particle Reynolds number values considered in this work. 
The correlation accurately recovers the evolution of the lift coefficient in all regimes, for all particle Reynolds numbers and angles of attack. 
The maximum and median prediction errors between the correlation and the PR-DNS are \reviewerT{13.93\% and 0.72\%, respectively}.


\subsubsection{Oblate spheroid}

For the oblate spheroid, the shape-dependent function, $\mathcal{F}_\mathrm{L}^{\mathrm{Ma}}$, in Eq.~\ref{eq:General_Lift_Coefficient}, is given by
\begin{align}\label{eq:oblate_lift_compressible}
\begin{split}
\mathcal{F}_\mathrm{L}^{\mathrm{Ma}}\left(\mathrm{Re_p},\alpha,\mathrm{Ma}\right) = & 
    \frac{\vartheta_{1} \log(\mathrm{Re_p})}{\left[1 + \exp\left(-\vartheta_2 (\mathrm{Ma} - \vartheta_3)\right)\right]} \cos(\alpha)^{\vartheta_4} \sin(\alpha)^{\vartheta_5} \\
    & + \vartheta_6 \exp\left(-\frac{(\mathrm{Ma} - \vartheta_7)^2}{\vartheta_8}\right) \cos(\alpha) \sin(\alpha)\, .
\end{split}
\end{align}
The results of the evolution of the lift coefficient as a function of the angle of attack for the oblate spheroid, see the centre of Figure~\ref{fig:lift_prediction}, show that the values for the lift coefficients are almost symmetric with respect to the angle of attack $\alpha = 45\degree$.
This trend indicates that the shift in the angle of attack at which the maximum lift coefficient occurs is less influenced by the increase in the particle Reynolds number than for the rod-like and the prolate spheroid particles.
\reviewerT{The specific evolution of the lift coefficient that varies from the other shapes confirms the necessity to have a unique correlation to predict the lift coefficient. This unique modification relies on the change of the exponent in the cosine-sine multiplication, realized by the exponents $\vartheta_4$ and $\vartheta_5$.}
In general, the correlation agrees well with the PR-DNS results, as the maximum and median prediction errors between the correlation and the PR-DNS are \reviewerT{13.66\% and 0.19\%, respectively}.

\subsubsection{Rod-like particle}

To consider compressible effects in the prediction of the lift coefficient of an isolated rod-like particle, the shape-dependent function, $\mathcal{F}_\mathrm{L}^{\mathrm{Ma}}$, in Eq.~\ref{eq:General_Lift_Coefficient}, is given by
\begin{align}\label{eq:fibroid_lift_compressible}
\begin{split}
\mathcal{F}_\mathrm{L}^{\mathrm{Ma}}\left(\mathrm{Re_p},\alpha,\mathrm{Ma}\right) = & \frac{
\reviewerT{\left(\vartheta_1 / \log\left(\mathrm{Re_p}\right)\right)}^{\vartheta_2}}{\left[1 + \exp\left(\vartheta_3 \cdot (\mathrm{Ma} - \vartheta_4)\right)\right]} \cos\left(\Psi_1\left(\alpha, \mathrm{Re_p}\right)\right) \sin\left(\Psi_1\left(\alpha, \mathrm{Re_p}\right)\right)\\ & + \vartheta_5 \exp\left(\frac{-(\mathrm{Ma} - \vartheta_6)^2}{ \vartheta_7}\right) \left[
\cos\left(\Psi_2\left(\alpha, \mathrm{Re_p}\right)\right)
\sin\left(\Psi_2\left(\alpha, \mathrm{Re_p}\right)\right)\right]^{\vartheta_8}\, .
\end{split}
\end{align}
This expression is similar to the expression proposed to model the lift coefficient of the prolate spheroid, Eq.~\eqref{eq:fibroid_lift_compressible}, the correlation to predict the lift coefficient is split in the sigmoid term, for the transition from the subsonic to the supersonic flow regime, and the Gaussian function term for the bump in the transonic regime. These two terms are multiplied by a cosine-sine factor, where the angle of attack is modified using shifting functions to model the change in the maximum value of the lift coefficient as a function of the angle of attack \cite{Cheron2024}.
These functions are given by
\begin{equation}
    \Psi_1 = \cfrac{\pi}{2} \left(\alpha \cfrac{2}{\pi}\right)^{1 + 2.471 \log{\left(\mathrm{Re_p}\right)}^{-1.50}}
\end{equation}
for the first cosine-sine term in Eq.~\eqref{eq:fibroid_lift_compressible} and
\begin{equation}
    \Psi_2 = \cfrac{\pi}{2} \left(\alpha \cfrac{2}{\pi}\right)^{1 + \reviewerT{1630.79 \left(1 / \log{\left(\mathrm{Re_p}\right)}\right)^{6.00}}}
\end{equation}
for the second cosine-sine term in Eq.~\eqref{eq:fibroid_lift_compressible}.

The PR-DNS results along with the correlation for the prediction of the rod-like particle lift coefficient are shown in the right plot of Figure~\ref{fig:lift_prediction}, as a function of the Mach number, for all angles of attack and particle Reynolds number values considered in this work. 
The correlation accurately recovers the evolution of the lift coefficient in all regimes, for all particle Reynolds numbers and angles of attack. 
The maximum and median prediction errors between the correlation and the PR-DNS are \reviewerT{7.97\% and 0.15\%, respectively}.

\subsection{Torque coefficient}

\begin{figure}
  \begin{center}
    \includegraphics[width=1.0\linewidth]{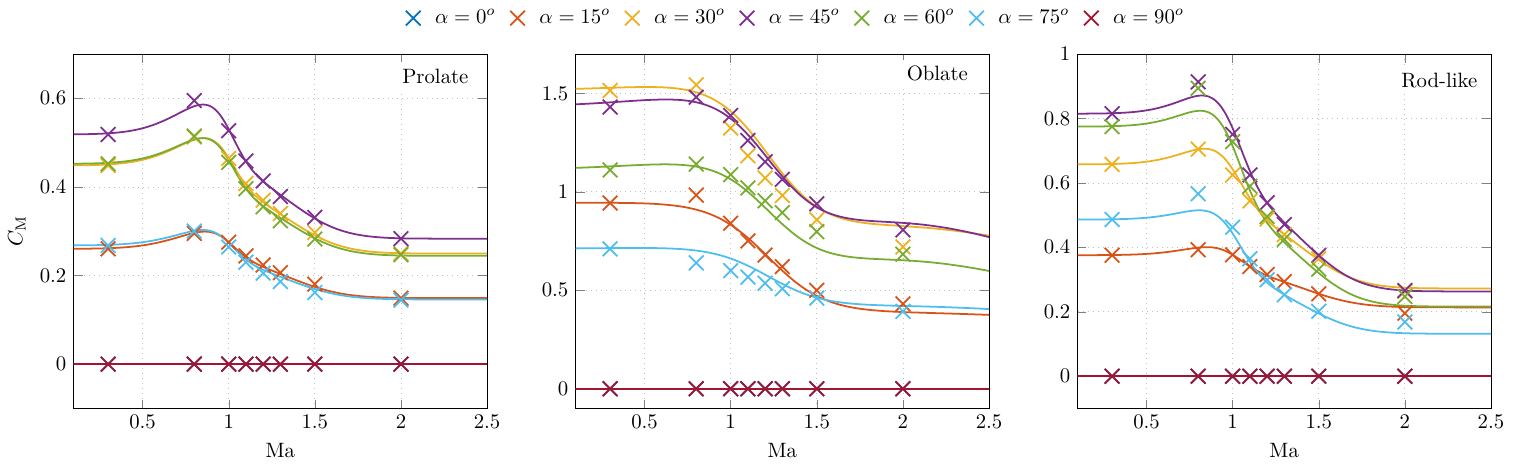}
    \includegraphics[width=1.0\linewidth]{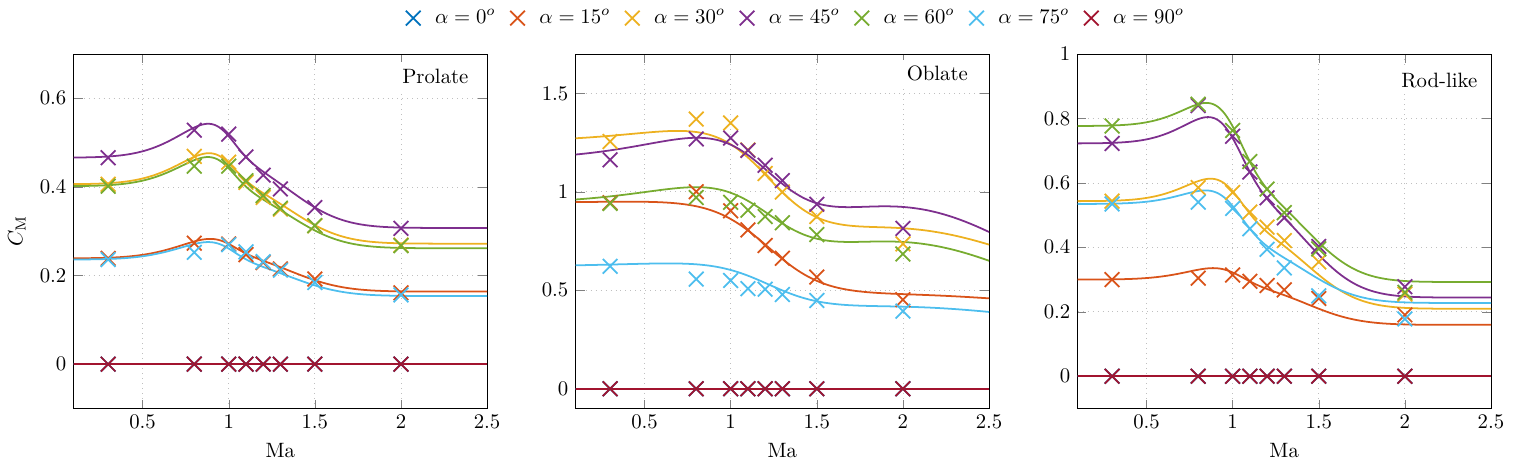}
  \end{center}
  \caption{The torque coefficient as a function of the Mach number for the three shapes studied, at different particle Reynolds numbers, for all considered angles of attack. From left column to right column: prolate spheroid, oblate spheroid, and rod-like particle. Top row: $\mathrm{Re_p} = 100$, bottom row: $\mathrm{Re_p} = 300$. The colour indicates the angle of attack, the marker indicates PR-DNS result, and the solid line shows the prediction of the torque coefficient \reviewerT{given by the derived correlations, Eqs.~\eqref{eq:correlation_torque_compressible}, ~\eqref{eq:correlation_torque_compressible_Psi1}, and~\eqref{eq:correlation_torque_compressible_Psi2}.}}
\label{fig:torque_prediction}
\end{figure}

As for the drag and lift coefficients, the proposed correlation for the torque coefficient of each studied particle shape is based on the torque coefficient in the incompressible flow regime and a compressible correction that is a function of the particle Reynolds number, the Mach number, and the angle of attack. 
The general expression is defined as
\begin{equation}\label{eq:General_Torque_Coefficient}
C_\mathrm{M}\left(\mathrm{Re_p},\alpha,\mathrm{Ma}\right) = C_\mathrm{M}\left(\mathrm{Re_p},\alpha,\mathrm{Ma}=0\right)  + \mathcal{F}_\mathrm{M}^{\mathrm{Ma}}\left(\mathrm{Re_p},\alpha,\mathrm{Ma}\right)
\end{equation}
where $C_\mathrm{M}$ is the torque coefficient of a specific particle given as a function of the particle Reynolds number, the Mach number, and the angle of attack. This coefficient is obtained from the summation of the torque coefficient for equivalent particle Reynolds number and angle of attack in incompressible flow conditions, and an additional shape-dependent function, $\mathcal{F}_\mathrm{M}^{\mathrm{Ma}}$, to account for the change in the torque coefficient in case of compressible flow.
For all shapes, the evolution of the torque coefficient as a function of the Mach number is very similar, so that the general expression of the shape-dependent function is unique, and given by
\reviewerT{ \begin{align}\label{eq:correlation_torque_compressible}
  \begin{split}
  \mathcal{F}_\mathrm{M}^{\mathrm{Ma}} & =  \frac{\vartheta_1 \mathrm{Re_p} + \vartheta_2  \left( \mathrm{Re_p} / \vartheta_3 \right)^{\vartheta_4}}{1 + \exp\left(-\vartheta_5 \left( \mathrm{Ma} - \vartheta_{6} \right)\right)} \cdot \left( \cos\left(\Psi_1\right) \sin\left(\Psi_1\right) \right)^{\vartheta_{7}} \\
    & + \left( \frac{\vartheta_{8} \alpha}{\vartheta_{10}}\right)^{\vartheta_9} \cdot \left(\frac{\vartheta_{11} \mathrm{Re_p}}{\vartheta_{13}}\right)^{\vartheta_{12}} \cdot \exp\left(
    \frac{-\left( \mathrm{Ma} - \vartheta_{14}\right)^2}{\vartheta_{15}}\right) \cdot \left( \cos\left(\Psi_2\right) \sin\left(\Psi_2\right) \right)^{\vartheta_{16}},
  \end{split}
\end{align}}
\reviewerT{with the functions to shift the maximum value of the cosine-sine product given as}
\begin{equation}\label{eq:correlation_torque_compressible_Psi1}
  \reviewerT{\Psi_1 = \cfrac{\pi}{2} \left(\alpha \cfrac{2}{\pi}\right)^{\vartheta_{17}},}
\end{equation}
\reviewerT{and}
\begin{equation}\label{eq:correlation_torque_compressible_Psi2}
  \reviewerT{\Psi_2 = \cfrac{\pi}{2} \left(\alpha \cfrac{2}{\pi}\right)^{\vartheta_{18}}.}
\end{equation} 
\reviewerT{As for the drag and lift correlations, the correlation to predict the torque coefficient is split in two main terms: the sigmoid term to model the change in the torque coefficient from subsonic to supersonice regimes, and a Gaussian function to model the smooth evolution of the coefficient from the subsonic to the supersonic flow regime. 
Both terms are multiplied by a shifted cosine-sine factor so that the torque coefficients remain zero for the angles of attack $\alpha = 0\degree$ and $\alpha = 90\degree$, and to model the shift in the maximum torque coefficient. The fitting coefficients, $\vartheta_i$, for the torque coefficients of all three considered particle shapes are given in Table \ref{table:torquecoefficients}.
The results of the PR-DNS along with the correlation for the torque coefficient of the prolate spheroid are shown in the left of Figure~\ref{fig:torque_prediction}, as a function of the Mach number, for all angles of attack and particle Reynolds number values considered in this work. 
The change from the subsonic to the supersonic flow regimes is accurately recovered for all considered angles of attack and particle Reynolds numbers by the sigmoid term, leading to maximum and median prediction errors between the correlation and the PR-DNS of 6.88\% and 0.44\%, respectively.
As for the prolate spheroid, the model prediction for the torque coefficient of the oblate spheroid is shown with the PR-DNS results in the centre of Figure~\ref{fig:torque_prediction}. The evolution of the torque coefficient of the oblate spheroid is very similar compared with the prolate spheroid and the correlation accurately recovers the smooth transition from the subsonic to the supersonic flow regime, leading to maximum and median prediction errors between the correlation and the PR-DNS of 12.96\% and 1.40\%, respectively.
For the rod-like particle, the model prediction for the torque coefficient is shown against the PR-DNS results in the right of Figure \ref{fig:torque_prediction}. In this case, the transition from subsonic to supersonic flow regime is much more significant compared to the other two particle shapes and the PR-DNS results in the transition region are more narrow, which makes it more difficult to find an accurate model prediction. Hence, the maximum prediction error between the correlation and the PR-DNS is 22.40\% because of a single outlier, but the median prediction error is 0.18\%, which shows that the correlation is over all in an excellent agreement with the PR-DNS results. 
}
\begin{table}
  \centering
  \reviewerT{
    
     \caption{\reviewerT{List of the fitting coefficients in Eqs.~\eqref{eq:correlation_torque_compressible}, ~\eqref{eq:correlation_torque_compressible_Psi1}, and~\eqref{eq:correlation_torque_compressible_Psi2}, used in the correlation to predict the change in the torque coefficient in case of compressible flow.}}\label{table:torquecoefficients}
    \begin{tabular}{l || c c c}
     & Prolate & Oblate & Rod-like \\
    \hline
    \hline
    $\vartheta_1$ & 2.28 $\times 10^{-4}$ & -4.43 $\times 10^{-3}$ & -2.32 $\times 10^{-5}$ \\ 
    $\vartheta_2$ & -0.444 & -0.759 & -1.15 \\ 
    $\vartheta_3$ & 157 & 203 & 127 \\ 
    $\vartheta_4$ & -0.223 & -1.03 & -0.134 \\ 
    $\vartheta_5$ & 15.1 & 5.93 & 12.3 \\ 
    $\vartheta_6$ & 1.01 & 1.22 & 1.03 \\ 
    $\vartheta_7$ & 0.993 & 1.11 & 1.04 \\ 
    $\vartheta_8$ & 4.96 $\times 10^{-2}$ & 3.18 $\times 10^{-9}$ & 13.8 \\ 
    $\vartheta_9$ & 1.44 & 8.95 $\times 10^{-2}$ & 1.88 \\ 
    $\vartheta_{10}$ & 7.42 $\times 10^{-3}$ & 3.59 $\times 10^{-2}$ & 3.64 \\ 
    $\vartheta_{11}$ & 3.81 $\times 10^{-2}$ & 3.89 & 0.520 \\ 
    $\vartheta_{12}$ & 1.68 $\times 10^{-2}$ & 0.569 & 0.158 \\ 
    $\vartheta_{13}$ & 5.86 $\times 10^{-9}$ & 10.8 & 1.22 $\times 10^{-6}$ \\ 
    $\vartheta_{14}$ & 1.10 & 1.98 & 1.18 \\ 
    $\vartheta_{15}$ & 9.29 $\times 10^{-3}$ & 1.04 & 3.71 $\times 10^{-3}$ \\ 
    $\vartheta_{16}$ & 1.05 & 2.50 & 1.16 \\ 
    $\vartheta_{17}$ & 0.164 & 1.17 & 0.157 \\ 
    $\vartheta_{18}$ & 1.04 & 0.809 & 1.32 \\ 
    \end{tabular}
  }
\end{table}

\section{Conclusions}
\label{Conclusion}
\reviewerT{
In this study, we have presented particle-resolved direct numerical simulations (PR-DNS) with three-dimensional body-fitted hexahedral meshes to investigate the aerodynamic force and torque coefficients of particles with different shapes, using a finite-volume-based solver for compressible flows. 
The employed computational domain is spherical, centred around the particle, facilitating the rotation of the particle without changing the computational mesh. 
This computational domain is represented by non-uniform and body-conforming meshes, refined near the particle.
PR-DNS are carried out for three particle shapes, namely a prolate spheroid with an aspect ratio of 5/2, an oblate spheroid with an aspect ratio of 5, and a rod-like particle with an aspect ratio of 5. The simulations are carried out at a reference temperature of 300 K and a reference pressure of $10^5$ Pa, maintaining a Prandtl number of 0.71.

This study has examined the behaviour of non-spherical particles in compressible flows, focusing on the steady-state velocity as well as the aerodynamic force coefficients (drag, lift, and  torque) for the three considered particle shapes at Mach numbers ranging from 0.3 to 2.0, angles of attack from 0\degree to 90\degree, and particle Reynolds numbers between 100 and 300.
The bow shock developing in the supersonic flow regime in front of the particle is largest for the oblate spheroid and smallest for the prolate spheroid.
In the transonic regime, the oblate spheroid exhibits an unstable wake behaviour when the angle of attack is larger than $0\degree$. The drag coefficients increase with the angle of attack and are highest for the oblate spheroid. All shapes show a rapid increase in drag in the transonic regime, followed by a decrease in the supersonic regime. The lift and  torque coefficients follow similar trends, with the oblate spheroid having the highest values.
We can, therefore, conclude that the shape of the particles impacts the flow behaviour and aerodynamic forces significantly, with oblate spheroids showing the most pronounced effects due to their geometric characteristics.

The correlations proposed in this work are valid for particle Reynolds numbers between 100 and 300, Mach numbers between 0 and 2, and angles of attack ranging from $0 \degree$ to $90 \degree$. The correlations accurately recover the change in the evolution of the aerodynamic force and torque coefficients in both transonic and supersonic regimes, for all considered particle Reynolds numbers and angles of attack. The largest maximum error of the correlations for the drag coefficient is 10.19\% for the prolate spheroid, while the median error for the three considered particle shapes is approximately 1-2\%. The proposed correlations for the lift coefficient have a slightly larger maximum error, with the prolate particle correlation having the highest maximum error of 13.93\%. However, the median error of the correlations for the lift coefficient is below 1\% which highlights the overall excellent accuracy. For the torque coefficient, the correlation of the rod-like particle has the largest maximum error with 22.4\%. Again, the median error of the torque correlations for all three particles shapes is approximately 1\%, which indicates that the relatively high maximum error comes from a single outlier, but the overall correlations are in excellent agreement.

The development and validation of aerodynamic force and torque correlations for non-spherical particles in compressible flows mark a significant advancement in the field of multiphase flow modelling. By providing accurate and tailored correlations, this work addresses the long-standing challenge of representing the complex behaviour of non-spherical particles under varying and compressible flow conditions. The new correlations will, for example, enhance the precision of point-particle simulations, enabling more reliable predictions of particle-laden flow dynamics for compressible flow regimes.

While our current study focuses on oblate spheroids, prolate spheroids, and rod-like particles, these shapes serve as idealized cases that span a range of aspect ratios and flow characteristics. For particles with shapes that deviate from these categories, various approaches could be considered.
One possibility could be the interpolation between the correlations. If the particle shape is approximately spheroidal or rod-like but does not fit perfectly within our definitions, users might interpolate between the aerodynamic force coefficients for oblate, prolate, and rod-like shapes based on the particle aspect ratio and geometry.
For irregular or non-standard shapes, the methodology used in this study (e.g., using numerical simulations or experimental data) could be extended to derive additional correlations for those shapes. Alternatively, a shape factor or a drag coefficient correction based on empirical or computational approaches could be applied.
Users may also combine our correlations with existing generalized drag or force models that account for irregular particle shapes through empirical shape factors or coefficients.
}
\section*{Data availability statement} 
The data that support the findings of this study are reproducible
and files to regenerate the data as well as an executable to implement the correlations to predict the drag, lift, and torque coefficients are openly available in the repository 
with DOI \href{https://doi.org/10.5281/zenodo.14283844}{10.5281/zenodo.14283844}
\section*{Acknowledgements}
\label{Acknowledgement}
This research was funded by the Deutsche Forschungsgemeinschaft (DFG, German Research Foundation) - Project-ID 447633787.

\appendix

\section{\reviewerT{Mesh convergence study}}
\label{Appendix A}
\begin{figure}[ht]
  \centering
  \includegraphics[width=0.5\linewidth]{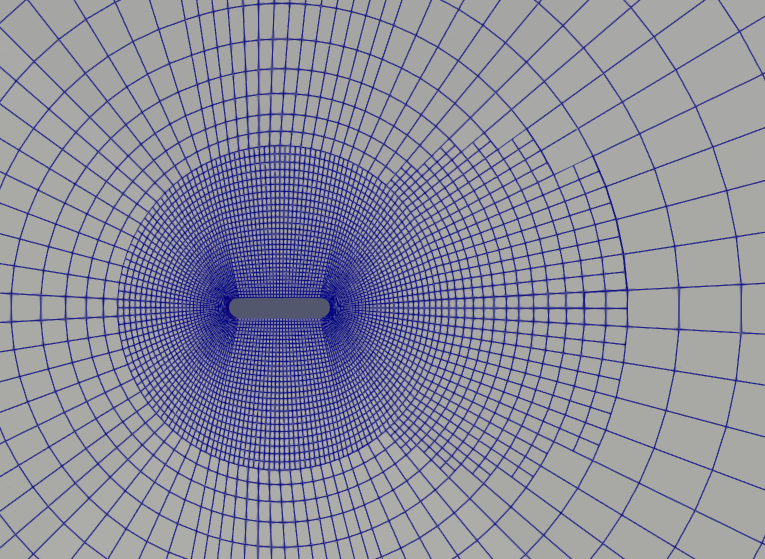}
   \caption{\reviewerT{Example of a static AMR mesh, showing the mesh refinement in the area of interest around the particle and its wake.}}
  \label{fig:Static_AMR}
  \end{figure}

\begin{figure}[H]
	\centering
    \subcaptionbox{$\mathrm{Re} = 300$, $\mathrm{Ma} = 0.3$, $\alpha = 0\degree$}{\includegraphics[scale=0.53]{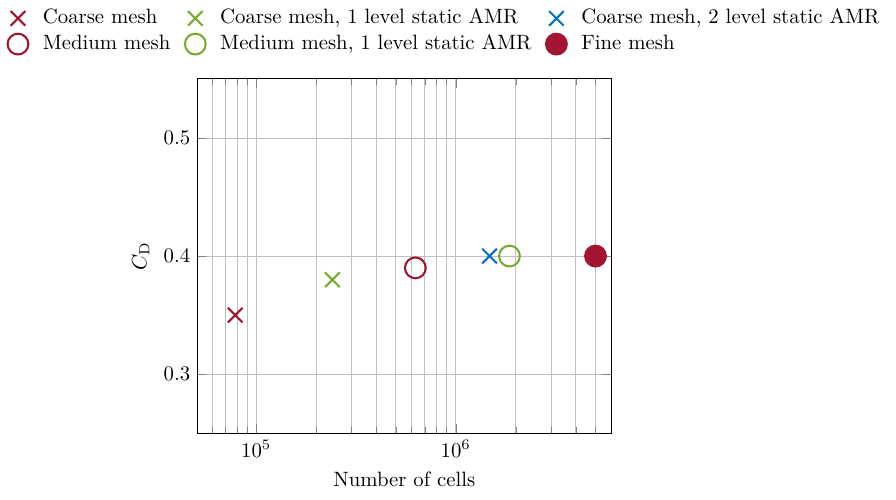}}
    \subcaptionbox{$\mathrm{Re} = 300$, $\mathrm{Ma} = 2.0$, $\alpha = 0\degree$}{\includegraphics[scale=0.53]{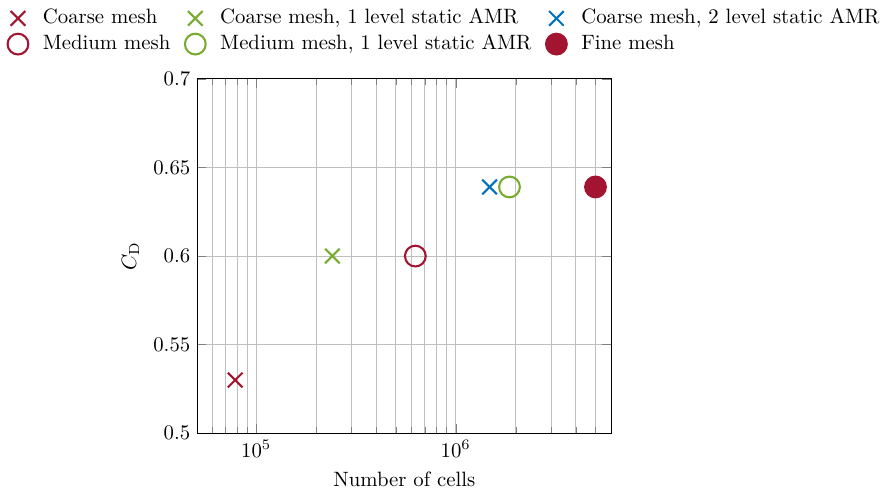}}
\caption{\reviewerT{Convergence behaviour of the drag coefficient for a prolate spheroid particle.}}
\label{fig:Convergence_Prolate}
\end{figure}

\begin{figure}[H]
	\centering
    \subcaptionbox{$\mathrm{Re} = 300$, $\mathrm{Ma} = 0.3$, $\alpha = 0\degree$}{\includegraphics[scale=0.53]{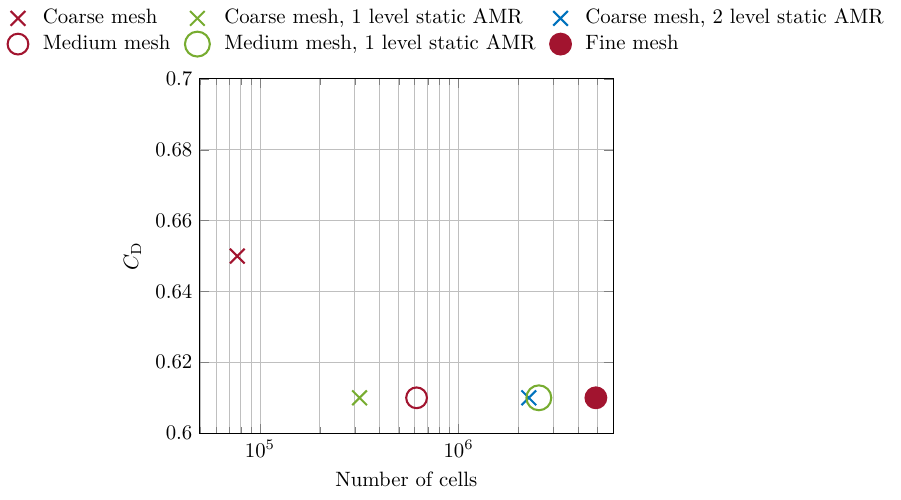}}
    \subcaptionbox{$\mathrm{Re} = 300$, $\mathrm{Ma} = 2.0$, $\alpha = 0\degree$}{\includegraphics[scale=0.53]{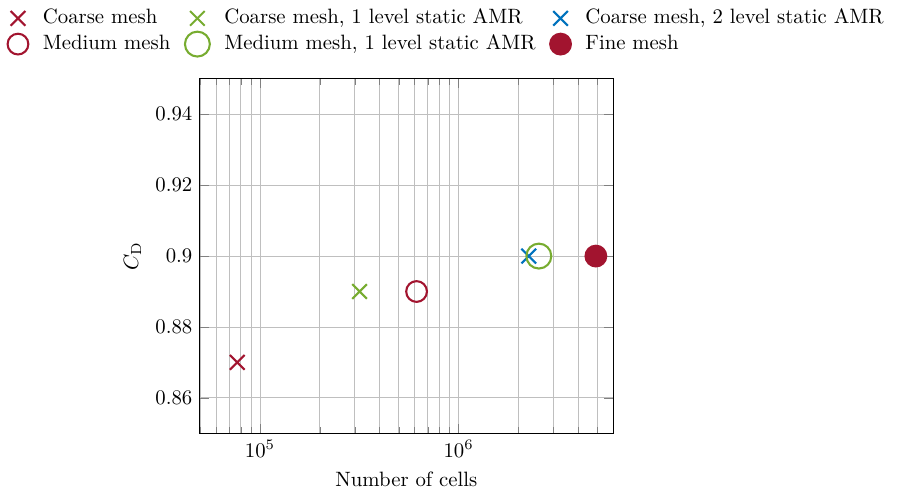}}
\caption{\reviewerT{Convergence behaviour of the drag coefficient for a oblate spheroid particle.}}
\label{fig:Convergence_Oblate}
\end{figure}

\begin{figure}[H]
	\centering
    \subcaptionbox{$\mathrm{Re} = 300$, $\mathrm{Ma} = 0.3$, $\alpha = 0\degree$}{\includegraphics[scale=0.53]{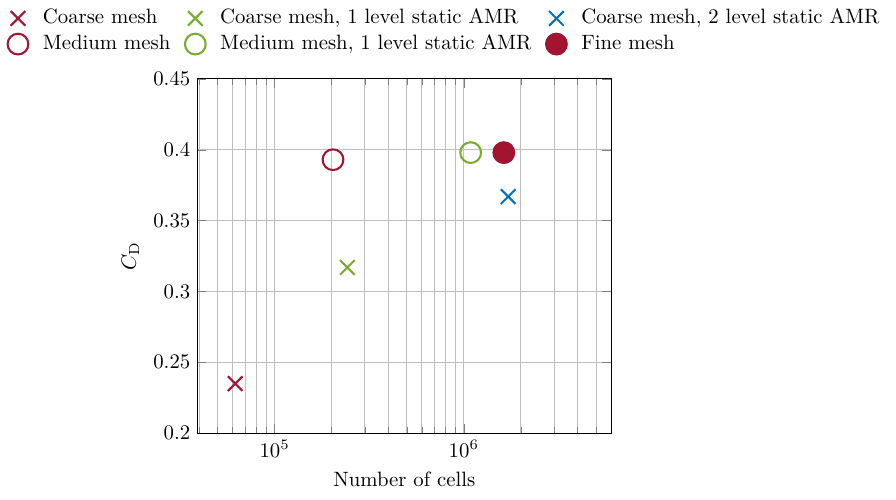}}
    \subcaptionbox{$\mathrm{Re} = 300$, $\mathrm{Ma} = 2.0$, $\alpha = 0\degree$}{\includegraphics[scale=0.53]{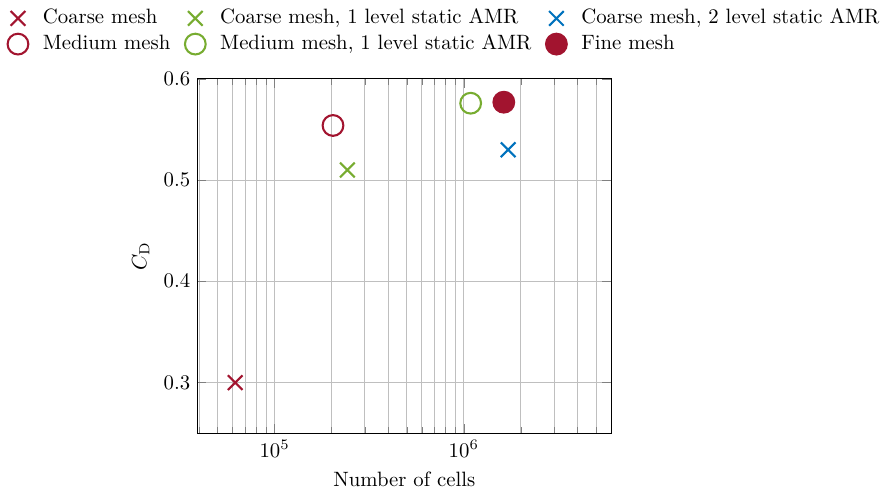}}
\caption{\reviewerT{Convergence behaviour of the drag coefficient for a rod-like particle.}}
\label{fig:Convergence_Rodlike}
\end{figure}

\reviewerT{In this appendix we give the mesh convergence studies for the three non-spherical particle shapes. For each shape the convergence behaviour of the drag coefficient is shown for $\mathrm{Re}=300$, $\alpha=0$ and $\mathrm{Ma}$ of 0.3 and 2. A coarse and a medium mesh with and without static adaptive mesh refinement (AMR) (see Figure \ref{fig:Static_AMR}) in the area of interest are compared to a fine mesh. 
In Figures \ref{fig:Convergence_Prolate} - \ref{fig:Convergence_Rodlike}, the crosses represent the results for a coarse base mesh with optional static AMR in the area of interest around the particle and its wake. One and two levels of static AMR refinement mean that the initial coarse cells in the area of interest are subdivided once or twice, resulting in a finer mesh where needed. The circles represent a medium mesh with optional static AMR and the filled circle represent a very fine mesh as a reference. 
}


\section{\reviewerO{Lift and torque coefficient diagrams}}
\label{Appendix B}
\begin{figure}[H]
	\centering
    \subcaptionbox{Prolate spheroid\label{fig:CL_ProlateSpheroid}}{\includegraphics[width=15cm, height=6cm]{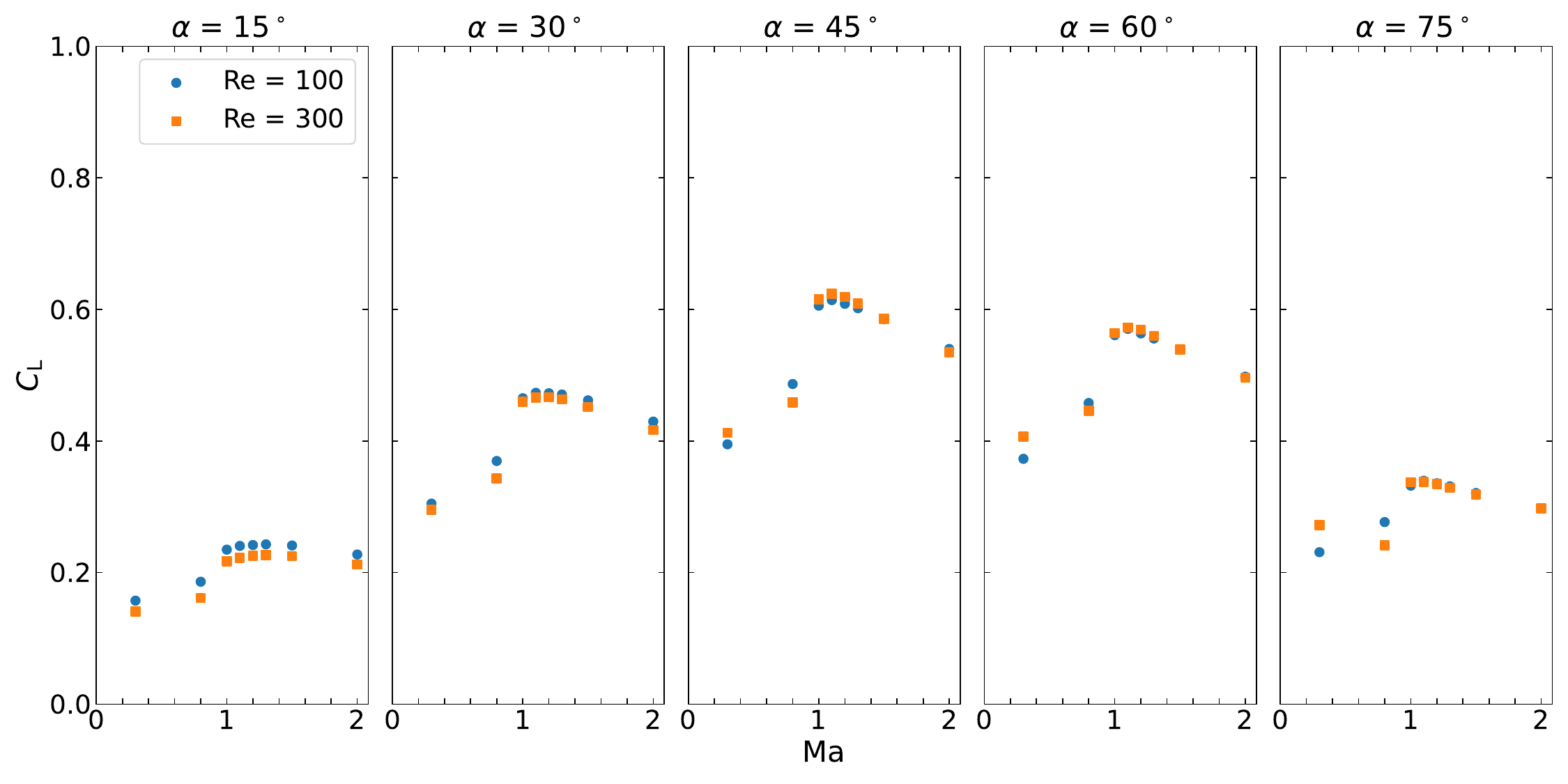}}\\
    \subcaptionbox{Oblate spheroid\label{fig:CL_OblateSpheroid}}{\includegraphics[width=15cm, height=6cm]{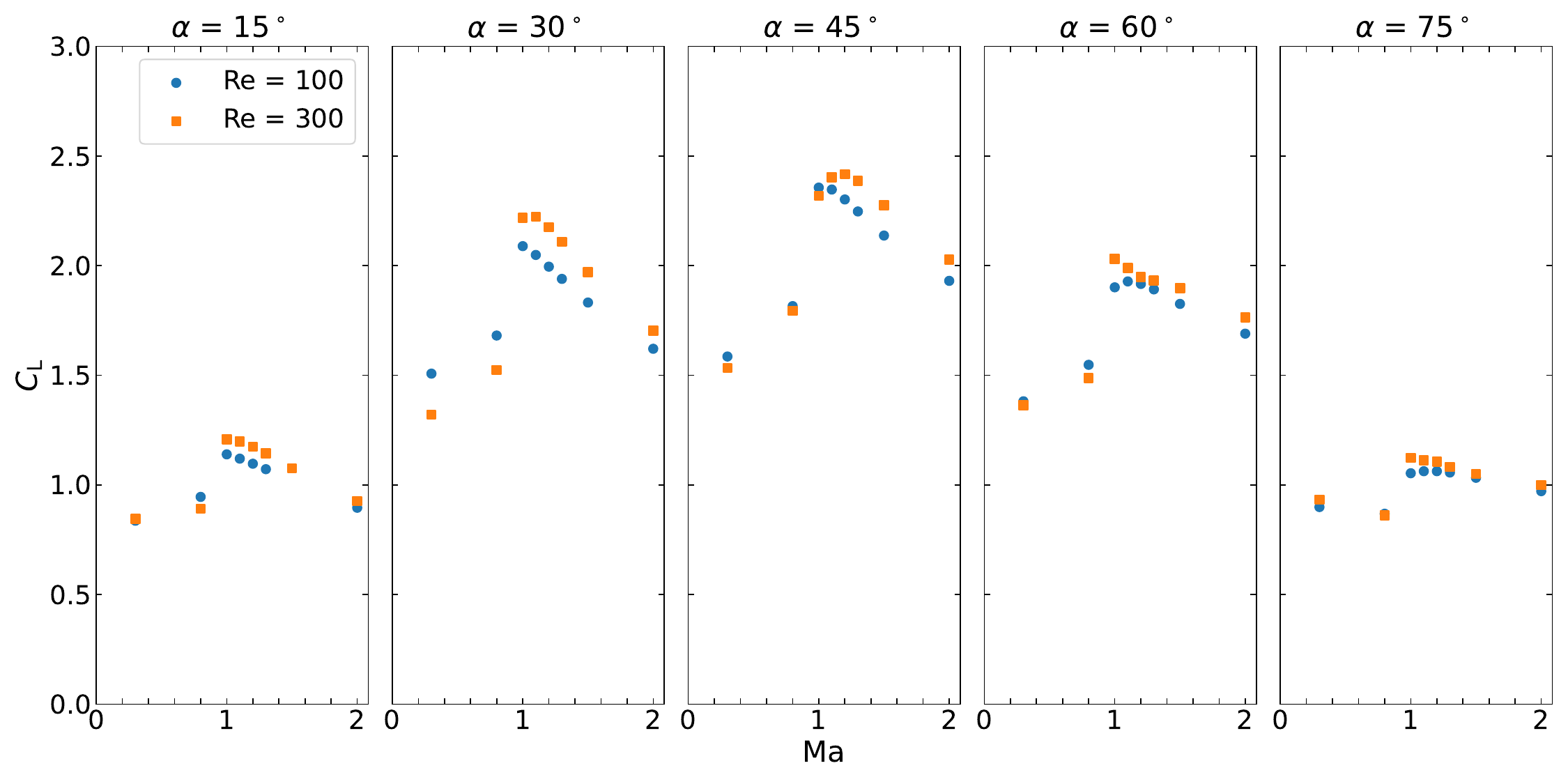}}\\
    \subcaptionbox{Rod-like particle\label{fig:CL_Fibroid}}{\includegraphics[width=15cm, height=6cm]{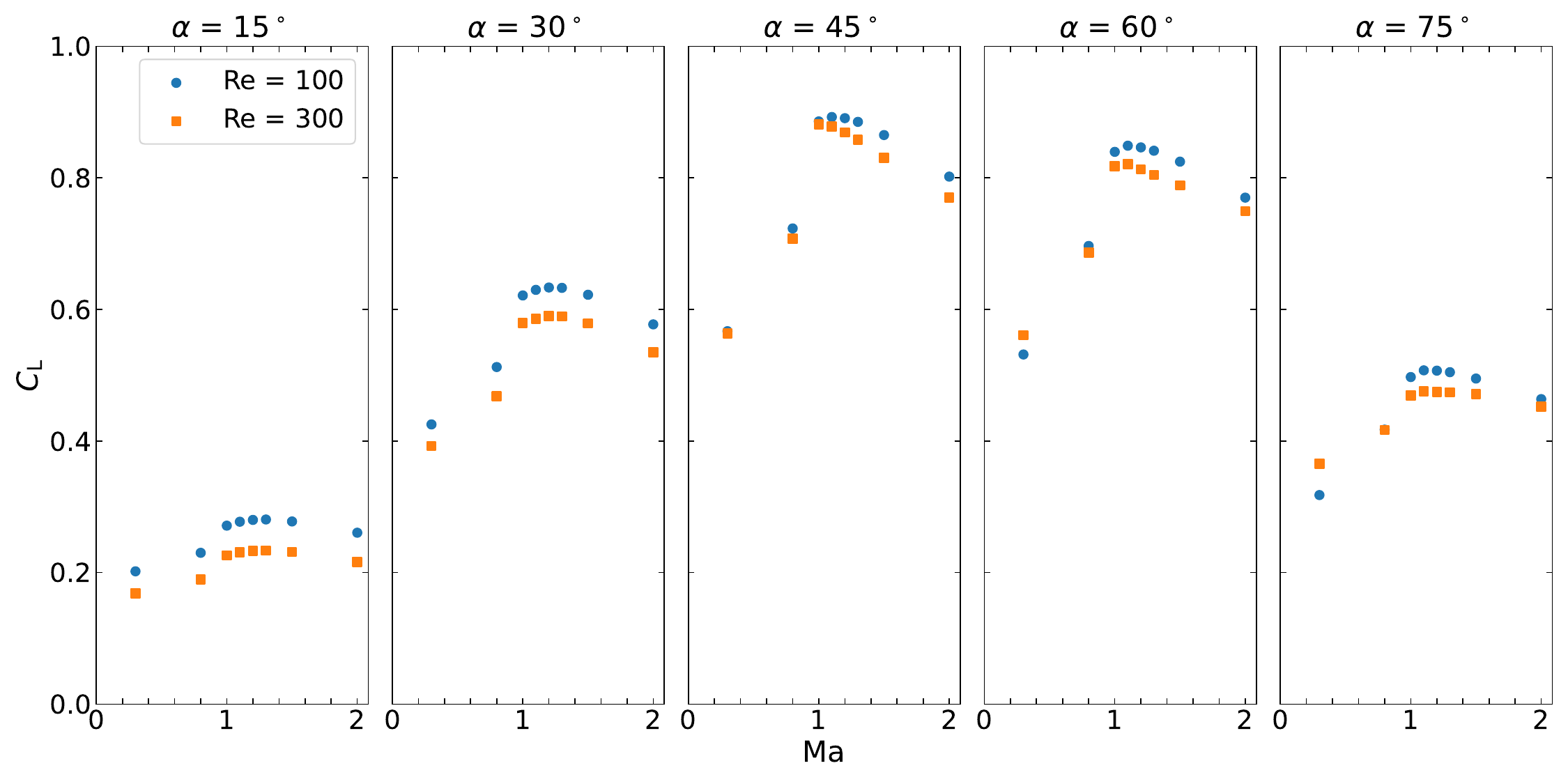}}  
\caption{Variation of the lift coefficient $C_{\mathrm{L}}$ with Mach number $\mathrm{Ma}$ for different angles of attack considered for the three considered non-spherical particle shapes.}
\label{fig:CL_over_M}
\end{figure}

\begin{figure}[H]
	\centering
    \subcaptionbox{Prolate spheroid\label{fig:CM_ProlateSpheroid}}{\includegraphics[width=15cm, height=6cm]{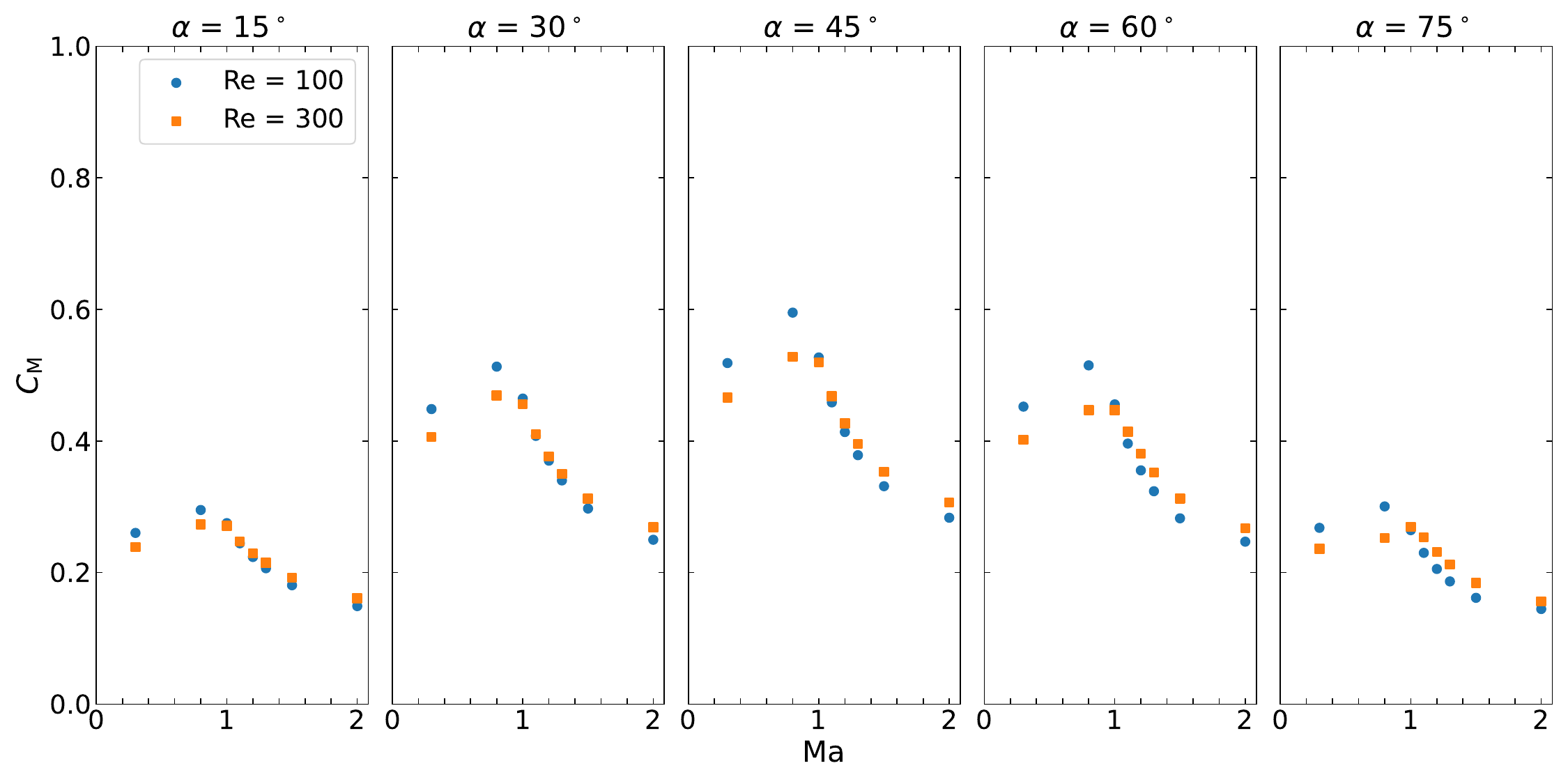}}\\
    \subcaptionbox{Oblate spheroid\label{fig:CM_OblateSpheroid}}{\includegraphics[width=15cm, height=6cm]{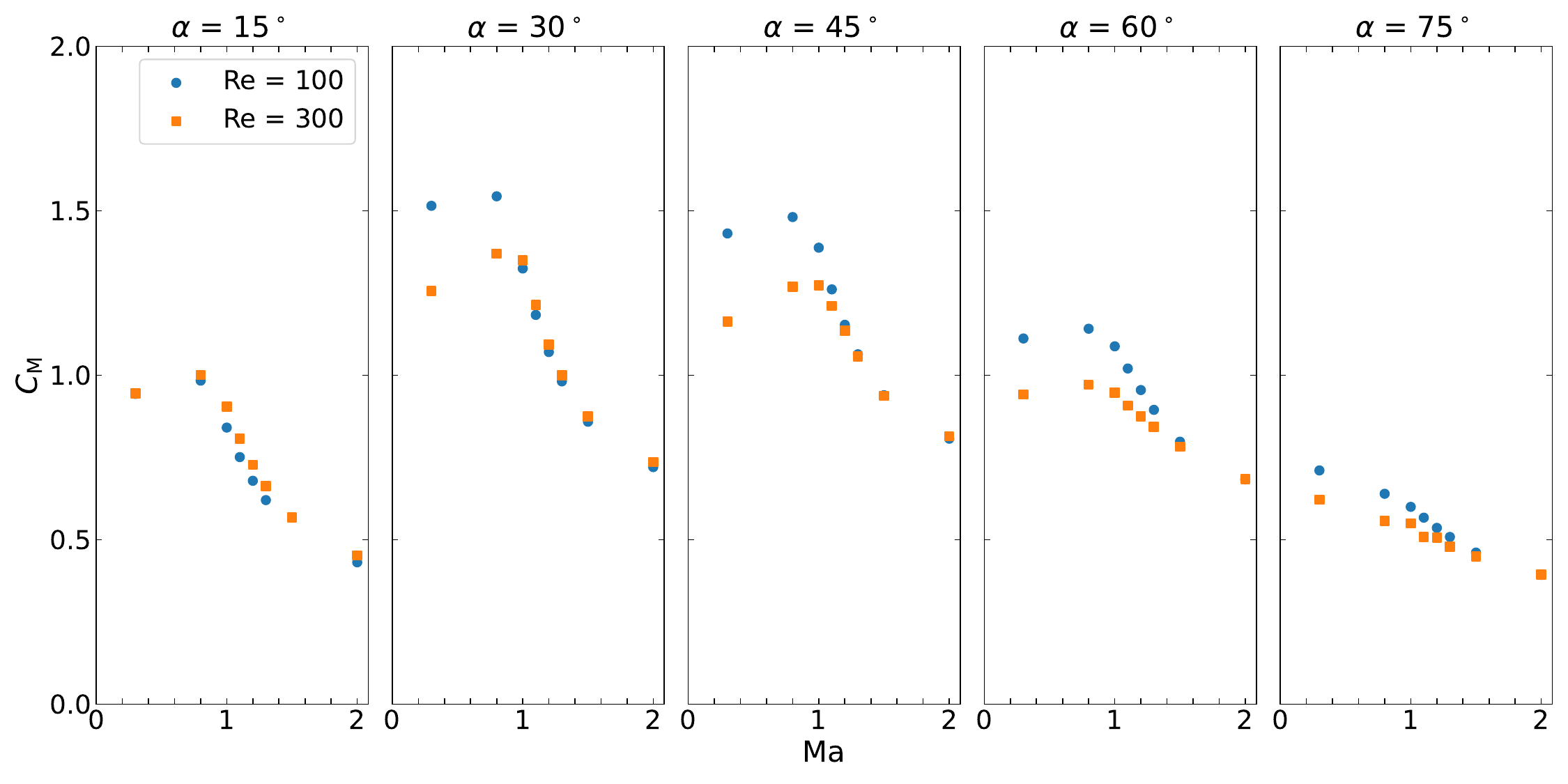}}\\
    \subcaptionbox{Rod-like particle\label{fig:CM_Fibroid}}{\includegraphics[width=15cm, height=6cm]{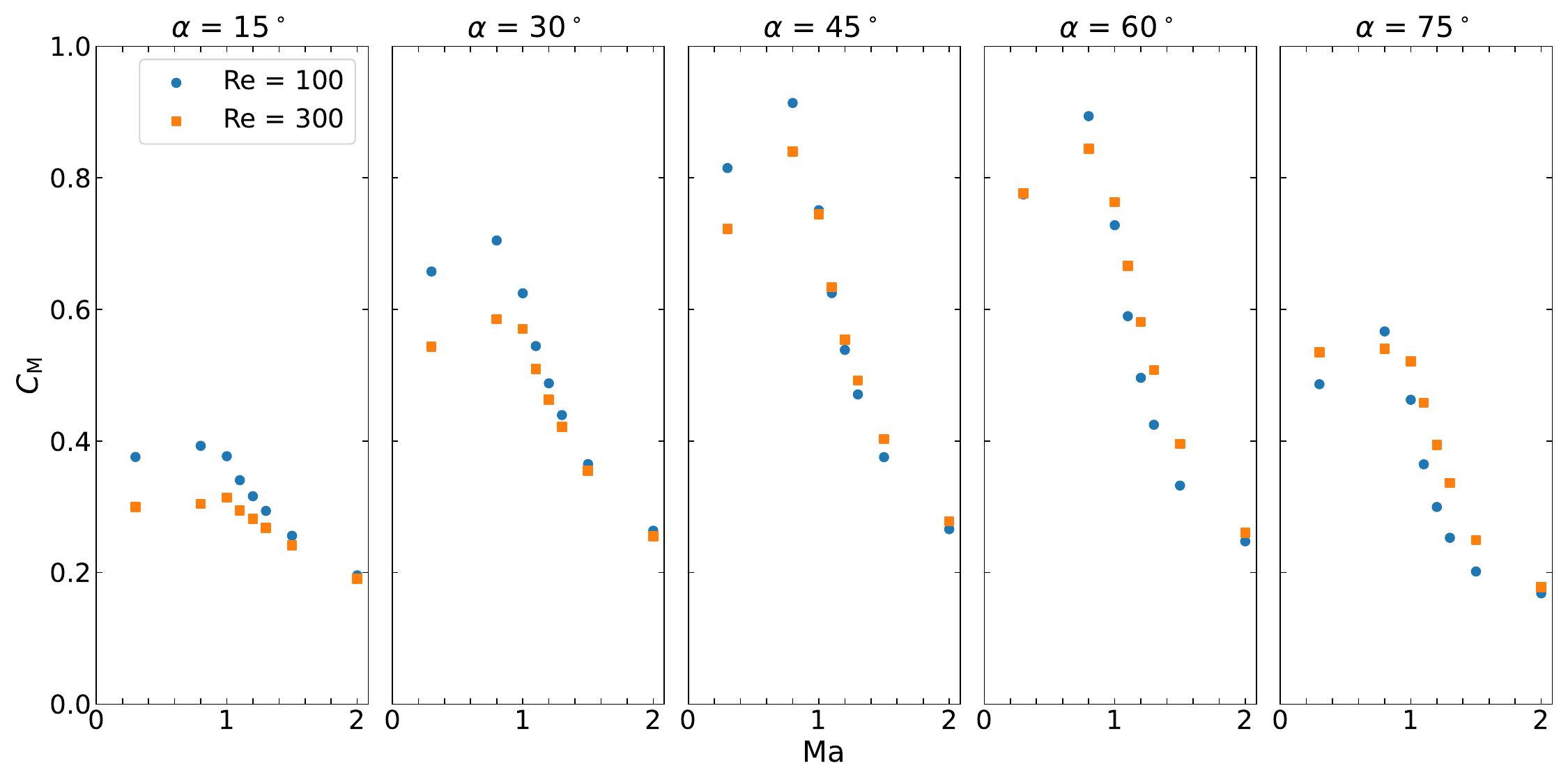}}  
\caption{Variation of the  torque coefficient $C_{\mathrm{M}}$ with Mach number $\mathrm{Ma}$ for different angles of attack considered for the three considered non-spherical particle shapes.}
\label{fig:CM_over_M}
\end{figure}

\reviewerO{In this appendix we give the diagrams for the variation of the lift and torque coefficients with the Mach number for different angles of attack.
Figures \ref{fig:CL_over_M} and \ref{fig:CM_over_M} present the lift coefficients, $C_{\mathrm{L}}$, and the  torque coefficients, $C_{\mathrm{M}}$, respectively, for all three non-spherical particle shapes as a function of the Mach number for all considered angles of attack. For angles of attack of 0\degree and 90\degree, the lift and torque coefficients are not shown, since they are zero for these flow configurations.
}

\clearpage
\section{\reviewerO{Summary of the full models to predict the drag, lift, and torque coefficients}}
\label{Appendix C}
\reviewerO{
In this appendix we give a summary of the full models for each non-spherical particle shape considered in this study. We exemplary choose the correlations for the incompressible limit of \citet{Zastawny2012}, but any other correlation (e.g., \cite{Sanjeevi2018, Cheron2024}) for the respective particle shape and aspect ratio in the incompressible limit may be used. These full models can be directly employed in point particle simulations.

\subsection{Drag coefficient}
The proposed correlations for the drag coefficient of each studied particle shape are based on drag coefficient correlations in the incompressible flow regime and a compressible correction. Both are a function
of the particle Reynolds number, the Mach number, and the angle of attack. 
The general expression is defined as 
\begin{equation}
C_\mathrm{D}\left(\mathrm{Re_p},\alpha,\mathrm{Ma}\right) = C_\mathrm{D}\left(\mathrm{Re_p},\alpha,\mathrm{Ma}=0\right) \left[1  + \mathcal{F}_\mathrm{D}^{\mathrm{Ma}}\left(\mathrm{Re_p},\alpha,\mathrm{Ma}\right)\right]
\end{equation}
where $C_\mathrm{D}$ is the drag coefficient of a specific particle given as a function of the particle Reynolds number, angle of attack, and Mach number. This coefficient depends on the drag coefficient for equivalent particle Reynolds number and angle of attack in incompressible flow conditions, and an additional shape-dependent function, $\mathcal{F}_\mathrm{D}^{\mathrm{Ma}}$ to account for the change in the coefficient as a result of the compressibility of the flow.

The full model for the drag coefficient of the prolate spheroid, as the combination of the correlation of \citet{Zastawny2012} for the incompressible limit and ours for the compressible limit, is
\begin{align}
  \begin{split}\label{eq:prolate_drag_full}
    C_\mathrm{D}\left(\mathrm{Re_p},\alpha,\mathrm{Ma}\right) & = \left[ \left( \frac{a_1}{\mathrm{Re_p}^{a_2}} + \frac{a_3}{\mathrm{Re_p}^{a_4}} \right) + \left( \left( \frac{a_5}{\mathrm{Re_p}^{a_6}} + \frac{a_7}{\mathrm{Re_p}^{a_8}} \right) - \left( \frac{a_1}{\mathrm{Re_p}^{a_2}} + \frac{a_3}{\mathrm{Re_p}^{a_4}} \right)  \right) \sin^{a_0}\left(\alpha\right) \right] \\
    & \left[ 1 + \frac{\left(\vartheta_1 \mathrm{Re_p} + \vartheta_2 \alpha^{\vartheta_3}\right)}{\left[1 + \exp\left(-\vartheta_4 (\mathrm{Ma} - \vartheta_5)\right)\right]} + \vartheta_6 \exp\left(-\frac{(\mathrm{Ma} - \vartheta_7)^2}{\vartheta_8}\right)   \right],
  \end{split}
\end{align}
the full model for the oblate spheroid is 
\begin{align}
  \begin{split}\label{eq:oblate_drag_full}
    C_\mathrm{D}\left(\mathrm{Re_p},\alpha,\mathrm{Ma}\right) & = \left[ \left( \frac{a_1}{\mathrm{Re_p}^{a_2}} + \frac{a_3}{\mathrm{Re_p}^{a_4}} \right) + \left( \left( \frac{a_5}{\mathrm{Re_p}^{a_6}} + \frac{a_7}{\mathrm{Re_p}^{a_8}} \right) - \left( \frac{a_1}{\mathrm{Re_p}^{a_2}} + \frac{a_3}{\mathrm{Re_p}^{a_4}} \right)  \right) \sin^{a_0}\left(\alpha\right) \right] \\
    & \left[ 1 +  \frac{\vartheta_1\big/{\log(\mathrm{Re_p})} \left(\vartheta_2{\alpha}\big/{\pi}\right)^{ \vartheta_3 \left({\mathrm{Re_p}}\right)^{\vartheta_4}} + {\log(\mathrm{Re_p})}\big/{\vartheta_5} - 1}{\left[1 + \exp\left(-\vartheta_6 (\mathrm{Ma} - \vartheta_7)\right)\right]} + \vartheta_8 \exp\left(-\frac{(\mathrm{Ma} - \vartheta_{9})^2}{\vartheta_{10}}\right)  \right],
  \end{split}
\end{align}
and the full model for the rod-like particle is 
\begin{align}
  \begin{split}\label{eq:rodlike_drag_full}
    C_\mathrm{D}\left(\mathrm{Re_p},\alpha,\mathrm{Ma}\right) & = \left[ \left( \frac{a_1}{\mathrm{Re_p}^{a_2}} + \frac{a_3}{\mathrm{Re_p}^{a_4}} \right) + \left( \left( \frac{a_5}{\mathrm{Re_p}^{a_6}} + \frac{a_7}{\mathrm{Re_p}^{a_8}} \right) - \left( \frac{a_1}{\mathrm{Re_p}^{a_2}} + \frac{a_3}{\mathrm{Re_p}^{a_4}} \right)  \right) \sin^{a_0}\left(\alpha\right) \right] \\
    & \left[ 1 + \frac{\vartheta_1 \cdot \mathrm{Re_p} + \vartheta_2 \cdot \left(\frac{\mathrm{Re_p}}{\vartheta_3}\right)^{\vartheta_4} \cdot \alpha^{\vartheta_5}}{\left[ 1.0 + \exp{\left(-\vartheta_6 \cdot (\mathrm{Ma} - \vartheta_7)\right)}\right]} 
    + \left(\frac{\vartheta_8 \cdot \alpha}{\mathrm{Re_p}}\right)^{\vartheta_9} \cdot \exp{\left(-\frac{(\mathrm{Ma} - \vartheta_{10})^2}{\vartheta_{11}}\right)}   \right].
  \end{split}
\end{align}
The fitting coefficients $a_i$ and $\vartheta_i$ for the drag coefficients of all three considered particle shapes are given in Table~\ref{table:Zastawny_dragcoefficients_appendix} and Table~\ref{table:dragcoefficients_appendix}.
\begin{table}[]
  \centering
  \reviewerO{
  \caption{List of the fitting coefficients in Eqs.~\eqref{eq:prolate_drag_full}, ~\eqref{eq:oblate_drag_full}, and~\eqref{eq:rodlike_drag_full}, used in the incompressible term based on the correlation of \citet{Zastawny2012}.}\label{table:Zastawny_dragcoefficients_appendix}
  \label{tab:drag_coefficient}
  \begin{tabular}{lccc}
  \hline
  Coefficient & Eq.~\eqref{eq:prolate_drag_full} & Eq.~\eqref{eq:oblate_drag_full} & Eq.~\eqref{eq:rodlike_drag_full} \\ \hline
  $a_0$ & 2.0   & 1.96  & 2.12  \\
  $a_1$ & 5.1   & 5.82  & 20.35 \\
  $a_2$ & 0.48  & 0.44  & 0.98  \\
  $a_3$ & 15.52 & 15.56 & 2.77  \\
  $a_4$ & 1.05  & 1.068 & 0.396 \\
  $a_5$ & 24.68 & 35.41 & 29.14 \\
  $a_6$ & 0.98  & 0.96  & 0.97  \\
  $a_7$ & 3.19  & 3.63  & 3.66  \\
  $a_8$ & 0.21  & 0.05  & 0.16  \\ \hline
  \end{tabular}
  }
\end{table}
\begin{table}
  \centering
  \reviewerO{
   \caption{List of the fitting coefficients in Eqs.~\eqref{eq:prolate_drag_full}, ~\eqref{eq:oblate_drag_full}, and~\eqref{eq:rodlike_drag_full}, used in the correlation to predict the change in the drag coefficient in the case of compressible flow.}\label{table:dragcoefficients_appendix}
  \begin{tabular}{l || c c c c c c c c c c c}
   & $\vartheta_1$ & $\vartheta_2$ & $\vartheta_3$ & $\vartheta_4$ & $\vartheta_5$ & $\vartheta_6$ & $\vartheta_7$ & $\vartheta_8$ & $\vartheta_9$ & $\vartheta_{10}$ & $\vartheta_{11}$ \\
  \hline
  \hline
  Eq.~\eqref{eq:prolate_drag_full} & 1.55 $\times$ $10^{-3}$ & 0.201 & 0.492 & 15.6 & 0.878 & 0.183 & 1.21 & 0.142 & - & - & - \\
  Eq.~\eqref{eq:oblate_drag_full} & 6.07 & 0.269 & 4.40 $\times$ $10^{-5}$ & 2.10 & 3.90 & 21.3 & 0.853 & 0.130 & 1.17 & 0.252 & - \\
  Eq.~\eqref{eq:rodlike_drag_full} & 1.14 $\times$ $10^{-3}$ & 0.121 & 2.64 & 0.0983 & 1.40 & 14.6 & 0.861 & 3.94 & 0.631 & 1.20 & 0.330
  \end{tabular}
  }
\end{table}

\subsection{Lift coefficient}
Following a similar approach as for the drag coefficients, the proposed correlation for the lift coefficient of each studied particle shape is based on the lift coefficient in the incompressible flow regime from \citet{Zastawny2012} and a compressible correction that is a function
of the particle Reynolds number, the Mach number, and the angle of attack. 
The general expression is defined as 
\begin{equation}\label{eq:General_Lift_Coefficient_appendix}
C_\mathrm{L}\left(\mathrm{Re_p},\alpha,\mathrm{Ma}\right) = C_\mathrm{L}\left(\mathrm{Re_p},\alpha,\mathrm{Ma}=0\right)  + \mathcal{F}_\mathrm{L}^{\mathrm{Ma}}\left(\mathrm{Re_p},\alpha,\mathrm{Ma}\right)
\end{equation}
where $C_\mathrm{L}$ is the lift coefficient of a specific particle given as a function of the particle Reynolds number, angle of attack, and Mach number. This coefficient is obtained from the summation of the lift coefficient for equivalent particle Reynolds number and angle of attack in incompressible flow conditions, and an additional shape-dependent function, $\mathcal{F}_\mathrm{L}^{\mathrm{Ma}}$, to account for the change in the lift coefficient in case of compressible flow.
The full model for the lift coefficient of the prolate spheroid is 
\begin{align}
  \begin{split}\label{eq:prolate_lift_full}
    C_\mathrm{L}\left(\mathrm{Re_p},\alpha,\mathrm{Ma}\right) & = \left[ \left( \frac{b_1}{\mathrm{Re_p}^{b_2}} + \frac{b_3}{\mathrm{Re_p}^{b_4}} \right) \sin\left(\alpha\right)^{b_5+b_6 \mathrm{Re_p}^{b_7}} \cos\left(\alpha\right)^{b_8+b_9\mathrm{Re_p}^{b_{10}}} \right] \\
    & + \frac{\vartheta_1 \log\left(\mathrm{Re_p}\right)^{\vartheta_2}}{\left[1 + \exp\left(\vartheta_3 (\mathrm{Ma} - \vartheta_4)\right)\right]} \cos\left(\Psi_1\left(\alpha, \mathrm{Re_p}\right)\right) \sin\left(\Psi_1\left(\alpha, \mathrm{Re_p}\right)\right) \\ 
    & + \vartheta_5 \exp\left(\frac{-(\mathrm{Ma} - \vartheta_6)^2}{ \vartheta_7}\right) \left[\cos\left(\Psi_2\left(\alpha, \mathrm{Re_p}\right)\right)\sin\left(\Psi_2\left(\alpha, \mathrm{Re_p}\right)\right)\right]^{\vartheta_8} \, ,
  \end{split}
\end{align}
with
\begin{equation}
  \Psi_1 = \cfrac{\pi}{2} \left(\alpha \cfrac{2}{\pi}\right)^{1 + 4.185 \times 10^{-5}\log{\left(\mathrm{Re_p}\right)}^{-10.5}}
\end{equation}
and
\begin{equation}
  \Psi_2 = \cfrac{\pi}{2} \left(\alpha \cfrac{2}{\pi}\right)^{1 + 2.146 \log{\left(\mathrm{Re_p}\right)}^{-1.444}}.
\end{equation}
The full model for the oblate spheroid is 
\begin{align}
  \begin{split}\label{eq:oblate_lift_full}
    C_\mathrm{L}\left(\mathrm{Re_p},\alpha,\mathrm{Ma}\right) & = \left[ \left( \frac{b_1}{\mathrm{Re_p}^{b_2}} + \frac{b_3}{\mathrm{Re_p}^{b_4}} \right) \sin\left(\alpha\right)^{b_5+b_6 \mathrm{Re_p}^{b_7}} \cos\left(\alpha\right)^{b_8+b_9\mathrm{Re_p}^{b_{10}}} \right] \\
    & + \frac{\vartheta_{1} \log(\mathrm{Re_p})}{\left[1 + \exp\left(-\vartheta_2 (\mathrm{Ma} - \vartheta_3)\right)\right]} \cos(\alpha)^{\vartheta_4} \sin(\alpha)^{\vartheta_5} \\
    & + \vartheta_6 \exp\left(-\frac{(\mathrm{Ma} - \vartheta_7)^2}{\vartheta_8}\right) \cos(\alpha) \sin(\alpha)\, ,
  \end{split}
\end{align}
and the full model for the rod-like particle is 
\begin{align}
  \begin{split}\label{eq:rodlike_lift_full}
    C_\mathrm{L}\left(\mathrm{Re_p},\alpha,\mathrm{Ma}\right) & = \left[ \left( \frac{b_1}{\mathrm{Re_p}^{b_2}} + \frac{b_3}{\mathrm{Re_p}^{b_4}} \right) \sin\left(\alpha\right)^{b_5+b_6 \mathrm{Re_p}^{b_7}} \cos\left(\alpha\right)^{b_8+b_9\mathrm{Re_p}^{b_{10}}} \right] \\
    & + \frac{\left(\vartheta_1 / \log\left(\mathrm{Re_p}\right)\right)^{\vartheta_2}}{\left[1 + \exp\left(\vartheta_3 \cdot (\mathrm{Ma} - \vartheta_4)\right)\right]} \cos\left(\Psi_1\left(\alpha, \mathrm{Re_p}\right)\right) \sin\left(\Psi_1\left(\alpha, \mathrm{Re_p}\right)\right)\\ 
    & + \vartheta_5 \exp\left(\frac{-(\mathrm{Ma} - \vartheta_6)^2}{ \vartheta_7}\right) \left[
      \cos\left(\Psi_2\left(\alpha, \mathrm{Re_p}\right)\right)
      \sin\left(\Psi_2\left(\alpha, \mathrm{Re_p}\right)\right)\right]^{\vartheta_8} \, ,
  \end{split}
\end{align}
with
\begin{equation}
  \Psi_1 = \cfrac{\pi}{2} \left(\alpha \cfrac{2}{\pi}\right)^{1 + 2.471 \log{\left(\mathrm{Re_p}\right)}^{-1.50}}
\end{equation}
and
\begin{equation}
  \Psi_2 = \cfrac{\pi}{2} \left(\alpha \cfrac{2}{\pi}\right)^{1 + 1630.79 \left(1 / \log{\left(\mathrm{Re_p}\right)}\right)^{6.00}}.
\end{equation}
The fitting coefficients $b_i$ and $\vartheta_i$ for the lift coefficients of all three considered particle shapes are given in Table~\ref{table:Zastawny_liftcoefficients_appendix} and Table~\ref{table:liftcoefficients_appendix}.

\begin{table}[]
  \centering
  \reviewerO{
  \caption{List of fitting coefficients in Eqs.~\eqref{eq:prolate_lift_full}, ~\eqref{eq:oblate_lift_full}, and~\eqref{eq:rodlike_lift_full}, used in the correlation to predict the change in the lift coefficient in the incompressible limit.}\label{table:Zastawny_liftcoefficients_appendix}
  \label{tab:lift_coefficient}
  \begin{tabular}{lccc}
  \hline
  Coefficient & Eq.~\eqref{eq:prolate_lift_full} & Eq.~\eqref{eq:oblate_lift_full} & Eq.~\eqref{eq:rodlike_lift_full} \\ \hline
  $b_1$  & 6.079  & 12.111 & 8.652  \\
  $b_2$  & 0.898  & 1.036  & 0.815  \\
  $b_3$  & 0.704  & 3.887  & 0.407  \\
  $b_4$  & -0.028 & 0.109  & -0.197 \\
  $b_5$  & 1.067  & 0.812  & 0.978  \\
  $b_6$  & 0.0025 & 0.249  & 0.036  \\
  $b_7$  & 0.818  & -0.198 & 0.451  \\
  $b_8$  & 1.049  & 5.821  & 1.359  \\
  $b_9$  & 0.0    & 4.717  & -0.43  \\
  $b_{10}$ & 0.0    & 0.007  & 0.007  \\ \hline
  \end{tabular}
  }
  \end{table}

  \begin{table}
    \centering
    \caption{\reviewerO{List of the fitting coefficients in Eqs.~\eqref{eq:prolate_lift_full}, ~\eqref{eq:oblate_lift_full}, and~\eqref{eq:rodlike_lift_full}, used in the correlation to predict the change in the lift coefficient in case of compressible flow.}}
    \label{table:liftcoefficients_appendix}
    \begin{tabular}{l || c c c c c c c c}
     \reviewerO{\,} & \reviewerO{$\vartheta_1$} & \reviewerO{$\vartheta_2$} & \reviewerO{$\vartheta_3$} & \reviewerO{$\vartheta_4$} & \reviewerO{$\vartheta_5$} & \reviewerO{$\vartheta_6$} & \reviewerO{$\vartheta_7$} & \reviewerO{$\vartheta_8$} \\
    \hline
    \hline
    \reviewerO{Eq.~\eqref{eq:prolate_lift_full}} & \reviewerO{1.17} & \reviewerO{-0.735} & \reviewerO{-21.8} & \reviewerO{0.870} & \reviewerO{0.277} & \reviewerO{1.12} & \reviewerO{0.217} & \reviewerO{2.00}\\
    \reviewerO{Eq.~\eqref{eq:oblate_lift_full}} & \reviewerO{0.535} & \reviewerO{130} & \reviewerO{0.807} & \reviewerO{1.84} & \reviewerO{1.56} & \reviewerO{-0.996} & \reviewerO{2.25} & \reviewerO{0.516} \\
    \reviewerO{Eq.~\eqref{eq:rodlike_lift_full}} & \reviewerO{2.43} & \reviewerO{1.20} & \reviewerO{-24.8} & \reviewerO{0.807} & \reviewerO{2.43} & \reviewerO{1.24} & \reviewerO{0.237} & \reviewerO{4.22} \\
    \end{tabular}
  \end{table}

\subsection{Torque coefficient}
As for the drag and lift coefficients, the proposed correlation for the torque coefficient of each studied particle shape is based on the torque coefficient in the incompressible flow regime (here exemplary \citet{Zastawny2012}) and a compressible correction that is a function of the particle Reynolds number, the Mach number, and the angle of attack. 
The general expression is defined as
\begin{equation}\label{eq:General_Torque_Coefficient_appendix}
C_\mathrm{M}\left(\mathrm{Re_p},\alpha,\mathrm{Ma}\right) = C_\mathrm{M}\left(\mathrm{Re_p},\alpha,\mathrm{Ma}=0\right)  + \mathcal{F}_\mathrm{M}^{\mathrm{Ma}}\left(\mathrm{Re_p},\alpha,\mathrm{Ma}\right)
\end{equation}
where $C_\mathrm{M}$ is the torque coefficient of a specific particle given as a function of the particle Reynolds number, the Mach number, and the angle of attack. This coefficient is obtained from the summation of the torque coefficient for equivalent particle Reynolds number and angle of attack in incompressible flow conditions, and an additional shape-dependent function, $\mathcal{F}_\mathrm{M}^{\mathrm{Ma}}$, to account for the change in the torque coefficient in case of compressible flow.
The full model for the torque coefficient of all three considered particle shapes is 
\begin{align}
  \begin{split}\label{eq:torque_full}
    C_\mathrm{M}\left(\mathrm{Re_p},\alpha,\mathrm{Ma}\right) & = \left[ \left( \frac{c_1}{\mathrm{Re_p}^{c_2}} + \frac{c_3}{\mathrm{Re_p}^{c_4}} \right) \sin\left(\alpha\right)^{c_5+c_6 \mathrm{Re_p}^{c_7}} \cos\left(\alpha\right)^{c_8+c_9\mathrm{Re_p}^{c_{10}}} \right] \\
    & + \frac{\vartheta_1 \mathrm{Re_p} + \vartheta_2  \left( \mathrm{Re_p} / \vartheta_3 \right)^{\vartheta_4}}{1 + \exp\left(-\vartheta_5 \left( \mathrm{Ma} - \vartheta_{6} \right)\right)} \cdot \left( \cos\left(\Psi_1\right) \sin\left(\Psi_1\right) \right)^{\vartheta_{7}} \\
    & + \left( \frac{\vartheta_{8} \alpha}{\vartheta_{10}}\right)^{\vartheta_9} \cdot \left(\frac{\vartheta_{11} \mathrm{Re_p}}{\vartheta_{13}}\right)^{\vartheta_{12}} \cdot \exp\left(
    \frac{-\left( \mathrm{Ma} - \vartheta_{14}\right)^2}{\vartheta_{15}}\right) \cdot \left( \cos\left(\Psi_2\right) \sin\left(\Psi_2\right) \right)^{\vartheta_{16}} \, ,
  \end{split}
\end{align}
with
\begin{equation}
  \Psi_1 = \cfrac{\pi}{2} \left(\alpha \cfrac{2}{\pi}\right)^{\vartheta_{17}}
\end{equation}
and
\begin{equation}
  \Psi_2 = \cfrac{\pi}{2} \left(\alpha \cfrac{2}{\pi}\right)^{\vartheta_{18}}.
\end{equation}
The fitting coefficients $c_i$ and $\vartheta_i$ for the torque coefficients of all three considered particle shapes are given in Table~\ref{table:Zastawny_torquecoefficients_appendix} and Table~\ref{table:torquecoefficients_appendix}.

\begin{table}[h!]
  \centering
  \reviewerO{
  \caption{List of fitting coefficients in Eqs.~\eqref{eq:torque_full}, used in the correlation to predict the change in the torque coefficient in the incompressible limit.}
  \label{table:Zastawny_torquecoefficients_appendix}
  \begin{tabular}{lccc}
  \hline
  Coefficient & Prolate spheroid & Oblate spheroid & Rod-like \\ \hline
  $c_1$  & 2.078  & 3.782  & 0.011  \\
  $c_2$  & 0.279  & 0.237  & -0.656 \\
  $c_3$  & 0.372  & 2.351  & 8.909  \\
  $c_4$  & 0.018  & 0.236  & 0.396  \\
  $c_5$  & 0.98   & -0.394 & 2.926  \\
  $c_6$  & 0.0    & 1.615  & -1.28  \\
  $c_7$  & 0.0    & -0.044 & 0.037  \\
  $c_8$  & 1.0    & -0.537 & -15.236 \\
  $c_9$  & 0.0    & 1.805  & 16.757 \\
  $c_{10}$ & 0.0    & -0.037 & -0.006 \\ \hline
  \end{tabular}
  }
  \end{table}

\begin{table}
  \centering
  \reviewerO{
    
     \caption{List of the fitting coefficients in Eq.~\eqref{eq:torque_full}, used in the correlation to predict the change in the torque coefficient in case of compressible flow.}\label{table:torquecoefficients_appendix}
    \begin{tabular}{l || c c c}
     & Prolate & Oblate & Rod-like \\
    \hline
    \hline
    $\vartheta_1$ & 2.28 $\times 10^{-4}$ & -4.43 $\times 10^{-3}$ & -2.32 $\times 10^{-5}$ \\ 
    $\vartheta_2$ & -0.444 & -0.759 & -1.15 \\ 
    $\vartheta_3$ & 157 & 203 & 127 \\ 
    $\vartheta_4$ & -0.223 & -1.03 & -0.134 \\ 
    $\vartheta_5$ & 15.1 & 5.93 & 12.3 \\ 
    $\vartheta_6$ & 1.01 & 1.22 & 1.03 \\ 
    $\vartheta_7$ & 0.993 & 1.11 & 1.04 \\ 
    $\vartheta_8$ & 4.96 $\times 10^{-2}$ & 3.18 $\times 10^{-9}$ & 13.8 \\ 
    $\vartheta_9$ & 1.44 & 8.95 $\times 10^{-2}$ & 1.88 \\ 
    $\vartheta_{10}$ & 7.42 $\times 10^{-3}$ & 3.59 $\times 10^{-2}$ & 3.64 \\ 
    $\vartheta_{11}$ & 3.81 $\times 10^{-2}$ & 3.89 & 0.520 \\ 
    $\vartheta_{12}$ & 1.68 $\times 10^{-2}$ & 0.569 & 0.158 \\ 
    $\vartheta_{13}$ & 5.86 $\times 10^{-9}$ & 10.8 & 1.22 $\times 10^{-6}$ \\ 
    $\vartheta_{14}$ & 1.10 & 1.98 & 1.18 \\ 
    $\vartheta_{15}$ & 9.29 $\times 10^{-3}$ & 1.04 & 3.71 $\times 10^{-3}$ \\ 
    $\vartheta_{16}$ & 1.05 & 2.50 & 1.16 \\ 
    $\vartheta_{17}$ & 0.164 & 1.17 & 0.157 \\ 
    $\vartheta_{18}$ & 1.04 & 0.809 & 1.32 \\ 
    \end{tabular}
  }
\end{table}
}
\clearpage
\bibliographystyle{elsarticle-harv}

\begin{thebibliography}{50}
    \expandafter\ifx\csname natexlab\endcsname\relax\def\natexlab#1{#1}\fi
    \providecommand{\url}[1]{\texttt{#1}}
    \providecommand{\href}[2]{#2}
    \providecommand{\path}[1]{#1}
    \providecommand{\DOIprefix}{doi:}
    \providecommand{\ArXivprefix}{arXiv:}
    \providecommand{\URLprefix}{URL: }
    \providecommand{\Pubmedprefix}{pmid:}
    \providecommand{\doi}[1]{\href{http://dx.doi.org/#1}{\path{#1}}}
    \providecommand{\Pubmed}[1]{\href{pmid:#1}{\path{#1}}}
    \providecommand{\bibinfo}[2]{#2}
    \ifx\xfnm\relax \def\xfnm[#1]{\unskip,\space#1}\fi
    \bibitem[{Anderson(2003)}]{Anderson2003}
    \bibinfo{author}{Anderson, J.D.}, \bibinfo{year}{2003}.
    \newblock \bibinfo{title}{Modern {Compressible} {Flow}: {With} a {Historical}
      {Perspective}}.
    \newblock \bibinfo{publisher}{McGraw-Hill New York}.
    \bibitem[{Bailey and Starr(1976)}]{Bailey1976}
    \bibinfo{author}{Bailey, A.B.}, \bibinfo{author}{Starr, R.F.},
      \bibinfo{year}{1976}.
    \newblock \bibinfo{title}{Sphere drag at transonic speeds and high {Reynolds}
      numbers}.
    \newblock \bibinfo{journal}{AIAA Journal} \bibinfo{volume}{14},
      \bibinfo{pages}{1631--1631}.
    \newblock \URLprefix \url{https://arc.aiaa.org/doi/10.2514/3.7262},
      \DOIprefix\doi{10.2514/3.7262}. \bibinfo{note}{publisher: American Institute
      of Aeronautics and Astronautics}.
    \bibitem[{Bashforth(1870)}]{Bashforth1870}
    \bibinfo{author}{Bashforth, F.}, \bibinfo{year}{1870}.
    \newblock \bibinfo{title}{Reports on experiments made with the {Bashforth}
      chronograph to determine the resistance of the air to the motion of
      projectiles}.
    \newblock \bibinfo{publisher}{H.M. Stationery Office}.
    \newblock \bibinfo{note}{Google-Books-ID: cqZYAAAAcAAJ}.
    \bibitem[{Bhagat and Goswami(2022)}]{Bhagat2022}
    \bibinfo{author}{Bhagat, A.M.}, \bibinfo{author}{Goswami, P.S.},
      \bibinfo{year}{2022}.
    \newblock \bibinfo{title}{Effect of rough wall on drag, lift, and torque on an
      ellipsoidal particle in a linear shear flow}.
    \newblock \bibinfo{journal}{Physics of Fluids} \bibinfo{volume}{34},
      \bibinfo{pages}{083312}.
    \newblock \URLprefix \url{https://doi.org/10.1063/5.0093232},
      \DOIprefix\doi{10.1063/5.0093232}.
    \bibitem[{Capecelatro(2022)}]{Capecelatro2021}
    \bibinfo{author}{Capecelatro, J.}, \bibinfo{year}{2022}.
    \newblock \bibinfo{title}{Modeling high-speed gas–particle flows relevant to
      spacecraft landings}.
    \newblock \bibinfo{journal}{International Journal of Multiphase Flow}
      \bibinfo{volume}{150}, \bibinfo{pages}{104008}.
    \newblock \URLprefix
      \url{https://www.sciencedirect.com/science/article/pii/S0301932222000301},
      \DOIprefix\doi{10.1016/j.ijmultiphaseflow.2022.104008}.
    \bibitem[{Capecelatro et~al.(2022)Capecelatro, Longest, Boerman, Sulaiman and
      Sundaresan}]{Capecelatro2022}
    \bibinfo{author}{Capecelatro, J.}, \bibinfo{author}{Longest, W.},
      \bibinfo{author}{Boerman, C.}, \bibinfo{author}{Sulaiman, M.},
      \bibinfo{author}{Sundaresan, S.}, \bibinfo{year}{2022}.
    \newblock \bibinfo{title}{Recent developments in the computational simulation
      of dry powder inhalers}.
    \newblock \bibinfo{journal}{Advanced Drug Delivery Reviews} ,
      \bibinfo{pages}{114461}\URLprefix
      \url{https://www.sciencedirect.com/science/article/pii/S0169409X22003519},
      \DOIprefix\doi{10.1016/j.addr.2022.114461}.
    \bibitem[{Capecelatro and Wagner(2024)}]{Capecelatro2024}
    \bibinfo{author}{Capecelatro, J.}, \bibinfo{author}{Wagner, J.L.},
      \bibinfo{year}{2024}.
    \newblock \bibinfo{title}{Gas–{Particle} {Dynamics} in {High}-{Speed}
      {Flows}}.
    \newblock \bibinfo{journal}{Annual Review of Fluid Mechanics}
      \bibinfo{volume}{56}, \bibinfo{pages}{379--403}.
    \newblock \URLprefix
      \url{https://www.annualreviews.org/doi/10.1146/annurev-fluid-121021-015818},
      \DOIprefix\doi{10.1146/annurev-fluid-121021-015818}.
    \bibitem[{Carlson and Hoglund(1964)}]{Carlson1964}
    \bibinfo{author}{Carlson, D.J.}, \bibinfo{author}{Hoglund, R.F.},
      \bibinfo{year}{1964}.
    \newblock \bibinfo{title}{Particle drag and heat transfer in rocket nozzles}.
    \newblock \bibinfo{journal}{AIAA Journal} \bibinfo{volume}{2},
      \bibinfo{pages}{1980--1984}.
    \newblock \URLprefix \url{https://arc.aiaa.org/doi/10.2514/3.2714},
      \DOIprefix\doi{10.2514/3.2714}.
    \bibitem[{Chéron et~al.(2024)Chéron, Evrard and van Wachem}]{Cheron2024}
    \bibinfo{author}{Chéron, V.}, \bibinfo{author}{Evrard, F.},
      \bibinfo{author}{van Wachem, B.}, \bibinfo{year}{2024}.
    \newblock \bibinfo{title}{Drag, lift and torque correlations for axi-symmetric
      rod-like non-spherical particles in locally linear shear flows}.
    \newblock \bibinfo{journal}{International Journal of Multiphase Flow}
      \bibinfo{volume}{171}, \bibinfo{pages}{104692}.
    \newblock \URLprefix
      \url{https://www.sciencedirect.com/science/article/pii/S0301932223003129},
      \DOIprefix\doi{10.1016/j.ijmultiphaseflow.2023.104692}.
    \bibitem[{Chéron and van Wachem(2024)}]{Cheron2024b}
    \bibinfo{author}{Chéron, V.}, \bibinfo{author}{van Wachem, B.},
      \bibinfo{year}{2024}.
    \newblock \bibinfo{title}{Drag, lift, and torque correlations for axi-symmetric
      rod-like non-spherical particles in linear wall-bounded shear flow}.
    \newblock \bibinfo{journal}{International Journal of Multiphase Flow} ,
      \bibinfo{pages}{104906}\URLprefix
      \url{https://linkinghub.elsevier.com/retrieve/pii/S0301932224001836},
      \DOIprefix\doi{10.1016/j.ijmultiphaseflow.2024.104906}.
    \bibitem[{Clift and Gauvin(1971)}]{Clift1971}
    \bibinfo{author}{Clift, R.}, \bibinfo{author}{Gauvin, W.H.},
      \bibinfo{year}{1971}.
    \newblock \bibinfo{title}{Motion of entrained particles in gas streams}.
    \newblock \bibinfo{journal}{The Canadian Journal of Chemical Engineering}
      \bibinfo{volume}{49}, \bibinfo{pages}{439--448}.
    \newblock \URLprefix \url{http://doi.wiley.com/10.1002/cjce.5450490403},
      \DOIprefix\doi{10.1002/cjce.5450490403}.
    \bibitem[{Dabade et~al.(2016)Dabade, Marath and Subramanian}]{Dabade2016}
    \bibinfo{author}{Dabade, V.}, \bibinfo{author}{Marath, N.K.},
      \bibinfo{author}{Subramanian, G.}, \bibinfo{year}{2016}.
    \newblock \bibinfo{title}{The effect of inertia on the orientation dynamics of
      anisotropic particles in simple shear flow}.
    \newblock \bibinfo{journal}{Journal of Fluid Mechanics} \bibinfo{volume}{791},
      \bibinfo{pages}{631--703}.
    \newblock \URLprefix
      \url{https://www.cambridge.org/core/journals/journal-of-fluid-mechanics/article/abs/effect-of-inertia-on-the-orientation-dynamics-of-anisotropic-particles-in-simple-shear-flow/946CB5A981ABAADF546483CCFE7AD0C8},
      \DOIprefix\doi{10.1017/jfm.2016.14}.
    \bibitem[{Denner et~al.(2020)Denner, Evrard and van Wachem}]{Denner2020}
    \bibinfo{author}{Denner, F.}, \bibinfo{author}{Evrard, F.},
      \bibinfo{author}{van Wachem, B.}, \bibinfo{year}{2020}.
    \newblock \bibinfo{title}{Conservative finite-volume framework and
      pressure-based algorithm for flows of incompressible, ideal-gas and real-gas
      fluids at all speeds}.
    \newblock \bibinfo{journal}{Journal of Computational Physics}
      \bibinfo{volume}{409}, \bibinfo{pages}{109348}.
    \newblock \URLprefix
      \url{https://linkinghub.elsevier.com/retrieve/pii/S0021999120301224},
      \DOIprefix\doi{10.1016/j.jcp.2020.109348}.
    \bibitem[{Fillingham et~al.(2021)Fillingham, Vaddi, Bruning, Israel and
      Novosselov}]{Fillingham2021}
    \bibinfo{author}{Fillingham, P.}, \bibinfo{author}{Vaddi, R.S.},
      \bibinfo{author}{Bruning, A.}, \bibinfo{author}{Israel, G.},
      \bibinfo{author}{Novosselov, I.V.}, \bibinfo{year}{2021}.
    \newblock \bibinfo{title}{Drag, lift, and torque on a prolate spheroid resting
      on a smooth surface in a linear shear flow}.
    \newblock \bibinfo{journal}{Powder Technology} \bibinfo{volume}{377},
      \bibinfo{pages}{958--965}.
    \newblock \URLprefix
      \url{https://www.sciencedirect.com/science/article/pii/S0032591020309062},
      \DOIprefix\doi{10.1016/j.powtec.2020.09.042}.
    \bibitem[{Fröhlich et~al.(2020)Fröhlich, Meinke and Schröder}]{Frohlich2020}
    \bibinfo{author}{Fröhlich, K.}, \bibinfo{author}{Meinke, M.},
      \bibinfo{author}{Schröder, W.}, \bibinfo{year}{2020}.
    \newblock \bibinfo{title}{Correlations for inclined prolates based on highly
      resolved simulations}.
    \newblock \bibinfo{journal}{Journal of Fluid Mechanics} \bibinfo{volume}{901},
      \bibinfo{pages}{A5}.
    \newblock \URLprefix
      \url{https://www.cambridge.org/core/product/identifier/S0022112020004826/type/journal_article},
      \DOIprefix\doi{10.1017/jfm.2020.482}.
    \bibitem[{Happel and Brenner(1981)}]{Happel1981}
    \bibinfo{author}{Happel, J.}, \bibinfo{author}{Brenner, H.},
      \bibinfo{year}{1981}.
    \newblock \bibinfo{title}{Low {Reynolds} number hydrodynamics}.
      volume~\bibinfo{volume}{1} of \textit{\bibinfo{series}{Mechanics of fluids
      and transport processes}}.
    \newblock \bibinfo{publisher}{Springer Netherlands},
      \bibinfo{address}{Dordrecht}.
    \newblock \URLprefix \url{http://link.springer.com/10.1007/978-94-009-8352-6},
      \DOIprefix\doi{10.1007/978-94-009-8352-6}.
    \bibitem[{Henderson(1976)}]{Henderson1976}
    \bibinfo{author}{Henderson, C.B.}, \bibinfo{year}{1976}.
    \newblock \bibinfo{title}{Drag coefficients of spheres in continuum and
      rarefied flows}.
    \newblock \bibinfo{journal}{AIAA Journal} \bibinfo{volume}{14},
      \bibinfo{pages}{707--708}.
    \newblock \URLprefix \url{https://arc.aiaa.org/doi/10.2514/3.61409},
      \DOIprefix\doi{10.2514/3.61409}.
    \bibitem[{Hogan et~al.(2015)Hogan, Taberner, Jones and Hunter}]{Hogan2015}
    \bibinfo{author}{Hogan, N.C.}, \bibinfo{author}{Taberner, A.J.},
      \bibinfo{author}{Jones, L.A.}, \bibinfo{author}{Hunter, I.W.},
      \bibinfo{year}{2015}.
    \newblock \bibinfo{title}{Needle-free delivery of macromolecules through the
      skin using controllable jet injectors}.
    \newblock \bibinfo{journal}{Expert Opinion on Drug Delivery}
      \bibinfo{volume}{12}, \bibinfo{pages}{1637--1648}.
    \newblock \URLprefix
      \url{http://www.tandfonline.com/doi/full/10.1517/17425247.2015.1049531},
      \DOIprefix\doi{10.1517/17425247.2015.1049531}.
    \bibitem[{Hölzer and Sommerfeld(2008)}]{Holzer2008}
    \bibinfo{author}{Hölzer, A.}, \bibinfo{author}{Sommerfeld, M.},
      \bibinfo{year}{2008}.
    \newblock \bibinfo{title}{New simple correlation formula for the drag
      coefficient of non-spherical particles}.
    \newblock \bibinfo{journal}{Powder Technology} \bibinfo{volume}{184},
      \bibinfo{pages}{361--365}.
    \newblock \URLprefix
      \url{http://linkinghub.elsevier.com/retrieve/pii/S0032591007004792},
      \DOIprefix\doi{10.1016/j.powtec.2007.08.021}.
    \bibitem[{Jacobs(1929)}]{Jacobs1929}
    \bibinfo{author}{Jacobs, E.N.}, \bibinfo{year}{1929}.
    \newblock \bibinfo{title}{Sphere {Drag} {Tests} in the {Variable} {Density}
      {Wind} {Tunnel}}.
    \newblock \URLprefix
      \url{https://digital.library.unt.edu/ark:/67531/metadc54020/}.
      \bibinfo{note}{number: NACA-TN-312}.
    \bibitem[{Johansson(2019)}]{Johansson2019}
    \bibinfo{author}{Johansson, R.}, \bibinfo{year}{2019}.
    \newblock \bibinfo{title}{Numerical {Python}: {Scientific} {Computing} and
      {Data} {Science} {Applications} with {Numpy}, {SciPy} and {Matplotlib}}.
    \newblock \bibinfo{publisher}{Apress}, \bibinfo{address}{Berkeley, CA}.
    \newblock \URLprefix \url{http://link.springer.com/10.1007/978-1-4842-4246-9},
      \DOIprefix\doi{10.1007/978-1-4842-4246-9}.
    \bibitem[{Kaskas(1964)}]{Kaskas1964}
    \bibinfo{author}{Kaskas, A.}, \bibinfo{year}{1964}.
    \newblock \bibinfo{title}{Berechnung der stationären und instationären
      {Bewegung} von {Kugeln} in ruhenden und strömenden {Medien}}.
    \newblock Master's thesis. Lehrstuhl fuer Thermodynamik und Verfahrenstechnik
      der T. U. Berlin. \bibinfo{address}{Germany}.
    \newblock \URLprefix \url{https://cir.nii.ac.jp/crid/1572824500138829056}.
    \bibitem[{Loth(2008)}]{Loth2008}
    \bibinfo{author}{Loth, E.}, \bibinfo{year}{2008}.
    \newblock \bibinfo{title}{Lift of a solid spherical particle subject to
      vorticity and/or spin}.
    \newblock \bibinfo{journal}{AIAA journal} \bibinfo{volume}{46},
      \bibinfo{pages}{801--809}.
    \newblock \URLprefix
      \url{http://cat.inist.fr/?aModele=afficheN&amp;cpsidt=20273885},
      \DOIprefix\doi{10.2514/1.29159}.
    \bibitem[{Loth et~al.(2021)Loth, Tyler~Daspit, Jeong, Nagata and
      Nonomura}]{Loth2021}
    \bibinfo{author}{Loth, E.}, \bibinfo{author}{Tyler~Daspit, J.},
      \bibinfo{author}{Jeong, M.}, \bibinfo{author}{Nagata, T.},
      \bibinfo{author}{Nonomura, T.}, \bibinfo{year}{2021}.
    \newblock \bibinfo{title}{Supersonic and {Hypersonic} {Drag} {Coefficients} for
      a {Sphere}}.
    \newblock \bibinfo{journal}{AIAA Journal} ,
      \bibinfo{pages}{3261--3274}\URLprefix
      \url{https://arc.aiaa.org/doi/10.2514/1.J060153},
      \DOIprefix\doi{10.2514/1.J060153}.
    \bibitem[{Mando and Rosendahl(2010)}]{Mando2010}
    \bibinfo{author}{Mando, M.}, \bibinfo{author}{Rosendahl, L.},
      \bibinfo{year}{2010}.
    \newblock \bibinfo{title}{On the motion of non-spherical particles at high
      {Reynolds} number}.
    \newblock \bibinfo{journal}{Powder Technology} \bibinfo{volume}{202},
      \bibinfo{pages}{1--13}.
    \newblock \URLprefix \url{http://dx.doi.org/10.1016/j.powtec.2010.05.001
      http://www.sciencedirect.com/science/article/pii/S0032591010002391},
      \DOIprefix\doi{10.1016/j.powtec.2010.05.001}.
    \bibitem[{Michaelides et~al.(2022)Michaelides, Sommerfeld and van
      Wachem}]{Wachem2022}
    \bibinfo{author}{Michaelides, E.E.}, \bibinfo{author}{Sommerfeld, S.},
      \bibinfo{author}{van Wachem, B.}, \bibinfo{year}{2022}.
    \newblock \bibinfo{title}{Multiphase {Flows} with {Droplets} and {Particles},
      {Third} {Edition}}.
    \newblock \bibinfo{edition}{3} ed., \bibinfo{publisher}{CRC Press},
      \bibinfo{address}{Boca Raton}.
    \newblock \DOIprefix\doi{10.1201/9781003089278}.
    \bibitem[{Nagata et~al.(2020a)Nagata, Noguchi, Nonomura, Ohtani and
      Asai}]{Nagata2020}
    \bibinfo{author}{Nagata, T.}, \bibinfo{author}{Noguchi, A.},
      \bibinfo{author}{Nonomura, T.}, \bibinfo{author}{Ohtani, K.},
      \bibinfo{author}{Asai, K.}, \bibinfo{year}{2020}a.
    \newblock \bibinfo{title}{Experimental investigation of transonic and
      supersonic flow over a sphere for {Reynolds} numbers of 103–105 by
      free-flight tests with schlieren visualization}.
    \newblock \bibinfo{journal}{Shock Waves} \bibinfo{volume}{30},
      \bibinfo{pages}{139--151}.
    \newblock \URLprefix \url{https://doi.org/10.1007/s00193-019-00924-0},
      \DOIprefix\doi{10.1007/s00193-019-00924-0}.
    \bibitem[{Nagata et~al.(2020b)Nagata, Nonomura, Takahashi and
      Fukuda}]{Nagata2020a}
    \bibinfo{author}{Nagata, T.}, \bibinfo{author}{Nonomura, T.},
      \bibinfo{author}{Takahashi, S.}, \bibinfo{author}{Fukuda, K.},
      \bibinfo{year}{2020}b.
    \newblock \bibinfo{title}{Direct numerical simulation of subsonic, transonic
      and supersonic flow over an isolated sphere up to a {Reynolds} number of
      1000}.
    \newblock \bibinfo{journal}{Journal of Fluid Mechanics} \bibinfo{volume}{904},
      \bibinfo{pages}{A36}.
    \newblock \URLprefix
      \url{https://www.cambridge.org/core/journals/journal-of-fluid-mechanics/article/direct-numerical-simulation-of-subsonic-transonic-and-supersonic-flow-over-an-isolated-sphere-up-to-a-reynolds-number-of-1000/24251E57F95CBFC00737F3EB3BB396BA},
      \DOIprefix\doi{10.1017/jfm.2020.629}. \bibinfo{note}{publisher: Cambridge
      University Press}.
    \bibitem[{Nagata et~al.(2016)Nagata, Nonomura, Takahashi, Mizuno and
      Fukuda}]{Nagata2016}
    \bibinfo{author}{Nagata, T.}, \bibinfo{author}{Nonomura, T.},
      \bibinfo{author}{Takahashi, S.}, \bibinfo{author}{Mizuno, Y.},
      \bibinfo{author}{Fukuda, K.}, \bibinfo{year}{2016}.
    \newblock \bibinfo{title}{Investigation on subsonic to supersonic flow around a
      sphere at low {Reynolds} number of between 50 and 300 by direct numerical
      simulation}.
    \newblock \bibinfo{journal}{Physics of Fluids} \bibinfo{volume}{28},
      \bibinfo{pages}{056101}.
    \newblock \URLprefix \url{http://aip.scitation.org/doi/10.1063/1.4947244},
      \DOIprefix\doi{10.1063/1.4947244}.
    \bibitem[{Nagata et~al.(2018)Nagata, Nonomura, Takahashi, Mizuno and
      Fukuda}]{Nagata2018}
    \bibinfo{author}{Nagata, T.}, \bibinfo{author}{Nonomura, T.},
      \bibinfo{author}{Takahashi, S.}, \bibinfo{author}{Mizuno, Y.},
      \bibinfo{author}{Fukuda, K.}, \bibinfo{year}{2018}.
    \newblock \bibinfo{title}{Direct numerical simulation of flow past a
      transversely rotating sphere up to a {Reynolds} number of 300 in compressible
      flow}.
    \newblock \bibinfo{journal}{Journal of Fluid Mechanics} \bibinfo{volume}{857},
      \bibinfo{pages}{878--906}.
    \newblock \URLprefix
      \url{https://www.cambridge.org/core/product/identifier/S0022112018007565/type/journal_article},
      \DOIprefix\doi{10.1017/jfm.2018.756}.
    \bibitem[{Ouchene(2020)}]{Ouchene2020}
    \bibinfo{author}{Ouchene, R.}, \bibinfo{year}{2020}.
    \newblock \bibinfo{title}{Numerical simulation and modeling of the hydrodynamic
      forces and torque acting on individual oblate spheroids}.
    \newblock \bibinfo{journal}{Physics of Fluids} \bibinfo{volume}{32},
      \bibinfo{pages}{073303}.
    \newblock \URLprefix \url{https://pubs.aip.org/aip/pof/article/1060398},
      \DOIprefix\doi{10.1063/5.0011618}.
    \bibitem[{Ouchene et~al.(2016)Ouchene, Khalij, Arcen and
      Tanière}]{Ouchene2016}
    \bibinfo{author}{Ouchene, R.}, \bibinfo{author}{Khalij, M.},
      \bibinfo{author}{Arcen, B.}, \bibinfo{author}{Tanière, A.},
      \bibinfo{year}{2016}.
    \newblock \bibinfo{title}{A new set of correlations of drag, lift and torque
      coefficients for non-spherical particles and large {Reynolds} numbers}.
    \newblock \bibinfo{journal}{Powder Technology} \bibinfo{volume}{303},
      \bibinfo{pages}{33--43}.
    \newblock \URLprefix
      \url{http://linkinghub.elsevier.com/retrieve/pii/S0032591016304594},
      \DOIprefix\doi{10.1016/j.powtec.2016.07.067}.
    \bibitem[{Ouchene et~al.(2015)Ouchene, Khalij, Tanière and
      Arcen}]{Ouchene2015}
    \bibinfo{author}{Ouchene, R.}, \bibinfo{author}{Khalij, M.},
      \bibinfo{author}{Tanière, A.}, \bibinfo{author}{Arcen, B.},
      \bibinfo{year}{2015}.
    \newblock \bibinfo{title}{Drag, lift and torque coefficients for ellipsoidal
      particles: {From} low to moderate particle {Reynolds} numbers}.
    \newblock \bibinfo{journal}{Computers \& Fluids} \bibinfo{volume}{113},
      \bibinfo{pages}{53--64}.
    \newblock \URLprefix
      \url{http://linkinghub.elsevier.com/retrieve/pii/S004579301400468X},
      \DOIprefix\doi{10.1016/j.compfluid.2014.12.005}.
    \bibitem[{Parmar et~al.(2010)Parmar, Haselbacher and Balachandar}]{Parmar2010}
    \bibinfo{author}{Parmar, M.}, \bibinfo{author}{Haselbacher, A.},
      \bibinfo{author}{Balachandar, S.}, \bibinfo{year}{2010}.
    \newblock \bibinfo{title}{Improved {Drag} {Correlation} for {Spheres} and
      {Application} to {Shock}-{Tube} {Experiments}}.
    \newblock \bibinfo{journal}{AIAA Journal} \bibinfo{volume}{48},
      \bibinfo{pages}{1273--1276}.
    \newblock \URLprefix \url{https://arc.aiaa.org/doi/10.2514/1.J050161},
      \DOIprefix\doi{10.2514/1.J050161}.
    \bibitem[{Roos and Willmarth(1971)}]{Roos1971}
    \bibinfo{author}{Roos, F.W.}, \bibinfo{author}{Willmarth, W.W.},
      \bibinfo{year}{1971}.
    \newblock \bibinfo{title}{Some experimental results on sphere and disk drag}.
    \newblock \bibinfo{journal}{AIAA Journal} \bibinfo{volume}{9},
      \bibinfo{pages}{285--291}.
    \newblock \URLprefix \url{https://arc.aiaa.org/doi/10.2514/3.6164},
      \DOIprefix\doi{10.2514/3.6164}.
    \bibitem[{Rosendahl(2000)}]{Rosendahl2000}
    \bibinfo{author}{Rosendahl, L.}, \bibinfo{year}{2000}.
    \newblock \bibinfo{title}{Using a multi-parameter particle shape description to
      predict the motion of non-spherical particle shapes in swirling flow}.
    \newblock \bibinfo{journal}{Applied Mathematical Modelling}
      \bibinfo{volume}{24}, \bibinfo{pages}{11--25}.
    \newblock \URLprefix
      \url{http://linkinghub.elsevier.com/retrieve/pii/S0307904X99000232},
      \DOIprefix\doi{10.1016/S0307-904X(99)00023-2}.
    \bibitem[{Salman and Verba(1988)}]{Salman1988}
    \bibinfo{author}{Salman, A.D.}, \bibinfo{author}{Verba, A.},
      \bibinfo{year}{1988}.
    \newblock \bibinfo{title}{New aproximate equations to estimate the drag
      coefficient of different particles of regular shape}.
    \newblock \bibinfo{journal}{Periodica Polytechnica Chemical Engineering}
      \bibinfo{volume}{32}, \bibinfo{pages}{261--276}.
    \newblock \URLprefix \url{https://pp.bme.hu/ch/article/view/2794}.
      \bibinfo{note}{number: 4}.
    \bibitem[{Sanjeevi et~al.(2022)Sanjeevi, Dietiker and Padding}]{Sanjeevi2022}
    \bibinfo{author}{Sanjeevi, S.K.P.}, \bibinfo{author}{Dietiker, J.F.},
      \bibinfo{author}{Padding, J.T.}, \bibinfo{year}{2022}.
    \newblock \bibinfo{title}{Accurate hydrodynamic force and torque correlations
      for prolate spheroids from {Stokes} regime to high {Reynolds} numbers}.
    \newblock \bibinfo{journal}{Chemical Engineering Journal}
      \bibinfo{volume}{444}, \bibinfo{pages}{136325}.
    \newblock \URLprefix
      \url{https://www.sciencedirect.com/science/article/pii/S1385894722018204},
      \DOIprefix\doi{10.1016/j.cej.2022.136325}.
    \bibitem[{Sanjeevi et~al.(2018)Sanjeevi, Kuipers and Padding}]{Sanjeevi2018}
    \bibinfo{author}{Sanjeevi, S.K.P.}, \bibinfo{author}{Kuipers, J.A.M.},
      \bibinfo{author}{Padding, J.T.}, \bibinfo{year}{2018}.
    \newblock \bibinfo{title}{Drag, lift and torque correlations for non-spherical
      particles from {Stokes} limit to high {Reynolds} numbers}.
    \newblock \bibinfo{journal}{International Journal of Multiphase Flow}
      \bibinfo{volume}{106}, \bibinfo{pages}{325--337}.
    \newblock \URLprefix
      \url{http://www.sciencedirect.com/science/article/pii/S0301932217307851},
      \DOIprefix\doi{10.1016/j.ijmultiphaseflow.2018.05.011}.
    \bibitem[{Short(1967)}]{Short1967}
    \bibinfo{author}{Short, B.J.}, \bibinfo{year}{1967}.
    \newblock \bibinfo{title}{Dynamic flight behavior of a ballasted sphere at
      {Mach} numbers from 0.4 to 14.5}.
    \newblock \bibinfo{type}{Technical Report} \bibinfo{number}{NASA-TN-D-4198}.
      NASA Ames Research Center.
    \newblock \URLprefix \url{https://ntrs.nasa.gov/citations/19670029052}.
      \bibinfo{note}{nTRS Author Affiliations: NASA Ames Research Center NTRS
      Document ID: 19670029052 NTRS Research Center: Legacy CDMS (CDMS)}.
    \bibitem[{Singh et~al.(2022)Singh, Kroells, Li, Ching, Ihme, Hogan and
      Schwartzentruber}]{Singh2022}
    \bibinfo{author}{Singh, N.}, \bibinfo{author}{Kroells, M.},
      \bibinfo{author}{Li, C.}, \bibinfo{author}{Ching, E.}, \bibinfo{author}{Ihme,
      M.}, \bibinfo{author}{Hogan, C.J.}, \bibinfo{author}{Schwartzentruber, T.E.},
      \bibinfo{year}{2022}.
    \newblock \bibinfo{title}{General {Drag} {Coefficient} for {Flow} over
      {Spherical} {Particles}}.
    \newblock \bibinfo{journal}{AIAA Journal} \bibinfo{volume}{60},
      \bibinfo{pages}{587--597}.
    \newblock \URLprefix \url{https://doi.org/10.2514/1.J060648},
      \DOIprefix\doi{10.2514/1.J060648}. \bibinfo{note}{publisher: American
      Institute of Aeronautics and Astronautics \_eprint:
      https://doi.org/10.2514/1.J060648}.
    \bibitem[{Spearman and Braswell(1993)}]{Spearman1993}
    \bibinfo{author}{Spearman, M.L.}, \bibinfo{author}{Braswell, D.O.},
      \bibinfo{year}{1993}.
    \newblock \bibinfo{title}{Aerodynamics of a sphere and an oblate spheroid for
      {Mach} numbers from 0.6 to 10.5 including some effects of test conditions}.
    \newblock \bibinfo{type}{Technical Report} \bibinfo{number}{NAS 1.15:109016}.
      NASA Langley Research Center.
    \newblock \URLprefix \url{https://ntrs.nasa.gov/citations/19940008699}.
    \bibitem[{Stokes(1851)}]{Stokes1851}
    \bibinfo{author}{Stokes, G.G.}, \bibinfo{year}{1851}.
    \newblock \bibinfo{title}{On the effect of the internal friction of fluids on
      the motion of pendulums}.
    \newblock \bibinfo{journal}{Transactions of the Cambridge Philosophical
      Society} \bibinfo{volume}{9}, \bibinfo{pages}{8}.
    \bibitem[{Tenneti et~al.(2011)Tenneti, Garg and Subramaniam}]{Tenneti2011}
    \bibinfo{author}{Tenneti, S.}, \bibinfo{author}{Garg, R.},
      \bibinfo{author}{Subramaniam, S.}, \bibinfo{year}{2011}.
    \newblock \bibinfo{title}{Drag law for monodisperse gas–solid systems using
      particle-resolved direct numerical simulation of flow past fixed assemblies
      of spheres}.
    \newblock \bibinfo{journal}{International Journal of Multiphase Flow}
      \bibinfo{volume}{37}, \bibinfo{pages}{1072--1092}.
    \newblock \URLprefix
      \url{http://www.sciencedirect.com/science/article/pii/S0301932211001170},
      \DOIprefix\doi{10.1016/j.ijmultiphaseflow.2011.05.010}.
    \bibitem[{Theofanous et~al.(2018)Theofanous, Mitkin and Chang}]{Theofanous2018}
    \bibinfo{author}{Theofanous, T.G.}, \bibinfo{author}{Mitkin, V.},
      \bibinfo{author}{Chang, C.H.}, \bibinfo{year}{2018}.
    \newblock \bibinfo{title}{Shock dispersal of dilute particle clouds}.
    \newblock \bibinfo{journal}{Journal of Fluid Mechanics} \bibinfo{volume}{841},
      \bibinfo{pages}{732--745}.
    \newblock \URLprefix
      \url{https://www.cambridge.org/core/journals/journal-of-fluid-mechanics/article/shock-dispersal-of-dilute-particle-clouds/FBE48DECAA77FE8A0DF13883837D9FB3},
      \DOIprefix\doi{10.1017/jfm.2018.110}. \bibinfo{note}{publisher: Cambridge
      University Press}.
    \bibitem[{Villafuerte(2015)}]{Villafuerte2015}
    \bibinfo{editor}{Villafuerte, J.} (Ed.), \bibinfo{year}{2015}.
    \newblock \bibinfo{title}{Modern {Cold} {Spray}}.
    \newblock \bibinfo{publisher}{Springer International Publishing},
      \bibinfo{address}{Cham}.
    \newblock \URLprefix \url{http://link.springer.com/10.1007/978-3-319-16772-5},
      \DOIprefix\doi{10.1007/978-3-319-16772-5}.
    \bibitem[{van Wachem et~al.(2015)van Wachem, Zastawny, Zhao and
      Mallouppas}]{vanWachem2015}
    \bibinfo{author}{van Wachem, B.}, \bibinfo{author}{Zastawny, M.},
      \bibinfo{author}{Zhao, F.}, \bibinfo{author}{Mallouppas, G.},
      \bibinfo{year}{2015}.
    \newblock \bibinfo{title}{Modelling of gas–solid turbulent channel flow with
      non-spherical particles with large {Stokes} numbers}.
    \newblock \bibinfo{journal}{International Journal of Multiphase Flow}
      \bibinfo{volume}{68}, \bibinfo{pages}{80--92}.
    \newblock \URLprefix
      \url{http://linkinghub.elsevier.com/retrieve/pii/S0301932214001931},
      \DOIprefix\doi{10.1016/j.ijmultiphaseflow.2014.10.006}.
    \bibitem[{Wadell(1934)}]{Wadell1934}
    \bibinfo{author}{Wadell, H.}, \bibinfo{year}{1934}.
    \newblock \bibinfo{title}{The coefficient of resistance as a function of
      {Reynolds} number for solids of various shapes}.
    \newblock \bibinfo{journal}{Journal of the Franklin Institute}
      \bibinfo{volume}{April}, \bibinfo{pages}{459--490}.
    \newblock \URLprefix
      \url{http://www.sciencedirect.com/science/article/pii/S0016003234905081}.
    \bibitem[{Xiao et~al.(2017)Xiao, Denner and van Wachem}]{Xiao2017}
    \bibinfo{author}{Xiao, C.N.}, \bibinfo{author}{Denner, F.},
      \bibinfo{author}{van Wachem, B.}, \bibinfo{year}{2017}.
    \newblock \bibinfo{title}{Fully-coupled pressure-based finite-volume framework
      for the simulation of fluid flows at all speeds in complex geometries}.
    \newblock \bibinfo{journal}{Journal of Computational Physics}
      \bibinfo{volume}{346}, \bibinfo{pages}{91--130}.
    \newblock \URLprefix
      \url{http://linkinghub.elsevier.com/retrieve/pii/S0021999117304540},
      \DOIprefix\doi{10.1016/j.jcp.2017.06.009}.
    \bibitem[{Zastawny et~al.(2012)Zastawny, Mallouppas, Zhao and van
      Wachem}]{Zastawny2012}
    \bibinfo{author}{Zastawny, M.}, \bibinfo{author}{Mallouppas, G.},
      \bibinfo{author}{Zhao, F.}, \bibinfo{author}{van Wachem, B.},
      \bibinfo{year}{2012}.
    \newblock \bibinfo{title}{Modelling of gas-solid turbulent flows with
      non-spherical particles}, in: \bibinfo{booktitle}{8th {International}
      {Conference} on {Multiphase} {Flow} ({ICMF} 2013)},
      \bibinfo{publisher}{Elsevier Ltd}. pp. \bibinfo{pages}{227--239}.
    \newblock \URLprefix
      \url{http://linkinghub.elsevier.com/retrieve/pii/S0301932211002047},
      \DOIprefix\doi{10.1016/j.ijmultiphaseflow.2011.09.004}.
    
    \end{thebibliography}

\end{document}